\documentclass[11pt]{article}
\usepackage{amsmath}
\usepackage{amssymb}
\usepackage{times}
\usepackage{color}
\usepackage{bm}
\usepackage{mathrsfs}
\usepackage{epsfig,graphics}
\usepackage{epstopdf}
\usepackage[breaklinks]{hyperref}
\hypersetup{colorlinks, linkcolor=blue, filecolor=blue, urlcolor=blue, citecolor=blue}

\begin{document}

\pagestyle{plain}
\hsize = 6.5 in
\vsize = 8.5 in
\hoffset = -0.75 in
\voffset = -0.5 in
\baselineskip = 0.29 in

\newcommand{\DF}[2]{{\displaystyle\frac{#1}{#2}}}

\title{Potential landscape and flux field theory for turbulence and nonequilibrium fluid systems\footnote{This manuscript has been accepted for publication on \emph{Annals of Physics}, {\bf 389}, 63-101 (2018). DOI: \href{https://doi.org/10.1016/j.aop.2017.12.001}{10.1016/j.aop.2017.12.001}} $^{,}$ \footnote{\copyright 2017. This manuscript version is made available under the CC-BY-NC-ND 4.0 license $\hspace{135pt}$ \href{http://creativecommons.org/licenses/by-nc-nd/4.0/}{http://creativecommons.org/licenses/by-nc-nd/4.0/}}}
\author{Wei Wu$^1$, Feng Zhang$^1$ and Jin Wang$^{1, 2, }\footnote{Email: jin.wang.1@stonybrook.edu}$\\[15pt]
$^1$\small{State Key Laboratory of Electroanalytical Chemistry}\\
\small{Changchun Institute of Applied Chemistry, Chinese Academy of Sciences}\\
\small{Changchun, 130022, China}\\[10pt]
$^2$\small{Department of Physics and Department of Chemistry}\\
\small{State University of New York at Stony Brook}\\
\small{Stony Brook, NY 11794, USA}\\[10pt]}

\maketitle

\begin{abstract}

Turbulence is a paradigm for far-from-equilibrium systems without time reversal symmetry. To capture the nonequilibrium irreversible nature of turbulence and investigate its implications, we develop a potential landscape and flux field theory for turbulent flow and more general nonequilibrium fluid systems governed by stochastic Navier-Stokes equations. We find that equilibrium fluid systems with time reversibility are characterized by a detailed balance constraint that quantifies the detailed balance condition. In nonequilibrium fluid systems with nonequilibrium steady states, detailed balance breaking leads directly to a pair of interconnected consequences, namely, the non-Gaussian potential landscape and the irreversible probability flux, forming a `nonequilibrium trinity'. The nonequilibrium trinity characterizes the nonequilibrium irreversible essence of fluid systems with intrinsic time irreversibility and is manifested in various aspects of these systems. The nonequilibrium stochastic dynamics of fluid systems including turbulence with detailed balance breaking is shown to be driven by both the non-Gaussian potential landscape gradient and the irreversible probability flux, together with the reversible convective force and the stochastic stirring force. We reveal an underlying connection of the energy flux essential for turbulence energy cascade to the irreversible probability flux and the non-Gaussian potential landscape generated by detailed balance breaking. Using the energy flux as a center of connection, we demonstrate that the four-fifths law in fully developed turbulence is a consequence and reflection of the nonequilibrium trinity. We also show how the nonequilibrium trinity can affect the scaling laws in turbulence.

\vskip 0.3cm


\end{abstract}





\section{Introduction}

The true nature of turbulence remains elusive despite more than a hundred years of devotion of countless geniuses since the pioneering work of Osborne Reynolds who investigated experimentally the transition from laminar to turbulent flow \cite{Reynolds}. The energy cascade picture in turbulent flows proposed by Richardson  \cite{energycascade} is arguably the most important physical picture of turbulence, in which energy flows from the large scale where energy is injected, transferred through the intermediate scales by the nonlinear convective force,  down to the small scale where energy is dissipated by molecular viscosity. The modern viewpoint of turbulence started from Kolmogorov's ground-breaking theory of scaling laws \cite{Kolmogorov1941a,Kolmogorov4/5} based on the hypotheses of universality and self-similarity \cite{Landau,Kolmogorov1962} to quantify the energy cascade process. Later experimental observations demonstrated that self-similarity of the energy cascade in turbulence is broken due to the intermittency phenomena \cite{Batchelor1949}, which has inspired intensive investigations on this subject \cite{Legacy,Sreenivasan1997}. Now the modern turbulence research has developed into a vast field with a variety of branches.

As is well known, turbulence is a far-from-equilibrium phenomenon without time reversal symmetry \cite{TurbulenceIrreversibility1,TurbulenceIrreversibility2}. The nonequilibrium irreversible character of turbulence plays an important role in various facets of the turbulence phenomenon. In particular, the energy cascade process, with a directional flow of energy through scales, is a manifestation and reflection of the nonequilibrium irreversible nature of turbulence. It is therefore natural to approach the turbulence problem from the perspective of nonequilibrium statistical mechanics \cite{FDT0,FDT1,NSMTurbulence1,NSMTurbulence2,NSMTurbulence3}. The statistical properties of turbulence in connection with the deviation from equilibrium has been investigated from the angles of fluctuation-dissipation theorem (FDT) \cite{FDT2,FDT3,FDT4}, fluctuation theorem \cite{FT,FT1,FT2,FT3}, large deviation theory \cite{LDT1,LDT2,LDT3}, and time asymmetry in Lagrangian statistics \cite{TimeAsymmetry1,Xu2014,TimeAsymmetry2,TimeAsymmetry3,TimeAsymmetry4} among others.

However, a precise characterization of the nonequilibrium steady state with intrinsic time irreversibility for turbulent fluids governed by stochastic Navier-Stokes equations, based on the concept of detailed balance and its violation (i.e., detailed balance breaking) in stochastic dynamical systems \cite{Risken,Gardiner,Zia}, is still lacking to the best knowledge of the authors. In this respect the work on time asymmetry in Lagrangian statistics is relevant, which studied the manifestation of detailed balance breaking in the motion of fluid particles in the turbulent flow. The objective of the present work is to develop theoretically a systematic characterization of the intrinsic nonequilibrium irreversible nature of turbulent flow as a whole and investigate its manifestation in some major aspects of the turbulence phenomenon, within the framework of the potential landscape and flux field theory.

The potential landscape and flux field theory, a generalization of the potential landscape and flux framework to spatially extended systems (fields), is a theoretical framework that belongs to the larger field of stochastic approaches to nonequilibrium statistical mechanics. It is particularly suited for the study of the global dynamics and nonequilibrium thermodynamics of stochastic field systems governed by the Langevin and Fokker-Planck field equations \cite{WWJW2013,WWJW20141,WWJW20142}. The  potential landscape and flux framework, which has its historical origin in the energy landscape theory in protein folding dynamics, was initially developed for nonequilibrium biological systems and has been applied extensively in that area and beyond \cite{PotentialFluxOriginal,PFBiologyReview,JW2015AP,Chao}. Compared with the energy landscape theory, the potential landscape and flux framework places more emphasis on the essential role played by the probability flux that signifies detailed balance breaking in nonequilibrium steady states, within the discovery of a dual potential-flux form of the driving force that determines the underlying nonequilibrium irreversible dynamics \cite{PotentialFluxOriginal}. Based on the potential-flux form of the driving force, the global dynamics and nonequilibrium  thermodynamics for open stochastic dynamical systems have been quantified in the potential landscape and flux framework \cite{PFBiologyReview,JW2015AP}.

To provide sufficient context, we elaborate more on the connection of the potential landscape and flux framework to the concept of detailed balance and, more importantly, to its violation, detailed balance breaking. Historically, the concept of detailed balance originated from statistical physics and was first formulated by Ludwig Boltzmann in proving his famous H-theorem in the kinetic theory of gas. In general, the principle of detailed balance states that at the equilibrium state each elementary process is balanced by its time reversed process. It is a reflection of the microscopic time reversibility that characterizes the equilibrium nature of the steady state. For stochastic dynamical systems governed by Langevin and Fokker-Planck equations, the detailed balance condition takes on more explicit forms \cite{Risken,Gardiner}. The essential feature of detailed balance is the vanishing of the steady-state irreversible probability flux. A direct consequence of the vanishing steady-state irreversible probability flux is that the irreversible driving force of the system has a potential gradient form in the state space, where the potential landscape $\Phi$ is defined in terms of the steady-state probability distribution $P_s$ as $\Phi=-\ln P_s$ or $P_s=e^{-\Phi}$.

What becomes more interesting is when the steady state of the system is a nonequilibrium state, for which the detailed balance condition is violated. Nonequilibrium steady states can be maintained by open systems that constantly exchange matter, energy or information with its environment, a typical scenario for dissipative structures and living organisms. A distinguishing feature of these systems is the presence of nonvanishing steady-state flux of matter, energy or information, which is reflected on the dynamical level by the nonvanishing steady-state irreversible probability flux that signifies the violation of detailed balance. In other words, detailed balance breaking characterizes the time irreversibility in nonequilibrium steady states, indicated by the steady-state irreversible probability flux associated with the steady-state flux of matter, energy or information. It has become increasingly clear that the steady-state probability flux plays an indispensable role in characterizing nonequilibrium steady states \cite{Risken,Gardiner,Zia,PotentialFluxOriginal}. The nonvanishing irreversible probability flux that signifies detailed balance breaking leads to a dual potential-flux form of the driving force for nonequilibrium systems, where the irreversible driving force has an additional contribution from the irreversible probability flux besides the gradient of the potential landscape in the state space. This dual potential-flux form of the driving force provides the basis for the study of global dynamics and nonequilibrium thermodynamics in the potential landscape and flux framework.

The potential landscape and flux field theory has extended the range of application of the potential landscape and flux framework to spatially extended systems (fields) and deepened the understanding of certain aspects in the theoretical framework \cite{WWJW2013,WWJW20141,WWJW20142}. The global dynamics of nonequilibrium spatially extended systems has been investigated on the basis of the potential-flux form of the nonequilibrium driving force \cite{WWJW20141}. A set of nonequilibrium thermodynamic equations applicable to both spatially homogeneous and inhomogeneous systems has been established within this framework \cite{WWJW20142}, which has synthesized and generalized much of the work based on a stochastic approach to nonequilibrium thermodynamics.

It may be a legitimate question why this theoretical framework is relevant to and suitable for the study of fluid systems. A possible misconception is that since the dynamics for incompressible fluids, which we shall focus on in this article, is fundamentally vortical (solenoidal), a theory with `potential' as one of its major components (the other component is the `flux') may not be relevant. The first point that has to be understood is that the `potential' in this theoretical framework refers to a potential defined in the \emph{state space} (i.e. phase space) rather than in the physical space (unless the system state is simply the physical position). For incompressible fluid  systems the state space is the space of velocity field configurations $\mathbf{u}(\mathbf{x})$, which is a function space. The potential landscape in the state space, $\Phi[\mathbf{u}]$, is a functional of the velocity field. This is not a potential in the physical space and thus it does not exclude solenoidal dynamics in the physical space for incompressible fluids. We caution readers not to confuse properties in the state space with those in the physical space.

Furthermore, in addition to the `potential' component in this theoretical framework, the `flux' component signifying detailed balance breaking makes the theory particularly suited for the investigation of spatially extended systems with nonequilibrium steady states, including fluid systems. The fluid systems considered in this article are not isolated systems as they exchange energy with the environments (internal and external environments) in the form of energy injection and energy dissipation. This energy exchange with the environments in general allows the fluid system to sustain a nonequilibrium steady state. Turbulent fluid systems with forcing are paradigmatic of the nonequilibrium steady state scenario. They can be naturally approached in the potential landscape and flux field theoretical framework.

However, we must stress that the present work is not just a direct application of the previous potential landscape and flux field framework to fluid systems. This work extends the previous theoretical framework in at least two aspects. One aspect is the inclusion of odd variable under time reversal (the velocity field) \cite{Risken,Gardiner,EvenOdd2,QianHong,EvenOdd1}, which was not considered in the previous framework and has to be dealt with in fluid systems. This leads to the distinction of reversible and irreversible driving forces and probability fluxes that require explicit consideration of their time reversal properties, which was not necessary in the previous framework that deals with only even variables. The other aspect is the concept of nonequilibrium trinity (detailed balance breaking, non-Gaussian potential landscape, and irreversible probability flux) born in this work, which we propose to be a proper characterization of the nonequilibrium irreversible essence of fluid systems with nonequilibrium steady states (forced turbulence in particular). We speculate that the concept of nonequilibrium trinity is not limited to fluid systems and can be extended, with necessary modifications, to more general nonequilibrium stochastic dynamical systems.

The rest of this article is structured as follows. In Section $2$ we lay out the deterministic and stochastic field dynamics for incompressible fluids governed by Navier-Stokes equations. Particular attention will be given to the time reversal properties of the driving forces and the probability fluxes in the dynamical equations. In Section $3$ we establish the detailed balance constraint for equilibrium fluid systems and the nonequilibrium trinity for nonequilibrium fluid systems. Their implications on the structure of the stochastic fluid dynamics are also discussed. In particular, we show that the nonequilibrium stochastic fluid dynamics including turbulence with detailed balance breaking is driven by both the non-Gaussian potential landscape gradient and the irreversible probability flux, together with the reversible convective force and the stochastic stirring force. In Section $4$ we investigate energy balance, energy cascade and turbulence in the context of the potential landscape and flux field theory. The connections of the nonequilibrium trinity to the energy flux associated with turbulence energy cascade, the four-fifths law for fully developed turbulence and the scalings laws are revealed. Finally, the conclusion is given in Section $5$.

\section{Field dynamics for fluid systems}

We consider incompressible fluids in the three dimensional physical space. To avoid complications at the boundary in theoretical analysis, we consider fluids confined in a cubic box with side $L$ satisfying periodic boundary conditions. Equivalently, the fluid system is defined in a three dimensional torus $\mathbf{T}^3$. The large system size limit $L\rightarrow \infty$ will also be considered at a later stage, where the system is defined in the entire 3D physical space. It is worth noting that although periodic boundary conditions in the \emph{state space} can accommodate nonequilibrium steady states driven by a nonconservative force \cite{PeriodicBoundary}, the periodic boundary conditions here are applied in the \emph{physical space} and thus not directly related to the sustainment of nonequilibrium steady states.

\subsection{Deterministic field dynamics for fluid systems}

Consider an incompressible fluid with constant density (set to unity) without the influence of external forces, governed by the Navier-Stokes equation:
\begin{equation}\label{NSE}
\partial_t \mathbf{u}+\mathbf{u}\cdot\nabla\mathbf{u}=\nu \Delta\mathbf{u}-\nabla p,
\end{equation}
with the incompressibility condition
\begin{equation}\label{Incomp}
\nabla\cdot\mathbf{u}=0.
\end{equation}
Here $\mathbf{u}=\mathbf{u}(\mathbf{x},t)$ is the velocity field of the fluid at time $t$. Its time rate of change $\partial_t \mathbf{u}$ is determined by the forces in the fluid system. Conforming to the convention that forces are on the r.h.s. of the dynamical equation, we identify $-\mathbf{u}\cdot\nabla\mathbf{u}$ (notice the negative sign) as the nonlinear convective force, $\nu \Delta\mathbf{u}$ as the viscous force, and $-\nabla p$ as the pressure force. These `forces' have the dimension of acceleration as well as force density since the mass density has been set to the dimensionless unity.

The pressure $p$ is not independent of the velocity field $\mathbf{u}$. They are related by $\Delta p=-\nabla\cdot(\mathbf{u}\cdot\nabla\mathbf{u})$ from Eqs. (\ref{NSE}) and (\ref{Incomp}). This Poisson equation for $p$ in $\mathbf{T}^3$ can be solved with Fourier analysis. As a result, the pressure force can be expressed as $-\nabla p=\mathbf{\Pi}^{g}(\nabla)\cdot(\mathbf{u}\cdot\nabla\mathbf{u})$, where $\mathbf{\Pi}^{g}(\nabla)$ is the gradient projection operator in $\mathbf{T}^3$ whose action on a vector field produces its gradient (irrotational) component. Heuristically, $\mathbf{\Pi}^{g}(\nabla)\simeq\nabla\Delta^{-1}\nabla$. However, $\Delta$ is not exactly invertible under periodic boundary conditions as it has a zero eigen-value. The precise definition of $\mathbf{\Pi}^{g}(\nabla)$ is as follows.  For a vector field $\mathbf{v}(\mathbf{x})$ in $\mathbf{T}^3$, $\mathbf{\Pi}^{g}(\nabla)\cdot\mathbf{v}(\mathbf{x})=\int \mathbf{G}^g(\mathbf{x}-\mathbf{x}')\cdot\mathbf{v}(\mathbf{x}')d\mathbf{x}'$, where the matrix-valued integral kernel $\mathbf{G}^g(\mathbf{x}-\mathbf{x}')=\sum'_{\mathbf{k}}\left(\mathbf{k}\mathbf{k}/k^2\right)e^{i\mathbf{k}\cdot(\mathbf{x}-\mathbf{x}')}/L^3$. Here the wavevector $\mathbf{k}=2\pi\mathbf{n}/L$ for $\mathbf{n}$ with integer components, $k=|\mathbf{k}|$ and the sum $\sum'_{\mathbf{k}}$ excludes $\mathbf{k}=0$. The property $\nabla \times \mathbf{G}^g(\mathbf{x}-\mathbf{x}')=0$, as is easily verified, ensures that the projected vector field is irrotational.

The explicit expression of the pressure force, $-\nabla p=\int \mathbf{G}^g(\mathbf{x}-\mathbf{x}')\cdot[\mathbf{u}(\mathbf{x}')\cdot\nabla'\mathbf{u}(\mathbf{x}')]d\mathbf{x}'$, shows that it is a nonlocal force in the physical space, since it involves a spatial integral of the velocity field. This nonlocality in the physical space stems from the idealized condition that the fluid is incompressible, under which the velocity variation in one location instantaneously impacts the whole velocity field.
Moreover, $-\nabla p=\mathbf{\Pi}^{g}(\nabla)\cdot(\mathbf{u}\cdot\nabla\mathbf{u})$ means the pressure force counterbalances the gradient component of the nonlinear convective force, leaving only its solenoidal component, as a result of the incompressibility of the fluid.

Plugging the expression of the pressure force into Eq.(\ref{NSE}), we arrive at the Navier-Stokes equation in the solenoidal form \cite{McComb,Navier-StokesFunctionalAnalysis}:
\begin{equation}\label{NSESF}
\partial_t \mathbf{u}=\mathbf{\Pi}^{s}(\nabla)\cdot\left(-\mathbf{u}\cdot\nabla\mathbf{u}\right)+\nu \Delta\mathbf{u},
\end{equation}
where $\mathbf{\Pi}^{s}(\nabla)\cdot\left(-\mathbf{u}\cdot\nabla\mathbf{u}\right)=-\mathbf{u}\cdot\nabla\mathbf{u}-\nabla p$ is the solenoidal convective force which represents the combined effect of the convective force and the pressure force as a consequence of the incompressibility of the fluid. $\mathbf{\Pi}^{s}(\nabla)$ is the solenoidal projection operator in $\mathbf{T}^3$ whose action on a vector field gives its solenoidal component. Heuristically, $\mathbf{\Pi}^{s}(\nabla)\simeq \mathbf{I}-\nabla\Delta^{-1}\nabla$ where  $\mathbf{I}$ is the $3\times 3$ identity matrix. More precisely, for a vector field $\mathbf{v}(\mathbf{x})$ in $\mathbf{T}^3$, $\mathbf{\Pi}^{s}(\nabla)\cdot\mathbf{v}(\mathbf{x})=\int \mathbf{G}^s(\mathbf{x}-\mathbf{x}')\cdot\mathbf{v}(\mathbf{x}')d\mathbf{x}'$, where $\mathbf{G}^s(\mathbf{x}-\mathbf{x}')=\sum'_{\mathbf{k}}\left(\mathbf{I}-\mathbf{k}\mathbf{k}/k^2\right)e^{i\mathbf{k}\cdot(\mathbf{x}-\mathbf{x}')}/L^3$ and has the property $\nabla\cdot \mathbf{G}^s(\mathbf{x}-\mathbf{x}')=0$. Note that due to the nontrivial topology of the torus $\mathbf{T}^3$, related to the presence of a zero eigenvalue of $\Delta$ under periodic boundary conditions, a third projection operator $\mathbf{\Pi}^{0}(\nabla)$ is needed to complete the projection operators in $\mathbf{T}^3$, which is defined by $\mathbf{\Pi}^{0}(\nabla)\cdot\mathbf{v}(\mathbf{x})=\int \mathbf{v}(\mathbf{x}) d\mathbf{x}/L^3$. For the convective force, this last component vanishes.

The solenoidal Navier-Stokes equation in Eq. (\ref{NSESF}) has some properties that make it a more convenient starting point for further treatment. For instance, it preserves the incompressibility condition $\nabla\cdot\mathbf{u}=0$, which means the velocity field will remain solenoidal if it is initially so. Moreover, as there is no external force, the total momentum $\int \mathbf{u}(\mathbf{x})d\mathbf{x}$ is also conserved by the dynamics, which can be brought to zero through a Galilean transformation. Thus we only need to consider solenoidal velocity fields with zero total momentum in $\mathbf{T}^3$ (satisfying periodic boundary conditions). The state of the fluid system at each moment is described by such a velocity field. The collection of these velocity fields (subject to some technical conditions \cite{Navier-StokesFunctionalAnalysis}) form a function space, which is the state space of the fluid system denoted as $\Omega$.

The state of the fluid system evolves with time as a result of the driving force governing the dynamics of the system. The driving force field can be identified from Eq. (\ref{NSESF}) as $\mathbf{F}(\mathbf{x})[\mathbf{u}]= \mathbf{\Pi}^{s}(\nabla)\cdot\left(-\mathbf{u}\cdot\nabla\mathbf{u}\right)+\nu \Delta\mathbf{u}$, which consists of the solenoidal convective force and the viscous force. We denote the convective force as $\mathbf{F}^{con}(\mathbf{x})[\mathbf{u}]=-\mathbf{u}\cdot\nabla\mathbf{u}$, the solenoidal convective force as $\mathbf{F}^{scon}(\mathbf{x})[\mathbf{u}]=\mathbf{\Pi}^{s}(\nabla)\cdot\left(-\mathbf{u}\cdot\nabla\mathbf{u}\right)$, and the viscous force as $\mathbf{F}^{vis}(\mathbf{x})[\mathbf{u}]=\nu \Delta\mathbf{u}$.

An important difference between the solenoidal convective force and the viscous force is that the former is reversible while the latter is irreversible. This difference is demonstrated in their different behaviors under the time reversal transformation $t\rightarrow -t$  in relation to the solenoidal Navier-Stokes equation in Eq. (\ref{NSESF}). We know that the velocity $\mathbf{u}$ changes sign under time reversal (i.e., it is odd or has odd parity with respect to time reversal). Thus $\partial_t \mathbf{u}$ on the l.h.s. of the equation is even under time reversal since both $\mathbf{u}$ and $\partial_t$ change sign. On the r.h.s. of the equation, the solenoidal convective force is also even under time reversal (same as $\partial_t \mathbf{u}$) as it is quadratic in $\mathbf{u}$. Yet the viscous force is odd under time reversal (opposite to $\partial_t \mathbf{u}$) as it is linear in $\mathbf{u}$. Therefore, the part of the dynamical equation associated with the solenoidal convective force remains unchanged when time is reversed, while the part associated with the viscous force changes sign. This demonstrates that the solenoidal convective force is reversible while the viscous force is irreversible, in agreement with the intuitive understanding of their different physical natures (the solenoidal convective force is conservative while the viscous force is dissipative). To stress the time reversal properties, we will also use the notation $\mathbf{F}^{rev}$ instead of $\mathbf{F}^{scon}$ and $\mathbf{F}^{irr}$ instead of $\mathbf{F}^{vis}$.

\subsection{Langevin stochastic field dynamics for fluid systems}

When stochastic fluctuations are present in fluid systems, a stochastic description is needed instead of a deterministic one. In general, stochastic fluctuations in fluid systems may have an internal origin or an external origin (or both). Internal stochastic fluctuations may arise from the thermal fluctuations in the internal environment of the fluid system constituted by the microscopic molecular degrees of freedom of the fluid \cite{Landau,HydrodynamicFluctuations}. (Note that the fluid `system' we refer to in this article, whose state is described by the velocity field, does not include these microscopic molecular degrees of freedom that are regarded as the internal `environment' of the system.) On the other hand, external stochastic fluctuations may originate from the action of an external agent or the interaction of the fluid system with an external environment, as in the modeling of some stochastic `stirring' mechanisms. The distinction of these two sources of stochastic fluctuations is not necessary for most discussions in this article. Therefore, we shall treat them together without specifying the nature of the sources of stochastic fluctuations unless it becomes necessary.

Taking into account stochastic fluctuations, we consider the Navier-Stokes equation in Eq. (\ref{NSE}) with an additional stochastic forcing term:
\begin{equation}\label{SFNSE}
\partial_t \mathbf{u}+\mathbf{u}\cdot\nabla\mathbf{u}=\nu \Delta\mathbf{u}-\nabla p+\bm{\xi},
\end{equation}
where $\bm{\xi}(\mathbf{x},t)$ is the stochastic force field. This equation is still subject to the incompressibility condition $\nabla\cdot\mathbf{u}=0$.

The stochastic force is characterized as follows. We assume that the total stochastic force, $\int \bm{\xi}(\mathbf{x},t)d\mathbf{x}$, determining the dynamics of the center of mass of the fluid, vanishes exactly (not only on average) so that the total momentum of the fluid is still conserved. As for the statistical properties of the stochastic force, we assume that $\bm{\xi}$ is a Gaussian stochastic field with zero mean $\langle\bm{\xi}(\mathbf{x},t)\rangle=0$ and has the correlation $\langle\bm{\xi}(\mathbf{x},t)\bm{\xi}(\mathbf{x}',t')\rangle=2\mathbf{D}(\mathbf{x}-\mathbf{x}')\delta(t-t')$. The correlator $\mathbf{D}(\mathbf{x}-\mathbf{x}')$ characterizes the spatial correlation of the stochastic force, which has been assumed to be independent of the velocity field  $\mathbf{u}$ (i.e., $\bm{\xi}$ is an additive noise) and only dependent on the spatial difference $\mathbf{x}-\mathbf{x}'$ (i.e., $\bm{\xi}$ is statistically homogeneous in space).

Similar to the deterministic dynamics, with the aid of the incompressibility condition, the pressure force in Eq. (\ref{SFNSE}) can be expressed as $-\nabla p=-\mathbf{\Pi}^g(\nabla)\cdot(-\mathbf{u}\cdot\nabla\mathbf{u}+\bm{\xi})$. This means the gradient components of both the convective force and the stochastic force are counterbalanced by the pressure force, leaving only their solenoidal components. The resulting stochastic Navier-Stokes equation in the solenoidal form reads:
\begin{equation}\label{SNSESF}
\partial_t \mathbf{u}=\mathbf{\Pi}^{s}(\nabla)\cdot\left(-\mathbf{u}\cdot\nabla\mathbf{u}\right)+\nu \Delta\mathbf{u}+\bm{\xi}^s,
\end{equation}
where $\bm{\xi}^s=\mathbf{\Pi}^{s}(\nabla)\cdot\bm{\xi}$ is the solenoidal stochastic force. The solenoidal stochastic Navier-Stokes dynamics in Eq. (\ref{SNSESF}) is governed by both the deterministic driving force, consisting of the reversible solenoidal convective force and the irreversible viscous force, and the solenoidal stochastic force.

The solenoidal stochastic force $\bm{\xi}^s$ has the same statistical properties as $\bm{\xi}$, Gaussian with zero mean, except that its correlation is $\langle\bm{\xi}^s(\mathbf{x},t)\bm{\xi}^s(\mathbf{x}',t')\rangle=2\mathbf{D}^s(\mathbf{x}-\mathbf{x}')\delta(t-t')$. The spatial correlator $\mathbf{D}^s(\mathbf{x}-\mathbf{x}')$ is related to $\mathbf{D}(\mathbf{x}-\mathbf{x}')$ by
\begin{equation}\label{RDMSDM}
D_{ij}^s(\mathbf{x}-\mathbf{x}')=\Pi^s_{ik}(\nabla)\Pi^s_{jl}(\nabla')D_{kl}(\mathbf{x}-\mathbf{x}'),
\end{equation}
where repeated indexes are summed over. It is easy to see that $\nabla\cdot\mathbf{D}^s(\mathbf{x}-\mathbf{x}')=0$ as a result of $\nabla\cdot\bm{\xi}^s=0$.

The spatial correlator $\mathbf{D}^s(\mathbf{x}-\mathbf{x}')$ also serves as the diffusion matrix in the state space in the context of the Fokker-Planck field dynamics that will be discussed in a moment. It has the properties $D^s_{ij}(\mathbf{x}-\mathbf{x}')=D^s_{ji}(\mathbf{x}'-\mathbf{x})$ and $\iint \mathbf{v}(\mathbf{x})\cdot\mathbf{D}^s(\mathbf{x}-\mathbf{x}')\cdot\mathbf{v}(\mathbf{x}')d\mathbf{x}d\mathbf{x}'\geq 0$ for $\mathbf{v}(\mathbf{x})$ in the state space $\Omega$. Hence, $\mathbf{D}^s(\mathbf{x}-\mathbf{x}')$ can be viewed as an infinite-dimensional symmetric nonnegative-definite matrix in the state space, with its row labeled by $(i,\mathbf{x})$ and column labeled by $(j,\mathbf{x}')$. Accordingly, the vector field $\mathbf{v}(\mathbf{x})$, with component $v_i(\mathbf{x})$, can be considered as an infinite-dimensional vector in the state space indexed by $(i,\mathbf{x})$. We shall in addition assume that the diffusion matrix $\mathbf{D}^s(\mathbf{x}-\mathbf{x}')$ is invertible in the state space, so that the linear equation $\int\mathbf{D}^s(\mathbf{x}-\mathbf{x}')\cdot\mathbf{v}(\mathbf{x}')d\mathbf{x}'=\mathbf{w}(\mathbf{x})$ can be inverted to give $\mathbf{v}(\mathbf{x})=\int \mathbf{D}_{s}^{-1}(\mathbf{x}-\mathbf{x}')\cdot\mathbf{w}(\mathbf{x}')d\mathbf{x}'$ , for any $\mathbf{v}(\mathbf{x})$ and $\mathbf{w}(\mathbf{x})$ in the state space. Formally, $\mathbf{D}_{s}^{-1}(\mathbf{x}-\mathbf{x}')$ is defined by the equation $\int\mathbf{D}^s(\mathbf{x}-\mathbf{x}'')\cdot\mathbf{D}_{s}^{-1}(\mathbf{x}''-\mathbf{x}')d\mathbf{x}''=\mathbf{G}^s(\mathbf{x}-\mathbf{x}')$,
where $\mathbf{G}^s(\mathbf{x}-\mathbf{x}')$ is the integral kernel of the solenoidal projection operator $\mathbf{\Pi}^s(\nabla)$, which plays the role of the identity matrix in the state space of solenoidal vector fields.

\subsection{Fokker-Planck field dynamics for fluid systems}

The solenoidal stochastic Navier-Stokes equation in Eq. (\ref{SNSESF}) describes a Langevin-type stochastic field dynamics tracing a stochastic trajectory in the state space (velocity field configuration space). The corresponding ensemble dynamics governing the evolution of the probability distribution functional of the velocity field is described by the functional Fokker-Planck equation (FFPE) \cite{Gardiner,WWJW20141,WWJW20142}:
\begin{equation}\label{FFPENSE}
\begin{split}
\partial_t P_t[\mathbf{u}]=&-\int d\mathbf{x}\,\delta_{\mathbf{u}(\mathbf{x})}\cdot\Big(\left[\mathbf{\Pi}^{s}(\nabla)\cdot\left(-\mathbf{u}\cdot\nabla\mathbf{u}\right)+\nu \Delta\mathbf{u}\right]P_t[\mathbf{u}]\Big)\\
&+\int d\mathbf{x}\,\delta_{\mathbf{u}(\mathbf{x})}\cdot\left(\int d\mathbf{x}'\,\mathbf{D}^s(\mathbf{x}-\mathbf{x}')\cdot\delta_{\mathbf{u}(\mathbf{x}')}P_t[\mathbf{u}]\right),
\end{split}
\end{equation}
where $P_t[\mathbf{u}]\equiv P[\mathbf{u(\mathbf{x})},t]$ is the probability distribution functional defined on the state space $\Omega$ and $\delta_{\mathbf{u}(\mathbf{x})}\equiv \delta/\delta \mathbf{u}(\mathbf{x})$ is the short notation for the vector-valued functional derivative. Note that the functional derivative here is \emph{restricted} to the state space (solenoidal velocity fields with zero total momentum). Because of this constraint the basic rule of functional derivative is $\delta u_i(\mathbf{x})/\delta u_j(\mathbf{x}')=G^s_{ij}(\mathbf{x}-\mathbf{x}')$, where $G^s_{ij}(\mathbf{x}-\mathbf{x}')$ plays the same role as $\delta_{ij}\delta(\mathbf{x}-\mathbf{x}')$ without constraint.

The FFPE in Eq. (\ref{FFPENSE}) can be expressed in the form of a continuity equation in the \emph{state space}, $\partial_t P_t[\mathbf{u}]=-\int d\mathbf{x}\,\delta_{\mathbf{u}(\mathbf{x})}\cdot\mathbf{J}_t(\mathbf{x})[\mathbf{u}]$, representing conservation of probability. That means the change of the probability in a local region of the state space is due to the probability flow in and out of that region. The probability flow in the state space is characterized by the probability flux field functional, which is identified from the FFPE as
\begin{equation}\label{PFFFNSE}
\mathbf{J}_t(\mathbf{x})[\mathbf{u}]=\left[\mathbf{\Pi}^{s}(\nabla)\cdot\left(-\mathbf{u}\cdot\nabla\mathbf{u}\right)+\nu \Delta\mathbf{u}\right]P_t[\mathbf{u}]-\int d\mathbf{x}'\,\mathbf{D}^s(\mathbf{x}-\mathbf{x}')\cdot\delta_{\mathbf{u}(\mathbf{x}')}P_t[\mathbf{u}].
\end{equation}
$\mathbf{J}_t(\mathbf{x})[\mathbf{u}]$ depends explicitly on the spatial coordinate $\mathbf{x}$, i.e., it is a field in the physical space; it also depends on the velocity field $\mathbf{u}(\mathbf{x})$ as a whole, i.e., it is a functional. The probability flux field functional in Eq. (\ref{PFFFNSE}) has been grouped into two parts. The first part describes a drift process in the state space, where the drift velocity is given by the deterministic driving force in the Langevin stochastic field dynamics in Eq. (\ref{SNSESF}), consisting of the reversible solenoidal convective force and the irreversible viscous force. The other part describes a diffusion process in the state space, where the diffusion matrix is given by the spatial correlator of the stochastic force in the Langevin stochastic field dynamics in Eq. (\ref{SNSESF}).

The steady state of the system in which the statistics do not vary with time is of particular interest. The steady-state probability distribution functional $P_s[\mathbf{u}]$ satisfies $\partial_t P_s[\mathbf{u}]=0$ and is determined by the steady-state FFPE, with the r.h.s. of Eq. (\ref{FFPENSE}) set to zero. Accordingly, the steady-state probability flux field functional is given by
\begin{equation}\label{SSPFFFNSE}
\mathbf{J}_s(\mathbf{x})[\mathbf{u}]=\left[\mathbf{\Pi}^{s}(\nabla)\cdot\left(-\mathbf{u}\cdot\nabla\mathbf{u}\right)+\nu \Delta\mathbf{u}\right]P_s[\mathbf{u}]-\int d\mathbf{x}'\,\mathbf{D}^s(\mathbf{x}-\mathbf{x}')\cdot\delta_{\mathbf{u}(\mathbf{x}')}P_s[\mathbf{u}].
\end{equation}
In terms of $\mathbf{J}_s(\mathbf{x})[\mathbf{u}]$ the steady-state FFPE is simply $\int d\mathbf{x}\,\delta_{\mathbf{u}(\mathbf{x})}\cdot\mathbf{J}_s(\mathbf{x})[\mathbf{u}]=0$. Notice that the operator $\int d\mathbf{x}\,\delta_{\mathbf{u}(\mathbf{x})}\cdot$ is the functional divergence operator in the state space. This means $\mathbf{J}_s(\mathbf{x})[\mathbf{u}]$ is functional-divergence-free and can thus be considered as a solenoidal field in the \emph{state space} of velocity fields. In addition, given that the deterministic driving force is solenoidal in the physical space and that $\nabla\cdot \mathbf{D}^s(\mathbf{x}-\mathbf{x}')=0$, one can see from Eq. (\ref{SSPFFFNSE}) that $\mathbf{J}_s(\mathbf{x})[\mathbf{u}]$ is also a solenoidal field in the \emph{physical space} satisfying $\nabla\cdot\mathbf{J}_s(\mathbf{x})[\mathbf{u}]=0$. In other words, $\mathbf{J}_s(\mathbf{x})[\mathbf{u}]$ is solenoidal in both the state space and the physical space. However, it is important not to confuse properties in the state space with those in the physical space. Quantities in the state space are functionals of the velocity field. Vector calculus in the state space is done with the functional derivative $\delta_{\mathbf{u}(\mathbf{x})}$ rather than the nabla operator $\nabla$ in the physical space.

The probability flux can be decomposed into a reversible part and an irreversible part according to their different time reversal behaviors in relation to the FFPE.  The FFPE in the form of the continuity equation reads $\partial_t P_t[\mathbf{u}]=-\int d\mathbf{x}\,\delta_{\mathbf{u}(\mathbf{x})}\cdot\mathbf{J}_t(\mathbf{x})[\mathbf{u}]$. Note that probability does not change sign under time reversal. Thus the l.h.s. of the equation is odd under time reversal due to the time derivative $\partial_t$. On the r.h.s. of the equation $\delta_{\mathbf{u}(\mathbf{x})}$ is also odd under time reversal. Therefore, the part of the probability flux with even parity will preserve the form of the equation under time reversal and is thus reversible; the part with odd parity is correspondingly irreversible.

Applying the above analysis we decompose the steady-state probability flux in Eq. (\ref{SSPFFFNSE}) according to different time reversal properties. The reversible steady-state probability flux field is identified as
\begin{equation}\label{SSRPFFFNSE}
\mathbf{J}_s^{rev}(\mathbf{x})[\mathbf{u}]=[\mathbf{\Pi}^{s}(\nabla)\cdot\left(-\mathbf{u}\cdot\nabla\mathbf{u}\right)] P_s[\mathbf{u}],
\end{equation}
which is the reversible solenoidal convective force (even under time reversal) times the steady-state probability distribution. The remaining is the irreversible steady-state probability flux field
\begin{equation}\label{SSIPFFFNSE}
\mathbf{J}_s^{irr}(\mathbf{x})[\mathbf{u}]=\left(\nu \Delta\mathbf{u}\right)P_s[\mathbf{u}]-\int d\mathbf{x}'\,\mathbf{D}^s(\mathbf{x}-\mathbf{x}')\cdot\delta_{\mathbf{u}(\mathbf{x}')}P_s[\mathbf{u}],
\end{equation}
which consists of two parts. The first part is the irreversible viscous force (odd under time reversal) times $P_s[\mathbf{u}]$, which we define as the steady-state viscous probability flux denoted by $\mathbf{J}_s^{vis}(\mathbf{x})[\mathbf{u}]=\left(\nu \Delta\mathbf{u}\right)P_s[\mathbf{u}]$. The second part is an operator $\int d\mathbf{x}'\,\mathbf{D}^s(\mathbf{x}-\mathbf{x}')\cdot\delta_{\mathbf{u}(\mathbf{x}')}$, associated with the stochastic force, acting on $P_s[\mathbf{u}]$, which is also odd under time reversal due to $\delta_{\mathbf{u}(\mathbf{x}')}$. We define this part as the steady-state stochastic probability flux denoted by $\mathbf{J}_s^{sto}(\mathbf{x})[\mathbf{u}]=-\int d\mathbf{x}'\,\mathbf{D}^s(\mathbf{x}-\mathbf{x}')\cdot\delta_{\mathbf{u}(\mathbf{x}')}P_s[\mathbf{u}]$.

Hence, the solenoidal convective force contributes to the reversible steady-state probability flux, while both the viscous force and the stochastic force contribute to the irreversible steady-state probability flux. Note that the steady-state probability distribution $P_s[\mathbf{u}]$, in general (e.g., for nonequilibrium steady states), does not coincide with $P_s[-\mathbf{u}]$ although both are non-negative. We can also define the corresponding transient probability fluxes, $\mathbf{J}_t^{rev}(\mathbf{x})[\mathbf{u}]$, $\mathbf{J}_t^{irr}(\mathbf{x})[\mathbf{u}]$, $\mathbf{J}_t^{vis}(\mathbf{x})[\mathbf{u}]$, and $\mathbf{J}_t^{sto}(\mathbf{x})[\mathbf{u}]$, by replacing $P_s[\mathbf{u}]$ with $P_t[\mathbf{u}]$.

\subsection{Symbolic representation}

The mathematics involving fields with infinite degrees of freedom is in general quite complex. To avoid getting lost in the complexity and assist with clarity in the algebra to come, we introduce a symbolic representation that keeps only  the essential features of the relevant quantities without the distraction of details. Although the symbolic representation has the advantage of being compact, its abstractness and lack of specificity may be considered weakness. Therefore, it will be used as a complement to concrete representations (e.g., the physical space representation and the wavevector space representation) rather than to replace them.

The velocity field $\mathbf{u}(\mathbf{x})$ describing the state of the fluid system is represented symbolically as $u$. The force field functional $\mathbf{F}(\mathbf{x})[\mathbf{u}]$ that drives the deterministic dynamics of the system is a vector field in the state space, which is represented symbolically as $F(u)$. Then the deterministic solenoidal Navier-Stokes equation in Eq. (\ref{NSESF}) has the symbolic form $\dot{u}=F(u)$, where $F(u)=F^{rev}(u)+F^{irr}(u)$. The symbolic representation of the solenoidal stochastic Navier-Stokes equation in Eq. (\ref{SNSESF}) reads
\begin{equation}
\dot{u}=F(u)+\xi^s,
\end{equation}
with $\langle \xi^s(t)\xi^s(t')\rangle=2D^s\delta(t-t')$. Here $\xi^s$ represents the solenoidal stochastic force field $\bm{\xi}^s(\mathbf{x},t)$ and $D^s$ represents the diffusion matrix $\mathbf{D}^s(\mathbf{x}-\mathbf{x}')$ in the state space.

In the symbolic representation the FFPE in Eq. (\ref{FFPENSE}) has the compact form:
\begin{equation}\label{SFFFPE}
\partial_t P=-\partial \cdot (FP-D^s\cdot \partial P),
\end{equation}
where $\partial$  represents symbolically the functional derivative $\delta_{\mathbf{u}(\mathbf{x})}$ and the dot product is that in the state space which sums over both the discrete index of the 3D vector and the continuous index $\mathbf{x}$ of the spatial position (a spatial integral). Formally, Eq. (\ref{SFFFPE}) is the same as an ordinary Fokker-Planck equation for systems with finite degrees of freedom, but here it actually represents a FFPE for systems with infinite degrees of freedom.

Furthermore, the steady-state probability flux field functional in Eq.(\ref{SSPFFFNSE}) has the symbolic representation
\begin{equation}
J_s=FP_s-D^s\cdot\partial P_s,
\end{equation}
which satisfies the divergence-free condition $\partial\cdot J_s=0$ in the state space, equivalent to the steady-state FFPE. It also consists of a reversible part and an irreversible part, $J_s=J_s^{rev}+J_s^{irr}$, which are given by
\begin{equation}\label{RFSF}
J^{rev}_s=F^{rev}P_s,
\end{equation}
and
\begin{equation}\label{IRFSF}
\qquad J^{irr}_s=F^{irr}P_s-D^s\cdot\partial P_s,
\end{equation}
representing symbolically Eqs. (\ref{SSRPFFFNSE}) and (\ref{SSIPFFFNSE}), respectively.

\section{Potential landscape and flux field theory for fluid systems}\label{PFFLNSE}

In this section we establish the detailed balance constraint that gives a precise formulation of the detailed balance condition for equilibrium fluid systems and the nonequilibrium trinity that characterizes nonequilibrium fluid systems with detailed balance breaking. Their implications on the structure of the stochastic fluid dynamics are also discussed.

\subsection{Equilibrium fluid systems with detailed balance}

Equilibrium fluid systems preserve detailed balance. For stochastic dynamical systems governed by Langevin and Fokker-Planck equations, the necessary and sufficient conditions for detailed balance, characterizing the time reversal symmetry in equilibrium steady states, are given by three conditions \cite{Risken,Gardiner}. When applied to the stochastic fluid systems considered in this article, these conditions read as follows in the symbolic form:
\begin{eqnarray}
&&D^s(\epsilon u)=\epsilon D^s(u)\epsilon,\label{TRPDM}\\
&&J^{irr}_s=0,\label{VIF}\\
&&\partial\cdot J^{rev}_s=0.\label{SSFPEDB}
\end{eqnarray}
The first condition is on the time reversal property of the diffusion matrix, where $\epsilon$ is associated with the time reversal parity of the state variables ($\epsilon=1$ for even variables and $\epsilon=-1$ for odd variables). For the fluid systems we consider, this condition is trivially satisfied since the diffusion matrix $D^s$ does not depend on $u$ and $\epsilon=-1$ as the velocity field $u$ is odd with respect to time reversal. The third condition is essentially the steady-state FFPE $\partial \cdot J_s=0$, taking into account the second condition $J_s^{irr}=0$ and the relation $J_s=J_s^{rev}+J_s^{irr}$. Therefore, the primary condition for detailed balance is the second one, that is, vanishing steady-state irreversible probability flux. But the third condition is also needed for completeness.

\subsubsection{Potential form of the irreversible viscous force}

The second condition for detailed balance, $J_s^{irr}=0$, reads more explicitly, $\mathbf{J}_s^{irr}(\mathbf{x})[\mathbf{u}]=0$. This should be understood as a statement that $\mathbf{J}_s^{irr}(\mathbf{x})[\mathbf{u}]$ vanishes at all spatial positions in the physical domain and for all velocity field configurations in the state space. Notice the structure of $\mathbf{J}_s^{irr}(\mathbf{x})[\mathbf{u}]$ in Eq. (\ref{SSIPFFFNSE}) or its symbolic form in Eq. (\ref{IRFSF}), which consists of two parts associated with the viscous force and the stochastic force, respectively. This means the contribution from the viscous force and that from the stochastic force must balance each other out on a detailed level, at all spatial positions and for all velocity fields in the state space, in order for $\mathbf{J}_s^{irr}(\mathbf{x})[\mathbf{u}]$ to vanish completely.

As a consequence of vanishing $\mathbf{J}_s^{irr}(\mathbf{x})[\mathbf{u}]$, we obtain from Eq. (\ref{SSIPFFFNSE}), by dividing $P_s$ on both sides of the equation, the following potential form (in the state space) of the irreversible viscous force:
\begin{equation}\label{NSEPC}
\nu \Delta\mathbf{u}=-\int d\mathbf{x}'\,\mathbf{D}^s(\mathbf{x}-\mathbf{x}')\cdot\delta_{\mathbf{u}(\mathbf{x}')}\Phi[\mathbf{u}],
\end{equation}
which is represented symbolically as $F^{irr}=-D^s\cdot\partial\Phi$. Here $\Phi[\mathbf{u}]=-\ln P_s[\mathbf{u}]$ is the potential landscape functional defined in terms of the steady-state probability distribution functional. It has to be stressed that this potential form is in the state space rather than in the physical space. In fact, the property $\nabla\cdot\mathbf{D}^s(\mathbf{x}-\mathbf{x}')=0$ shows that the viscous force is solenoidal in the physical space, in agreement with its expression $\nu\Delta\mathbf{u}$ where the velocity field $\mathbf{u}$ is solenoidal in the physical space.

For equilibrium fluid  systems with detailed balance, Eq. (\ref{NSEPC}) shows that the irreversible viscous force has a potential form in the state space,  expressed as the functional gradient of the potential landscape, modified by the diffusion matrix associated with the stochastic force. This potential form of the irreversible viscous force, determined by the potential landscape alone, without the involvement of the irreversible probability flux, is the manifestation of detailed balance on the level of the dynamical driving force in equilibrium fluid systems. It signifies the reversible nature of the equilibrium fluid dynamics, due to the absence of the irreversible probability flux that indicates time reversal asymmetry.

\subsubsection{Potential condition}\label{PotentialCondition}

We further investigate the implications of the potential form. Inverting Eq. (\ref{NSEPC}) we obtain
\begin{equation}\label{NSEPCPFL}
\delta_{\mathbf{u}(\mathbf{x})}\Phi[\mathbf{u}]=-\int d\mathbf{x}'\,\mathbf{D}_s^{-1}(\mathbf{x}-\mathbf{x}')\cdot[\nu \Delta'\mathbf{u}(\mathbf{x}')],
\end{equation}
which reads symbolically $\partial\Phi=-D_s^{-1}\cdot F^{irr}$. This form expresses the potential gradient in the state space in terms of the diffusion matrix (stochastic force) and the viscous force.

In general, the potential form cannot be true for arbitrary $D^s$ and $F^{irr}$. This is because the `curl-free' property of the potential gradient, $\partial\wedge\partial\Phi=0$, imposes the constraint $\partial\wedge [D_s^{-1}\cdot F^{irr}]=0$, where the wedge product is the generalization of the cross product in 3D to higher dimensions, which anti-symmetrizes the indexes of the two state-space vectors in the product. This constraint is called the \emph{potential condition} \cite{Risken,Gardiner}, which for the fluid system reads more specifically:
\begin{equation}\label{PC}
\begin{split}
\delta_{u_i(\mathbf{x})}&\left\{\int d\mathbf{x}''\,(\mathbf{D}_s^{-1})_{jk}(\mathbf{x}'-\mathbf{x}'')[\nu \Delta''u_k(\mathbf{x}'')]\right\}\\
&\hspace{40pt}=\delta_{u_j(\mathbf{x}')}\left\{\int d\mathbf{x}''\,(\mathbf{D}_s^{-1})_{ik}(\mathbf{x}-\mathbf{x}'')[\nu \Delta''u_k(\mathbf{x}'')]\right\},
\end{split}
\end{equation}
where the repeated index $k$ is summed.

However, the potential condition in Eq. (\ref{PC}) is satisfied automatically, without any additional restrictions except those already assumed for the diffusion matrix (independent of $\mathbf{u}$ and dependent on $\mathbf{x}-\mathbf{x}'$). Indeed, carrying out the functional derivatives and spatial integrals in Eq. (\ref{PC}), this condition reduces to $\Delta' (\mathbf{D}_s^{-1})_{ij}(\mathbf{x}-\mathbf{x}')=\Delta (\mathbf{D}_s^{-1})_{ji}(\mathbf{x}'-\mathbf{x})$. This is true given the symmetry of the diffusion matrix in the state space and thus also its inverse (invariant when switching the index $(i,\mathbf{x})$ with $(j,\mathbf{x}')$), together with its dependence on the spatial difference $\mathbf{x}-\mathbf{x}'$ (so that $\Delta'$ can be replaced by $\Delta$).

This means the r.h.s. of Eq. (\ref{NSEPCPFL}) always has a functional gradient potential form, for any diffusion matrix $\mathbf{D}^s(\mathbf{x}-\mathbf{x}')$. In fact, one can directly verify Eq. (\ref{NSEPCPFL}) for the following quadratic potential in the state space:
\begin{equation}\label{Phi0NSE}
\Phi_0[\mathbf{u}]=\DF{\nu}{2}\iint d\mathbf{x}d\mathbf{x}'\mathbf{u}(\mathbf{x})\cdot\left[-\Delta\mathbf{D}_s^{-1}(\mathbf{x}-\mathbf{x}')\right]\cdot\mathbf{u}(\mathbf{x}').
\end{equation}
This also means the irreversible viscous force always has a potential form in the state space
\begin{equation}\label{PFIFRDB}
\nu \Delta\mathbf{u}=-\int d\mathbf{x}'\,\mathbf{D}^s(\mathbf{x}-\mathbf{x}')\cdot\delta_{\mathbf{u}(\mathbf{x}')}\Phi_0[\mathbf{u}].
\end{equation}

However, this does \emph{not} imply that the steady-state irreversible probability flux always vanishes, that the detailed balance condition is always satisfied, or that the fluid system considered is always an equilibrium one. The reason is that $\Phi_0[\mathbf{u}]$ in Eq. (\ref{Phi0NSE}) is not necessarily related to $P_s[\mathbf{u}]$ in the form $\Phi_0[\mathbf{u}]=-\ln P_s[\mathbf{u}]$ (or $P_s[\mathbf{u}]=e^{-\Phi_0[\mathbf{u}]}$) as required for those implications to hold true. In other words, $\Phi_0[\mathbf{u}]$ does not always coincide with $\Phi[\mathbf{u}]=-\ln P_s[\mathbf{u}]$. The potential landscape $\Phi[\mathbf{u}]$, defined in terms of $P_s[\mathbf{u}]$, must satisfy another condition determined by the steady-state FFPE for $P_s[\mathbf{u}]$. For equilibrium systems with detailed balance, this is simply the third condition for detailed balance in Eq. (\ref{SSFPEDB}), $\partial\cdot J_s^{rev}=0$, as will be discussed in a moment.

It is important to realize the difference between the potential form in Eq. (\ref{PFIFRDB}) and that in Eq. (\ref{NSEPC}). The potential form in Eq. (\ref{PFIFRDB}) is not contingent upon the detailed balance condition. It is purely a consequence of the form of the viscous force and the diffusion matrix, regardless of whether detailed balance holds or not. In contrast, the potential form in Eq. (\ref{NSEPC}), with $\Phi[\mathbf{u}]=-\ln P_s[\mathbf{u}]$, is a primary condition for detailed balance, equivalent to $\mathbf{J}_s^{irr}(\mathbf{x})[\mathbf{u}]=0$. The potential form in Eq. (\ref{PFIFRDB}) will play a role later in the establishment of the nonequilibrium trinity for nonequilibrium fluid systems with detailed balance breaking.

\subsubsection{Orthogonality condition for detailed balance}

The third condition for detailed balance is $\partial\cdot J_s^{rev}=0$ in Eq. (\ref{SSFPEDB}). Noticing that $J_s^{rev}=F^{rev}P_s$, this condition can be reexpressed in terms of the potential landscape as $F^{rev}\cdot\partial \Phi=\partial\cdot F^{rev}$, or more explicitly,
\begin{equation}\label{FDFRPFNSE}
\int d\mathbf{x}\,[\mathbf{\Pi}^{s}(\nabla)\cdot\left(-\mathbf{u}\cdot\nabla\mathbf{u}\right)]\cdot\delta_{\mathbf{u}(\mathbf{x})}\Phi[\mathbf{u}]=\int d\mathbf{x}\,\delta_{\mathbf{u}(\mathbf{x})}\cdot[\mathbf{\Pi}^{s}(\nabla)\cdot\left(-\mathbf{u}\cdot\nabla\mathbf{u}\right)].
\end{equation}
The l.h.s. of this equation is the inner product in the state space between the solenoidal convective force and the functional gradient of the potential landscape. The r.h.s. of this equation is the functional divergence of the solenoidal convective force, which can be shown to vanish, that is, the solenoidal convective force is also solenoidal in the state space (see \ref{FunctionalDivergence}).

As a consequence, Eq. (\ref{FDFRPFNSE}) reduces to the orthogonality condition
\begin{equation}\label{FDFRPFNSE2}
\int d\mathbf{x}\,[\mathbf{\Pi}^{s}(\nabla)\cdot\left(-\mathbf{u}\cdot\nabla\mathbf{u}\right)]\cdot\delta_{\mathbf{u}(\mathbf{x})}\Phi[\mathbf{u}]=0,
\end{equation}
which is represented symbolically as $F^{rev}\cdot\partial \Phi=0$.
Geometrically, it means that for detailed balance to hold, the functional gradient of the potential landscape should be orthogonal to the solenoidal convective force in the state space. This condition relates the potential landscape to the solenoidal convective force, in addition to the potential form in Eq. (\ref{NSEPC}) which relates the potential landscape to the viscous force and the stochastic force (diffusion matrix). (Note that even though $\Phi_0[\mathbf{u}]$ in Eq. (\ref{Phi0NSE}) satisfies the potential condition, it does not necessarily fulfill Eq. (\ref{FDFRPFNSE2}). This demonstrates that $\Phi_0[\mathbf{u}]$ in general is not the same as $\Phi[\mathbf{u}]$. But they do coincide when $\Phi_0[\mathbf{u}]$ also satisfies  Eq. (\ref{FDFRPFNSE2}).)

\subsubsection{Detailed balance constraint}

Now we put together all the conditions for detailed balance, which include the potential form in Eq. (\ref{NSEPC}) (or its inverted form in Eq. (\ref{NSEPCPFL})) and the orthogonality condition in Eq. (\ref{FDFRPFNSE2}). Notice that $\Phi[\mathbf{u}]$ in these two conditions can be eliminated by plugging Eq. (\ref{NSEPCPFL}) into Eq. (\ref{FDFRPFNSE2}), resulting in a single condition. We thus finally arrive at the \emph{detailed balance constraint}:
\begin{equation}\label{DBCENSE}
\iint d\mathbf{x}d\mathbf{x}'\,[\mathbf{\Pi}^{s}(\nabla)\cdot\left(-\mathbf{u}\cdot\nabla\mathbf{u}\right)]\cdot\mathbf{D}_s^{-1}(\mathbf{x}-\mathbf{x}')\cdot[\nu \Delta'\mathbf{u}(\mathbf{x}')]=0,
\end{equation}
which can be represented symbolically as $F^{rev}\cdot D_s^{-1}\cdot F^{irr}=0$. This detailed balance constraint is the necessary and sufficient condition for detailed balance characterizing equilibrium fluid systems with time reversal symmetry, governed by the solenoidal stochastic Navier-Stokes dynamics in Eq. (\ref{SNSESF}) or the equivalent Fokker-Planck field dynamics in Eq. (\ref{FFPENSE}), under the assumptions made about the diffusion matrix.

The detailed balance constraint relates the solenoidal convective force, the viscous force, and the stochastic force (diffusion matrix) together in a very specific manner. The geometric interpretation of this constraint is that the solenoidal convective force and the viscous force are perpendicular to each other in the state space with respect to the metric defined by the inverse diffusion matrix associated with the stochastic force. Physically, the detailed balance constraint represents a specific form of mechanical balance in the driving forces of the fluid system dynamics.

For equilibrium fluid systems the form of the diffusion matrix, namely the spatial correlator of the stochastic force, is restricted by the detailed balance constraint. Only when the diffusion matrix satisfies the detailed balance constraint can the fluid system obey detailed balance and have an equilibrium steady state. In this case  we denote the diffusion matrix more specifically as $\mathbf{D}_{eq}^s(\mathbf{x}-\mathbf{x}')$. The constraint in Eq. (\ref{DBCENSE}) must hold for all velocity fields in the state space. In view of the nonlinearity of the convective force, it places a rather strong restriction on the possible form of the diffusion matrix. One may wonder whether there is any stochastic force at all with a diffusion matrix that can meet such a stringent condition. We shall show later in the study of an example that the detailed balance constraint can indeed be fulfilled by a particular form of diffusion matrix associated with a specific form of stochastic force.

\subsubsection{Equilibrium steady state}\label{ESS}

For fluid systems obeying detailed balance, the steady state of the system is an equilibrium state with time reversal symmetry. Once the detailed balance constraint is respected by the diffusion matrix $\mathbf{D}^s_{eq}(\mathbf{x}-\mathbf{x}')$, the equilibrium potential landscape, denoted as $\Phi_{eq}[\mathbf{u}]$, can then be solved. It is easy to see that $\Phi_{eq}[\mathbf{u}]$, which should satisfy both Eq. (\ref{NSEPC}) and Eq. (\ref{FDFRPFNSE2}), has the same form as $\Phi_0[\mathbf{u}]$ in Eq. (\ref{Phi0NSE}), but with $\mathbf{D}^s(\mathbf{x}-\mathbf{x}')$ restricted to $\mathbf{D}^s_{eq}(\mathbf{x}-\mathbf{x}')$. That is, $\Phi_{eq}[\mathbf{u}]$ has the quadratic form
\begin{equation}
\Phi_{eq}[\mathbf{u}]=\DF{\nu}{2}\iint d\mathbf{x}d\mathbf{x}'\mathbf{u}(\mathbf{x})\cdot\left[-\Delta {\mathbf{D}_{eq}^{s}}^{-1}(\mathbf{x}-\mathbf{x}')\right]\cdot\mathbf{u}(\mathbf{x}').
\end{equation}

Accordingly, the equilibrium probability distribution functional, denoted as $P_{eq}[\mathbf{u}]$, related to $\Phi_{eq}[\mathbf{u}]$ through $P_{eq}[\mathbf{u}]=e^{-\Phi_{eq}[\mathbf{u}]}$, is the Gaussian distribution functional
\begin{equation}
P_{eq}[\mathbf{u}]=\mathcal{N}\exp\left\{-\DF{\nu}{2}\iint d\mathbf{x}d\mathbf{x}'\mathbf{u}(\mathbf{x})\cdot\left[-\Delta {\mathbf{D}_{eq}^{s}}^{-1}(\mathbf{x}-\mathbf{x}')\right]\cdot\mathbf{u}(\mathbf{x}')\right\},
\end{equation}
where $\mathcal{N}$ is the normalization constant. One can directly verify that $P_{eq}[\mathbf{u}]$ satisfies the steady-state FFPE. It can also be verified that in this case the steady-state irreversible probability flux vanishes, in agreement with the time reversal symmetry of the equilibrium steady state preserving detailed balance.

\subsubsection{Equilibrium stochastic fluid dynamics}

The detailed balance conditions for equilibrium fluid systems have important implications on the equilibrium stochastic fluid dynamics. Given the potential form of the irreversible viscous force in Eq. (\ref{NSEPC}), for equilibrium fluid systems with detailed balance, the stochastic Navier-Stokes equation in Eq. (\ref{SNSESF}) can be reformulated in the following potential form (in the state space):
\begin{equation}\label{GLENSE}
\partial_t\mathbf{u}(\mathbf{x},t)=\mathbf{F}^{rev}(\mathbf{x})[\mathbf{u}]-\int d\mathbf{x}'\,\mathbf{D}^s_{eq}(\mathbf{x}-\mathbf{x}')\cdot\delta_{\mathbf{u}(\mathbf{x}')}\Phi_{eq}[\mathbf{u}]+\bm{\xi}^s(\mathbf{x},t),
\end{equation}
which reads symbolically $\dot{u}=F^{rev}-D_{eq}^s\cdot\partial \Phi_{eq}+\xi^s$. Here $\mathbf{F}^{rev}(\mathbf{x})[\mathbf{u}]$ represents the reversible solenoidal convective force, $\Phi_{eq}[\mathbf{u}]=-\ln P_{eq}[\mathbf{u}]$ is the equilibrium potential landscape, and $\langle\bm{\xi}^s(\mathbf{x},t)\bm{\xi}^s(\mathbf{x}',t')\rangle=2\mathbf{D}^s_{eq}(\mathbf{x}-\mathbf{x}')\delta(t-t')$ defines the diffusion matrix $\mathbf{D}^s_{eq}(\mathbf{x}-\mathbf{x}')$ that should satisfy the detailed balance constraint.

The equilibrium potential landscape $\Phi_{eq}[\mathbf{u}]$ serves as a bridge of connection between the stochastic trajectory level and the ensemble distribution level for equilibrium fluid systems with detailed balance. On the one hand, it acts as a potential (in the state space) whose functional gradient determines the irreversible viscous force in the Langevin stochastic field dynamics that governs the evolution of stochastic trajectories. On the other hand, it is connected to the equilibrium steady-state probability distribution functional of the Fokker-Planck field dynamics that governs the evolution of ensemble distributions.

It is evident from Eq. (\ref{GLENSE}) that the irreversible probability flux that signifies time irreversibility does not play a role in the equilibrium stochastic fluid dynamics. It is the equilibrium potential landscape, together with the solenoidal convective force and stochastic force, that governs the equilibrium dynamics of the stochastic fluid system with detailed balance.

In addition to the absence of the irreversible probability flux, there are another two important features in the formal structure of the equilibrium fluid dynamics in Eq. (\ref{GLENSE}), as manifestations of detailed balance that relates the three driving forces in the system. One feature is that it obeys a FDT between the viscous force and the stochastic force, which results from the potential form of the viscous force. More specifically, the diffusion matrix defined by the spatial correlator of the stochastic force also serves as a damping matrix in the viscous force, before the functional gradient of the equilibrium potential landscape. The other feature is the orthogonality (in the state space) between the reversible solenoidal convective force and the functional gradient of the potential landscape in the viscous force, given explicitly in Eq. (\ref{FDFRPFNSE2}) as the orthogonality condition for detailed balance.

A generic Langevin equation, with the same formal structure as the stochastic field equation in Eq. (\ref{GLENSE}), has been derived from classical mechanics for Hamiltonian systems with finite degrees of freedom using the projection operator technique \cite{OperatorProjectionMethod,GenericLangevinDerivation}. Eq. (\ref{GLENSE}) has the same structure as the counterpart of the generic Langevin equation for systems with infinite degrees of freedom (fields), specialized to equilibrium stochastic fluid  systems with detailed balance. But it has been obtained from a different route, namely, the detailed balance conditions derived from the Fokker-Planck field dynamics.

\subsubsection{Application in the Landau-Lifshitz-Navier-Stokes system}\label{Example}

We study a particular example to demonstrate what has been developed so far for equilibrium fluid systems with detailed balance. More specifically, we consider the thermodynamic equilibrium case of the fluctuating hydrodynamics first proposed by Landau and Lifshitz to incorporate hydrodynamic fluctuations \cite{Landau,Landau2}, referred to as the Landau-Lifshitz-Navier-Stokes system (LLNSS).

\emph{Navier-Stokes equation.} For an incompressible fluid, the Landau-Lifshitz-Navier-Stokes equation (LLNSE), adapted to the setting and notations in this article, has the form \cite{Landau,Landau2}
\begin{equation}\label{LLNSE}
\partial_t \mathbf{u}+\mathbf{u}\cdot\nabla\mathbf{u}=\nu \Delta\mathbf{u}-\nabla p+\bm{\xi},
\end{equation}
where $\bm{\xi}=\nabla\cdot\mathbf{\Xi}$ is the stochastic force and $\mathbf{\Xi}$ is the stochastic stress tensor (a symmetric matrix). $\mathbf{\Xi}$ is Gaussian with zero mean and, for incompressible fluids at thermodynamic equilibrium, has the following correlation as a manifestation of the FDT \cite{HydrodynamicFluctuations}:
\begin{equation}\label{Correlation}
\langle \Xi_{ij}(\mathbf{x},t)\Xi_{kl}(\mathbf{x}',t')\rangle=2\nu k_BT(\delta_{ik}\delta_{jl}+\delta_{il}\delta_{jk})\delta(\mathbf{x}-\mathbf{x}')\delta(t-t'),
\end{equation}
where  $k_B$ is the Boltzmann constant and $T$ (uniform and constant) is the equilibrium temperature of the fluid. The correlation of $\bm{\xi}$ can then be calculated as follows
\begin{equation}
\begin{split}
\langle\xi_i(\mathbf{x},t)\xi_k(\mathbf{x}',t')\rangle&=\partial_j\partial'_l\langle\Xi_{ij}(\mathbf{x},t)\Xi_{kl}(\mathbf{x}',t')\rangle\\
&=-2\nu k_BT(\delta_{ik}\Delta+\partial_i\partial_k)\delta(\mathbf{x}-\mathbf{x}')\delta(t-t').
\end{split}
\end{equation}
This means the diffusion matrix as the spatial correlator of $\bm{\xi}$  has the form
\begin{equation}\label{DBDM}
\mathbf{D}(\mathbf{x}-\mathbf{x}')=-\nu k_BT(\mathbf{I}\Delta+\nabla\nabla)\delta(\mathbf{x}-\mathbf{x}').
\end{equation}

\emph{Solenoidal Navier-Stokes equation.} The solenoidal form of the LLNSE (see Eq. (\ref{SNSESF}))  reads
\begin{equation}\label{SFLLNSE}
\partial_t \mathbf{u}=\mathbf{\Pi}^s(\nabla)\cdot(-\mathbf{u}\cdot\nabla\mathbf{u})+\nu \Delta\mathbf{u}+\bm{\xi}^s,
\end{equation}
where $\bm{\xi}^{s}=\mathbf{\Pi}^s(\nabla)\cdot\bm{\xi}$ is the solenoidal stochastic force.  According to Eq. (\ref{RDMSDM}), we obtain, after some algebra, the solenoidal diffusion matrix $\mathbf{D}^s(\mathbf{x}-\mathbf{x}')$ as the spatial correlator of $\bm{\xi}^{s}$:
\begin{equation}\label{EDMPF}
\mathbf{D}^s(\mathbf{x}-\mathbf{x}')=-\nu k_BT(\mathbf{I}\Delta-\nabla\nabla)\delta(\mathbf{x}-\mathbf{x}')=\mathbf{\Pi}^s(\nabla)\left[-\nu k_BT\Delta \delta(\mathbf{x}-\mathbf{x}')\right].
\end{equation}
The last step can be proven rigorously with Fourier analysis, but is understood most directly with the heuristic relation $\mathbf{\Pi}^s(\nabla)\simeq\mathbf{I}-\nabla\Delta^{-1}\nabla$.

\emph{Detailed balance constraint.} Now we show that the diffusion matrix in Eq. (\ref{EDMPF}) satisfies the detailed balance constraint in Eq. (\ref{DBCENSE}). It is not easy to calculate the inverse of the diffusion matrix $\mathbf{D}_s^{-1}(\mathbf{x}-\mathbf{x}')$ in the physical space directly. But we can get around this difficulty by the following calculation. For $\mathbf{u}(\mathbf{x})$ in the state space, we have $\int\mathbf{D}^s(\mathbf{x}-\mathbf{x}')\cdot\mathbf{u}(\mathbf{x}')d\mathbf{x}'=\int \mathbf{\Pi}^s(\nabla)\left[-\nu k_BT\Delta \delta(\mathbf{x}-\mathbf{x}')\right]\cdot\mathbf{u}(\mathbf{x}')d\mathbf{x}'=\mathbf{\Pi}^s(\nabla)\cdot\left[-\nu k_BT\Delta \mathbf{u}(\mathbf{x})\right]=-\nu k_BT\Delta\mathbf{u}(\mathbf{x})$, where we have used Eq. (\ref{EDMPF}) and that $\mathbf{u}(\mathbf{x})$ is solenoidal. Inverting this linear equation, $\int\mathbf{D}^s(\mathbf{x}-\mathbf{x}')\cdot\mathbf{u}(\mathbf{x}')d\mathbf{x}'=-\nu k_BT\Delta\mathbf{u}(\mathbf{x})$, we obtain
\begin{equation}\label{InverseDMEX}
\int\mathbf{D}_s^{-1}(\mathbf{x}-\mathbf{x}')\cdot[\nu \Delta'\mathbf{u}(\mathbf{x}')]d\mathbf{x}'=-\DF{1}{k_BT}\mathbf{u}(\mathbf{x}).
\end{equation}

As a result, the detailed balance constraint in Eq. (\ref{DBCENSE}) now reduces to
\begin{equation}
\int d\mathbf{x}\,[\mathbf{\Pi}^{s}(\nabla)\cdot\left(-\mathbf{u}\cdot\nabla\mathbf{u}\right)]\cdot\mathbf{u}(\mathbf{x})=0.
\end{equation}
This is true since $\int d\mathbf{x}\,[\mathbf{\Pi}^{s}(\nabla)\cdot\left(-\mathbf{u}\cdot\nabla\mathbf{u}\right)]\cdot\mathbf{u}(\mathbf{x})=\int d\mathbf{x}\,\left(-\mathbf{u}\cdot\nabla\mathbf{u}\right)\cdot \mathbf{\Pi}^{s}(\nabla)\cdot\mathbf{u}(\mathbf{x})=\int d\mathbf{x}\,\left(-\mathbf{u}\cdot\nabla\mathbf{u}\right)\cdot \mathbf{u}(\mathbf{x})=-\int d\mathbf{x}\,\nabla\cdot(|\mathbf{u}|^2\mathbf{u}/2)=-\oint d\mathbf{s}\cdot(\mathbf{u}|\mathbf{u}|^2/2)=0$, where we have used the fact that $\mathbf{\Pi}^{s}(\nabla)$ is a self-adjoint (Hermitian) operator in the state space and that $\mathbf{u}(\mathbf{x})$ is solenoidal, together with the Gauss theorem in the physical space and the periodic boundary conditions. Physically, this is simply the statement that the (solenoidal) convective force does no net work to the fluid as a whole; it does not change the total kinetic energy of the fluid.

Therefore, we have shown that the LLNSS at thermodynamic equilibrium, with the diffusion matrix in Eq. (\ref{DBDM}) and its solenoidal form in Eq. (\ref{EDMPF}), obeys the detailed balance constraint that we have derived for equilibrium fluid systems. This demonstrates that the detailed balance constraint can indeed capture the equilibrium nature of the fluid systems we considered.

\emph{Equilibrium steady state.} The steady state of the LLNSS with detailed balance is an equilibrium state. Comparing Eq. (\ref{InverseDMEX}) with Eq. (\ref{NSEPCPFL}), we see that $\delta_{\mathbf{u}(\mathbf{x})}\Phi_{eq}[\mathbf{u}]=(1/k_BT)\mathbf{u}(\mathbf{x})$. Thus the equilibrium potential landscape functional can be solved as
\begin{equation}
\Phi_{eq}[\mathbf{u}]=\DF{1}{k_BT}H[\mathbf{u}]=\DF{1}{2k_BT}\int \left|\mathbf{u}(\mathbf{x})\right|^2 d\mathbf{x},
\end{equation}
where $H[\mathbf{u}]=(1/2)\int \left|\mathbf{u}(\mathbf{x})\right|^2d\mathbf{x}$ is the Hamiltonian (total kinetic energy) of the fluid system.

Consequently, the equilibrium steady-state probability distribution functional, according to the relation $P_{eq}[\mathbf{u}]=e^{-\Phi_{eq}[\mathbf{u}]}$, is given by
\begin{equation}
P_{eq}[\mathbf{u}]=\mathcal{N}\exp\left\{-\DF{1}{k_BT}H[\mathbf{u}]\right\}=\mathcal{N}\exp\left\{-\DF{1}{2k_BT}\int \left|\mathbf{u}(\mathbf{x})\right|^2 d\mathbf{x}\right\},
\end{equation}
where $\mathcal{N}$ is the normalization constant. This Gaussian probability distribution functional $P_{eq}[\mathbf{u}]$ has the form of the canonical ensemble, in agreement with the equilibrium nature of the steady state. One can verify that $P_{eq}[\mathbf{u}]$  is indeed the steady-state solution to the FFPE in Eq. (\ref{FFPENSE}) for the diffusion matrix in Eq. (\ref{EDMPF}).

The steady-state irreversible probability flux $\mathbf{J}_s^{irr}(\mathbf{x})[\mathbf{u}]$ in this case vanishes completely, which is essentially equivalent to the condition we used to solve the potential landscape. The steady-state probability flux is therefore completely reversible, i.e., $\mathbf{J}_s(\mathbf{x})[\mathbf{u}]=\mathbf{J}_s^{rev}(\mathbf{x})[\mathbf{u}]$. This is also characteristic of the equilibrium steady state.

\emph{Equilibrium stochastic dynamics.} The solenoidal form of the LLNSE can be reformulated in terms of the Hamiltonian, $H[\mathbf{u}]=k_BT \Phi_{eq}[\mathbf{u}]$, in the following potential form in the state space:
\begin{equation}\label{GLENSEEX3}
\partial_t\mathbf{u}(\mathbf{x},t)=\mathbf{F}^{rev}(\mathbf{x})[\mathbf{u}] -\int \mathbf{M}^{s}(\mathbf{x}-\mathbf{x}')\cdot\delta_{\mathbf{u}(\mathbf{x}')}H[\mathbf{u}]d\mathbf{x}' +\bm{\xi}^s(\mathbf{x},t),
\end{equation}
where $\mathbf{M}^{s}(\mathbf{x}-\mathbf{x}')$ is the viscous damping matrix defined as
\begin{equation}
\mathbf{M}^{s}(\mathbf{x}-\mathbf{x}')=(1/k_BT)\mathbf{D}^{s}(\mathbf{x}-\mathbf{x}')=\mathbf{\Pi}^s(\nabla)\left[-\nu \Delta \delta(\mathbf{x}-\mathbf{x}')\right].
\end{equation}

The three driving forces in Eq. (\ref{GLENSEEX3}) are linked to each other by detailed balance at thermodynamic equilibrium. The FDT in Eq. (\ref{GLENSEEX3}), which relates the viscous force to the stochastic force, now has the form
\begin{equation}
\langle \bm{\xi}^s(\mathbf{x},t)\bm{\xi}^s(\mathbf{x}',t') \rangle = 2k_BT\mathbf{M}^{s}(\mathbf{x}-\mathbf{x}')\delta(t-t').
\end{equation}
In addition, Eq. (\ref{GLENSEEX3}) also satisfies the orthogonality condition, $\int \mathbf{F}^{rev}(\mathbf{x})[\mathbf{u}]\cdot\delta_{\mathbf{u}(\mathbf{x})}H[\mathbf{u}]d\mathbf{x}=0$, which links the solenoidal convective force to the functional gradient of the Hamiltonian that appears in the viscous force.

It is not yet clear to us how much room is left in the form of the diffusion matrix, other than that in the above particular example, which can fulfill the detailed balance constraint to accommodate an equilibrium steady state. We leave this issue open for further investigation.

\subsection{Nonequilibrium fluid systems with detailed balance breaking}

The equilibrium fluid system preserving detailed balance depends on the stochastic force with a spatial correlation (diffusion matrix) in balance with the irreversible viscous force and the reversible solenoidal convective force, characterized by the detailed balance constraint. When the spatial correlator (diffusion matrix) of the stochastic force violates the detailed balance constraint (i.e., detailed balance breaking), the three driving forces in the fluid system become mechanically imbalanced and the steady state of the fluid system becomes a nonequilibrium state without time reversal symmetry. Mathematically, detailed balance breaking as the violation of the detailed balance constraint is represented by
\begin{equation}\label{DBB}
B[\mathbf{u}]\equiv \iint d\mathbf{x}d\mathbf{x}'\,[\mathbf{\Pi}^{s}(\nabla)\cdot\left(-\mathbf{u}\cdot\nabla\mathbf{u}\right)]\cdot\mathbf{D}_s^{-1}(\mathbf{x}-\mathbf{x}')\cdot[\nu \Delta'\mathbf{u}(\mathbf{x}')]\neq 0,
\end{equation}
where we have introduced $B[\mathbf{u}]$, a scalar functional of the velocity field, to characterize detailed balance breaking. We shall refer to $B[\mathbf{u}]$ as the detailed balance breaking functional. Detailed balance is preserved when $B[\mathbf{u}]$ vanishes for \emph{all} velocity fields in the state space. Detailed balance is violated when $B[\mathbf{u}]$ is nonvanishing at least for \emph{some} velocity fields in the state space.

\subsubsection{Potential-flux decomposition of the irreversible viscous force}

A principal indicator of detailed balance breaking is the presence of nonvanishing steady-state irreversible probability flux $\mathbf{J}_{s}^{irr}(\mathbf{x})[\mathbf{u}]$, which signifies time irreversibility in nonequilibrium steady states.  As a consequence, for nonequilibrium fluid systems without detailed balance, we obtain from  Eq. (\ref{SSIPFFFNSE}), by dividing both sides with $P_s[\mathbf{u}]$, the \emph{force decomposition equation} in the potential landscape and flux field theory  \cite{WWJW20141,WWJW20142}:
\begin{equation}\label{FDENEQNSE}
\nu \Delta\mathbf{u} =-\int d\mathbf{x}'\,\mathbf{D}^s(\mathbf{x}-\mathbf{x}')\cdot\delta_{\mathbf{u}(\mathbf{x}')}\Phi[\mathbf{u}]+\mathbf{V}_s^{irr}(\mathbf{x})[\mathbf{u}],
\end{equation}
which reads symbolically $F^{irr}=-D^s\cdot \partial\Phi+V_s^{irr}$. Here $\Phi[\mathbf{u}]=-\ln P_s[\mathbf{u}]$ is the nonequilibrium potential landscape, connected to the steady-state probability distribution $P_s[\mathbf{u}]=e^{-\Phi[\mathbf{u}]}$, and $\mathbf{V}_s^{irr}(\mathbf{x})[\mathbf{u}]=\mathbf{J}_s^{irr}(\mathbf{x})[\mathbf{u}]/P_s[\mathbf{u}]$ is the steady-state irreversible probability flux velocity, related to the irreversible steady-state probability flux by $\mathbf{J}_s^{irr}(\mathbf{x})[\mathbf{u}]=P_s[\mathbf{u}]\mathbf{V}_s^{irr}(\mathbf{x})[\mathbf{u}]$. As with $\mathbf{J}_s^{irr}(\mathbf{x})[\mathbf{u}]$, nonvanishing $\mathbf{V}_s^{irr}(\mathbf{x})[\mathbf{u}]$ is also an indicator of detailed balance breaking.

In contrast with the potential form of the irreversible viscous force in Eq. (\ref{NSEPC}) for equilibrium fluid systems with detailed balance, for nonequilibrium fluid systems with detailed balance breaking, the irreversible viscous force has a potential-flux decomposed form consisting of two parts as seen in Eq. (\ref{FDENEQNSE}). The first part has a potential form, expressed as the functional gradient of the potential landscape, modified by the diffusion matrix characterizing the stochastic force. This part also exists in the equilibrium fluid system with detailed balance as shown in Eq. (\ref{NSEPC}). The other part is a flux force, expressed as the steady-state irreversible probability flux velocity, which indicates detailed balance breaking. This flux force that signifies detailed balance breaking is absent from the equilibrium fluid system with detailed balance. The potential-flux decomposition of the irreversible viscous force is the manifestation of detailed balance breaking on the level of the dynamical driving force in nonequilibrium fluid systems without time reversal symmetry.

\subsubsection{Potential-flux coupling}

We now show that the irreversible flux velocity field $\mathbf{V}_s^{irr}(\mathbf{x})[\mathbf{u}]$ is tightly coupled with the nonequilibrium potential landscape $\Phi[\mathbf{u}]$ and that they are related to the deviation of the steady-state probability distribution from being Gaussian. We first note that the force decomposition equation in Eq. (\ref{FDENEQNSE}) can also be viewed as an equation defining the expression of $\mathbf{V}_s^{irr}(\mathbf{x})[\mathbf{u}]$, which reads
\begin{equation}\label{EISSPFVNSE}
\mathbf{V}_s^{irr}(\mathbf{x})[\mathbf{u}]=\nu \Delta\mathbf{u}+\int d\mathbf{x}'\,\mathbf{D}^s(\mathbf{x}-\mathbf{x}')\cdot\delta_{\mathbf{u}(\mathbf{x}')}\Phi[\mathbf{u}].
\end{equation}
This equation expresses the irreversible flux velocity in terms of the irreversible viscous force, the diffusion matrix associated with the stochastic force, and also the potential landscape. We know that $\Phi[\mathbf{u}]$ is defined in terms of the steady-state probability distribution $P_s[\mathbf{u}]=e^{-\Phi[\mathbf{u}]}$, which is actually determined by the solenoidal convective force, the viscous force, and the stochastic force together from the steady-state FFPE (see Eq. (\ref{FFPENSE})). Therefore, the irreversible flux velocity field $\mathbf{V}_s^{irr}(\mathbf{x})[\mathbf{u}]$ is essentially a manifestation of all three forces rather than only two forces (the viscous and stochastic forces). This agrees with the understanding that nonvanishing $\mathbf{J}_s^{irr}[\mathbf{u}]$ (and thus $\mathbf{V}_s^{irr}[\mathbf{u}]$) stems from the violation of the detailed balance constraint that relates all three driving forces. Later we shall see explicitly how this is so.

Then we recall from the discussions of the potential condition in Section \ref{PotentialCondition} that, for a given diffusion matrix $\mathbf{D}^s(\mathbf{x}-\mathbf{x}')$, the viscous force always has a potential form (see Eqs. (\ref{NSEPCPFL}) and (\ref{Phi0NSE})):
\begin{equation}\label{GPFBNSE}
\nu\Delta\mathbf{u}=-\int d\mathbf{x}'\, \mathbf{D}^s(\mathbf{x}-\mathbf{x}')\cdot\delta_{\mathbf{u}(\mathbf{x}')}\Phi_0[\mathbf{u}],
\end{equation}
where
\begin{equation}\label{GPDBNSE}
\Phi_0[\mathbf{u}]=\DF{\nu}{2}\iint d\mathbf{x}d\mathbf{x}'\,\mathbf{u}(\mathbf{x})\cdot\left[-\Delta\mathbf{D}_s^{-1}(\mathbf{x}-\mathbf{x}')\right]\cdot\mathbf{u}(\mathbf{x}').
\end{equation}
This form is not restricted to equilibrium fluid systems with detailed balance. It still holds even if detailed balance is broken as is the case we are considering. Here $\Phi_0[\mathbf{u}]$ cannot be interpreted as the equilibrium potential landscape since the detailed balance constraint is violated. It is not the nonequilibrium potential landscape either, because the form in Eq. (\ref{GPFBNSE}) differs from the force decomposition equation in Eq. (\ref{FDENEQNSE}), where the irreversible flux velocity is nonvanishing for nonequilibrium systems without detailed balance. Therefore, $\Phi_0[\mathbf{u}]$ is not a potential landscape that can be directly related to the equilibrium or nonequilibrium steady-state probability distributions. But it serves as a good benchmark in the study of nonequilibrium fluid  systems without detailed balance, as it is directly determined by the linear part of the dynamics of the system (the viscous force and the stochastic force). We shall refer to it as the Gaussian potential landscape given its relation to the Gaussian probability distribution functional $P_0[\mathbf{u}]=e^{-\Phi_0[\mathbf{u}]}$ (up to a normalization constant).

We introduce the deviated potential landscape defined as
\begin{equation}
\Lambda[\mathbf{u}]=\Phi_0[\mathbf{u}]-\Phi[\mathbf{u}]=\ln (P_s[\mathbf{u}]/P_0[\mathbf{u}]),
\end{equation}
where $\Phi_0[\mathbf{u}]$ is the Gaussian potential landscape associated with the Gaussian distribution functional $P_0[\mathbf{u}]=e^{-\Phi_0[\mathbf{u}]}$ and $\Phi[\mathbf{u}]$ is the nonequilibrium potential landscape associated with the nonequilibrium steady-state probability distribution functional $P_s[\mathbf{u}]=e^{-\Phi[\mathbf{u}]}$. The deviated potential landscape $\Lambda[\mathbf{u}]$ directly characterizes the deviation of the nonequilibrium potential landscape $\Phi[\mathbf{u}]$ from the Gaussian potential landscape $\Phi_0[\mathbf{u}]$. It also characterizes the deviation of the nonequilibrium steady-state distribution $P_s[\mathbf{u}]$ from the Gaussian distribution $P_0[\mathbf{u}]$ since they are related by $P_s[\mathbf{u}]=P_0[\mathbf{u}]e^{\Lambda[\mathbf{u}]}$. If $\Lambda[\mathbf{u}]$ is a constant (independent of $\mathbf{u}$), then $P_s[\mathbf{u}]$ coincides with $P_0[\mathbf{u}]$ (up to a normalization constant); otherwise, it indicates that $P_s[\mathbf{u}]$ is deviated from  $P_0[\mathbf{u}]$. Therefore, the word `deviation' (or `deviated'), which indicates deviation from being Gaussian, carries a double meaning in this context, indicating the deviation of the nonequilibrium potential landscape $\Phi[\mathbf{u}]$ (the nonequilibrium steady-state probability distribution functional $P_s[\mathbf{u}]$) from being the Gaussian potential landscape $\Phi_0[\mathbf{u}]$ (the Gaussian probability distribution functional $P_0[\mathbf{u}]$).

Plugging Eq. (\ref{GPFBNSE}) into Eq. (\ref{EISSPFVNSE}), we obtain the following \emph{flux deviation relation}:
\begin{equation}\label{FDRNSE}
\mathbf{V}_s^{irr}(\mathbf{x})[\mathbf{u}]=-\int d\mathbf{x}'\, \mathbf{D}^s(\mathbf{x}-\mathbf{x}')\cdot\delta_{\mathbf{u}(\mathbf{x}')}\Lambda[\mathbf{u}].
\end{equation}
The flux deviation relation establishes a connection between the steady-state irreversible flux velocity $\mathbf{V}_s^{irr}[\mathbf{u}]$ that signifies detailed balance breaking in nonequilibrium steady states with intrinsic time irreversibility and the deviated potential landscape $\Lambda[\mathbf{u}]$ that characterizes the deviation of the nonequilibrium steady-state probability distribution (nonequilibrium potential landscape) from the Gaussian distribution (Gaussian potential landscape). Since $\mathbf{D}^s(\mathbf{x}-\mathbf{x}')$ is invertible in the state space, this equation shows that $\mathbf{V}_s^{irr}[\mathbf{u}]$ vanishing is equivalent to $\Lambda[\mathbf{u}]$ being constant. Hence, $\Lambda[\mathbf{u}]$ can also serve as an indicator of detailed balance breaking as $\mathbf{V}_s^{irr}[\mathbf{u}]$ is. If $\Lambda[\mathbf{u}]$ is a constant, which means the steady-state distribution $P_s[\mathbf{u}]$ (the potential landscape $\Phi[\mathbf{u}]$) coincides with the Gaussian distribution $P_0[\mathbf{u}]$ (the Gaussian potential landscape $\Phi_0[\mathbf{u}]$), then $\mathbf{V}_s^{irr}[\mathbf{u}]$ vanishes according to the flux deviation relation, which indicates detailed balance and time reversibility in equilibrium steady states. If $\Lambda[\mathbf{u}]$ is not a constant, which means the steady-state distribution $P_s[\mathbf{u}]$ (the potential landscape $\Phi[\mathbf{u}]$) deviates from the Gaussian distribution $P_0[\mathbf{u}]$ (the Gaussian potential landscape $\Phi_0[\mathbf{u}]$), then $\mathbf{V}_s^{irr}[\mathbf{u}]$ is nonvanishing according to the flux deviation relation, which indicates detailed balance breaking and time irreversibility in nonequilibrium steady states.

This shows that detailed balance breaking as the violation of the detailed balance constraint has two closely related characteristic consequences, namely, the irreversible probability flux velocity $\mathbf{V}_s^{irr}[\mathbf{u}]$ that characterizes time reversal symmetry breaking, and the deviated potential landscape $\Lambda[\mathbf{u}]$ that characterizes the deviation from the Gaussian probability distribution $P_0[\mathbf{u}]$ (the Gaussian potential landscape $\Phi_0[\mathbf{u}]$). The non-Gaussian potential landscape and the irreversible probability flux are tightly coupled to each other by the flux deviation relation; both are deeply rooted in detailed balance breaking characterized by the violation of the detailed balance constraint that represents a form of mechanical imbalance in the driving forces of the fluid system.

\subsubsection{Nonequilibrium trinity}

We establish the nonequilibrium trinity by showing explicitly how detailed balance breaking directly gives rise to the two interrelated consequences, namely, the non-Gaussian potential landscape and the irreversible probability flux. This can be shown clearly from the equation governing $\Lambda[\mathbf{u}]$, which can be obtained in principle from the steady-state FFPE for $P_s[\mathbf{u}]$ expressed in terms of $\Lambda[\mathbf{u}]$ through the relation $P_s[\mathbf{u}]=P_0[\mathbf{u}]e^{\Lambda[\mathbf{u}]}$.

But we adopt a more strategic approach to obtain the equation governing $\Lambda[\mathbf{u}]$. We first introduce the total and reversible steady-state probability flux velocities defined by $\mathbf{V}_s[\mathbf{u}]=\mathbf{J}_s[\mathbf{u}]/P_s[\mathbf{u}]$ and $\mathbf{V}_s^{rev}[\mathbf{u}]=\mathbf{J}_s^{rev}[\mathbf{u}]/P_s[\mathbf{u}]$, respectively, in accord with $\mathbf{V}_s^{irr}[\mathbf{u}]=\mathbf{J}_s^{irr}[\mathbf{u}]/P_s[\mathbf{u}]$ already introduced. They have the relation $\mathbf{V}_s[\mathbf{u}]=\mathbf{V}_s^{rev}[\mathbf{u}]+\mathbf{V}_s^{irr}[\mathbf{u}]$. The reversible flux velocity, according to $\mathbf{J}_s^{rev}[\mathbf{u}]$ in Eq. (\ref{SSRPFFFNSE}), is simply given by $\mathbf{V}_s^{rev}[\mathbf{u}]=\mathbf{F}^{rev}[\mathbf{u}]=\mathbf{\Pi}^{s}(\nabla)\cdot\left(-\mathbf{u}\cdot\nabla\mathbf{u}\right)$, i.e., the solenoidal convective force. The expression of the irreversible flux velocity $\mathbf{V}_s^{irr}[\mathbf{u}]$ has been given in Eq. (\ref{EISSPFVNSE}), which is also expressed in terms of $\Lambda[\mathbf{u}]$ through the flux deviation relation in Eq. (\ref{FDRNSE}). Thus $\mathbf{V}_s[\mathbf{u}]$ has the following expression in relation to $\Lambda[\mathbf{u}]$:
\begin{equation}\label{SSPFITOL}
\mathbf{V}_s(\mathbf{x})[\mathbf{u}]=\mathbf{\Pi}^{s}(\nabla)\cdot\left(-\mathbf{u}\cdot\nabla\mathbf{u}\right)-\int d\mathbf{x}' \mathbf{D}^s(\mathbf{x}-\mathbf{x}')\cdot\delta_{\mathbf{u}(\mathbf{x}')}\Lambda[\mathbf{u}],
\end{equation}
which reads symbolically $V_s=F^{rev}-D^s\cdot\partial\Lambda$.

Then we notice that the steady-state FFPE in the form $\int d\mathbf{x}\,\delta_{\mathbf{u}}\cdot\mathbf{J}_s[\mathbf{u}]=0$, where $\mathbf{J}_s[\mathbf{u}]=\mathbf{V}_s[\mathbf{u}]P_s[\mathbf{u}]$ and $P_s[\mathbf{u}]=e^{-\Phi[\mathbf{u}]}$, can be reexpressed in terms of $\Phi[\mathbf{u}]$ and $\mathbf{V}_s[\mathbf{u}]$ as follows:
\begin{equation}\label{LFRNSE}
\int d\mathbf{x}\,\mathbf{V}_s(\mathbf{x})[\mathbf{u}]\cdot\delta_{\mathbf{u}(\mathbf{x})}\Phi[\mathbf{u}]=\int d\mathbf{x}\, \delta_{\mathbf{u}(\mathbf{x})}\cdot\mathbf{V}_s(\mathbf{x})[\mathbf{u}],
\end{equation}
which has the symbolic representation $V_s\cdot\partial \Phi=\partial\cdot V_s$. The l.h.s. of the equation is the inner product in the state space between $\mathbf{V}_s[\mathbf{u}]$ and $\delta_{\mathbf{u}}\Phi[\mathbf{u}]$; the r.h.s. of the equation is the functional divergence of $\mathbf{V}_s[\mathbf{u}]$.

Plugging $\mathbf{V}_s(\mathbf{x})[\mathbf{u}]$ in Eq. (\ref{SSPFITOL}) and $\Phi[\mathbf{u}]=\Phi_0[\mathbf{u}]-\Lambda[\mathbf{u}]$ into Eq. (\ref{LFRNSE}) and using Eq. (\ref{GPFBNSE}) with its inverse, we finally obtain the equation governing $\Lambda[\mathbf{u}]$, which we term the \emph{nonequilibrium source equation}:
\begin{eqnarray}\label{DENSE}
&&\iint d\mathbf{x}d\mathbf{x}'\left(\delta_{\mathbf{u}(\mathbf{x})}\Lambda[\mathbf{u}]\right) \cdot \mathbf{D}^s(\mathbf{x}-\mathbf{x}')\cdot\left(\delta_{\mathbf{u}(\mathbf{x}')}\Lambda[\mathbf{u}]\right)\nonumber\\
&+&\iint d\mathbf{x}d\mathbf{x}' \delta_{\mathbf{u}(\mathbf{x})}\cdot\mathbf{D}^s(\mathbf{x}-\mathbf{x}')\cdot\delta_{\mathbf{u}(\mathbf{x}')}\Lambda[\mathbf{u}]\nonumber\\
&+&\int d\mathbf{x}\, \left[\nu\Delta\mathbf{u}-\mathbf{\Pi}^{s}(\nabla)\cdot\left(-\mathbf{u}\cdot\nabla\mathbf{u}\right)\right]\cdot\delta_{\mathbf{u}(\mathbf{x})}\Lambda[\mathbf{u}]\nonumber\\
&=&\iint d\mathbf{x}d\mathbf{x}'\,[\mathbf{\Pi}^{s}(\nabla)\cdot\left(-\mathbf{u}\cdot\nabla\mathbf{u}\right)]\cdot\mathbf{D}_s^{-1}(\mathbf{x}-\mathbf{x}')\cdot[\nu \Delta'\mathbf{u}(\mathbf{x}')],
\end{eqnarray}
which in the symbolic form reads
\begin{equation}
\partial\Lambda\cdot D^s\cdot\partial\Lambda+\partial\cdot D^s\cdot \partial \Lambda+(F^{irr}-F^{rev})\cdot\partial\Lambda=F^{rev}\cdot D_s^{-1}\cdot F^{irr}.
\end{equation}

This nonequilibrium source equation is essentially the steady-state FFPE reformulated in terms of $\Lambda[\mathbf{u}]$. We only need to observe two simple features in this seemingly formidable equation. The first feature is that all the three terms on the l.h.s. of this equation contain $\delta_{\mathbf{u}}\Lambda[\mathbf{u}]$, which characterizes the nonequilibrium quality of the system. The second feature is that the r.h.s. of this equation, acting as a source term to the equation, is exactly the detailed balance breaking functional $B[\mathbf{u}]$ introduced in Eq. (\ref{DBB}), which characterizes the violation of the detailed balance constraint when it is nonvanishing.

The significance of the nonequilibrium source equation is seen as follows. When the detailed balance constraint is obeyed, the source term $B[\mathbf{u}]$ on the r.h.s. of Eq. (\ref{DENSE}) vanishes, which allows for a constant solution $\Lambda[\mathbf{u}]=\text{const}$. If the steady-state distribution $P_s[\mathbf{u}]$ to the FFPE is unique under suitable conditions, then $\Lambda[\mathbf{u}]$ as a solution to Eq. (\ref{DENSE}) will also be unique since $\Lambda[\mathbf{u}]=\ln (P_s[\mathbf{u}]/P_0[\mathbf{u}])$. In that case, constant $\Lambda[\mathbf{u}]$ will be the only solution to Eq. (\ref{DENSE}) when detailed balance holds. It then follows  that the steady-state probability distribution $P_s[\mathbf{u}]=P_0[\mathbf{u}]e^{\Lambda[\mathbf{u}]}$ coincides with the Gaussian distribution $P_0[\mathbf{u}]$ and that the irreversible probability flux velocity $\mathbf{V}_s^{irr}[\mathbf{u}]$ vanishes according to the flux deviation relation in Eq. (\ref{FDRNSE}). In contrast, when the detailed balance constraint is violated, the source term $B[\mathbf{u}]$ on the r.h.s. of Eq. (\ref{DENSE}) is nonzero at least for some velocity fields, which generates a nonconstant solution $\Lambda[\mathbf{u}]$ to Eq. (\ref{DENSE}). Accordingly,  the steady-state distribution $P_s[\mathbf{u}]$ deviates from the Gaussian distribution $P_0[\mathbf{u}]$ since $P_s[\mathbf{u}]=P_0[\mathbf{u}]e^{\Lambda[\mathbf{u}]}$ and the irreversible probability flux velocity $\mathbf{V}_s^{irr}[\mathbf{u}]$ does not vanish given the flux deviation relation.

\begin{figure}[!ht]
\centering
\includegraphics[width=4in]{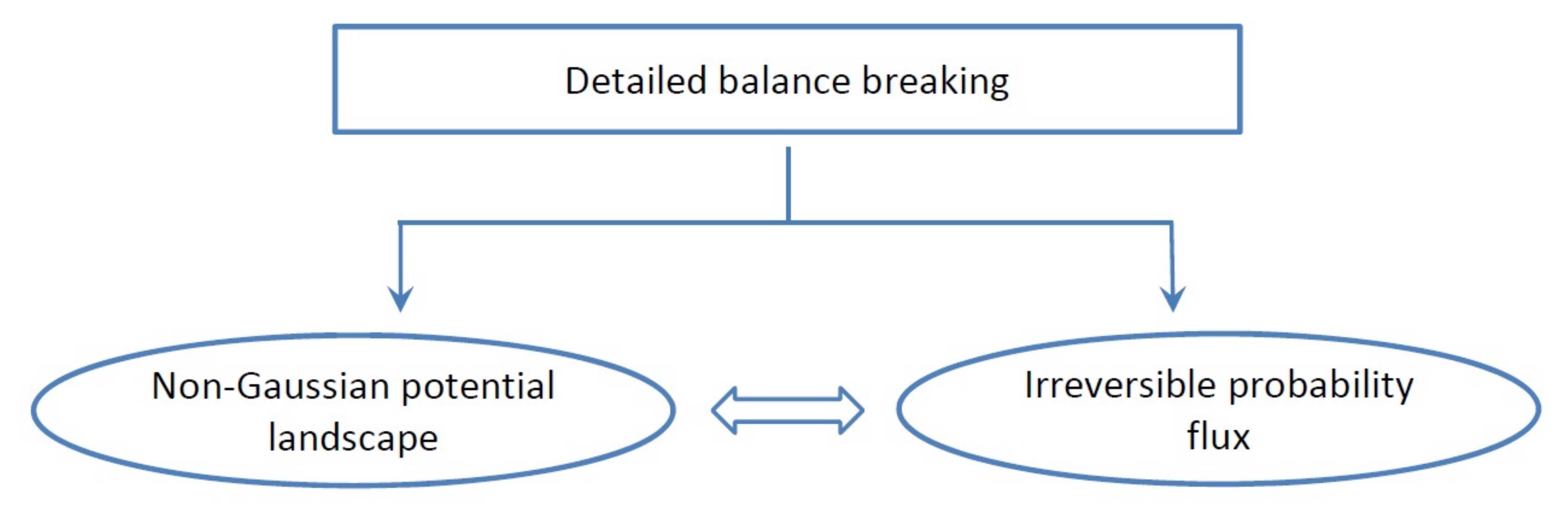}
\caption{A schematic representation of the nonequilibrium trinity.}\label{figure1}
\end{figure}

This demonstrates clearly how detailed balance breaking as the violation of the detailed balance constraint, representing a form of mechanical imbalance in the three driving forces of the fluid system, is the very source that drives the potential landscape (steady-state probability distribution) to deviate from being Gaussian and generates the steady-state irreversible probability flux velocity (steady-state irreversible probability flux), with these two consequential aspects connected to each other by the flux deviation relation. We have thus established the `nonequilibrium trinity', namely, detailed balance breaking, non-Gaussian potential landscape and irreversible probability flux, which captures the nonequilibrium irreversible nature of nonequilibrium fluid  systems with intrinsic time irreversibility. The nonequilibrium trinity is mathematically endorsed by the nonequilibrium source equation in Eq. (\ref{DENSE}) and the flux deviation relation in Eq. (\ref{FDRNSE}). A schematic representation of the nonequilibrium trinity is shown in Fig. \ref{figure1}.

\subsubsection{Nonequilibrium stochastic fluid dynamics}

Detailed balance breaking and the resulting nonequilibrium trinity have implications for the stochastic fluid dynamics. In view of the force decomposition equation in Eq. (\ref{FDENEQNSE}), where the irreversible viscous force has the potential-flux decomposition form as a result of detailed balance breaking, the solenoidal stochastic Navier-Stokes equation in Eq. (\ref{SNSESF}), for nonequilibrium fluid  systems without detailed balance, can be reformulated in the following potential-flux form:
\begin{equation}\label{PFFSNSE}
\partial_t\mathbf{u}(\mathbf{x},t)=\mathbf{F}^{rev}(\mathbf{x})[\mathbf{u}]-\int d\mathbf{x}'\,\mathbf{D}^s(\mathbf{x}-\mathbf{x}')\cdot\delta_{\mathbf{u}(\mathbf{x}')}\Phi[\mathbf{u}]+\mathbf{V}_s^{irr}(\mathbf{x})[\mathbf{u}]+\bm{\xi}^s(\mathbf{x},t),
\end{equation}
where $\Phi[\mathbf{u}]=-\ln P_s[\mathbf{u}]$, $\mathbf{V}_s^{irr}(\mathbf{x})[\mathbf{u}]=\mathbf{J}_s^{irr}(\mathbf{x})[\mathbf{u}]/P_s[\mathbf{u}]$, and $\langle\bm{\xi}^s(\mathbf{x},t)\bm{\xi}^s(\mathbf{x}',t')\rangle=2\mathbf{D}^s(\mathbf{x}-\mathbf{x}')\delta(t-t')$.

The potential-flux form of the stochastic Navier-Stokes equation shows that the nonequilibrium dynamics of stochastic fluid systems with detailed balance breaking is governed by both the potential landscape functional gradient and the irreversible probability flux velocity, together with the solenoidal convective force and the stochastic force. The nonequilibrium potential landscape $\Phi[\mathbf{u}]$ and the irreversible flux velocity $\mathbf{V}_s^{irr}(\mathbf{x})[\mathbf{u}]$ play a dual role in establishing a connection between the individual trajectory level and the collective ensemble level. On the one hand, they act together as the irreversible viscous force in the Langevin stochastic field dynamics that governs the evolution of individual stochastic trajectories. One the other hand, they are connected to the steady-state probability distribution and probability flux in the Fokker-Planck field dynamics that governs the evolution of the collective ensemble.

Compared with the potential form of the stochastic Navier-Stokes equation in Eq. (\ref{GLENSE}) for equilibrium fluid systems with detailed balance, the potential-flux form of the stochastic Navier-Stokes equation for nonequilibrium fluid systems without detailed balance has a different structure.  Most prominently, there is an additional driving force, the steady-state irreversible probability flux velocity $\mathbf{V}_s^{irr}(\mathbf{x})[\mathbf{u}]$, which originates from detailed balance breaking and signifies time irreversibility in the nonequilibrium steady state. Hence, the nonequilibrium stochastic fluid dynamics is additionally powered by detailed balance breaking, in comparison to the equilibrium stochastic fluid dynamics with detailed balance. This extra driving force from detailed balance breaking has never been identified before in nonequilibrium fluid dynamics and turbulence dynamics in particular until now. We shall demonstrate in the next section that the energy flux in turbulence energy cascade associated with the breaking up of large vortices into smaller ones is actually powered by this new driving force arising from detailed balance breaking, thus offering new insights into the turbulence dynamics.

Moreover, we note that Eq. (\ref{PFFSNSE}) still satisfies a FDT between the potential part of the viscous force and the stochastic force, where the spatial correlator of the stochastic force also serves as the damping matrix in front of the functional gradient of the nonequilibrium potential landscape. However, the orthogonality condition in Eq. (\ref{FDFRPFNSE2}) that holds in equilibrium fluid systems, stating that the solenoidal convective force is orthogonal in the state space to the functional gradient of the equilibrium potential landscape, is no longer valid for nonequilibrium fluid systems with detailed balance breaking. In other words, $\mathbf{F}^{rev}(\mathbf{x})[\mathbf{u}]$ is generally not orthogonal to $\delta_{\mathbf{u}(\mathbf{x})}\Phi[\mathbf{u}]$ in the state space for nonequilibrium fluid systems.

\begin{figure}[!ht]
\centering
\includegraphics[width=4in]{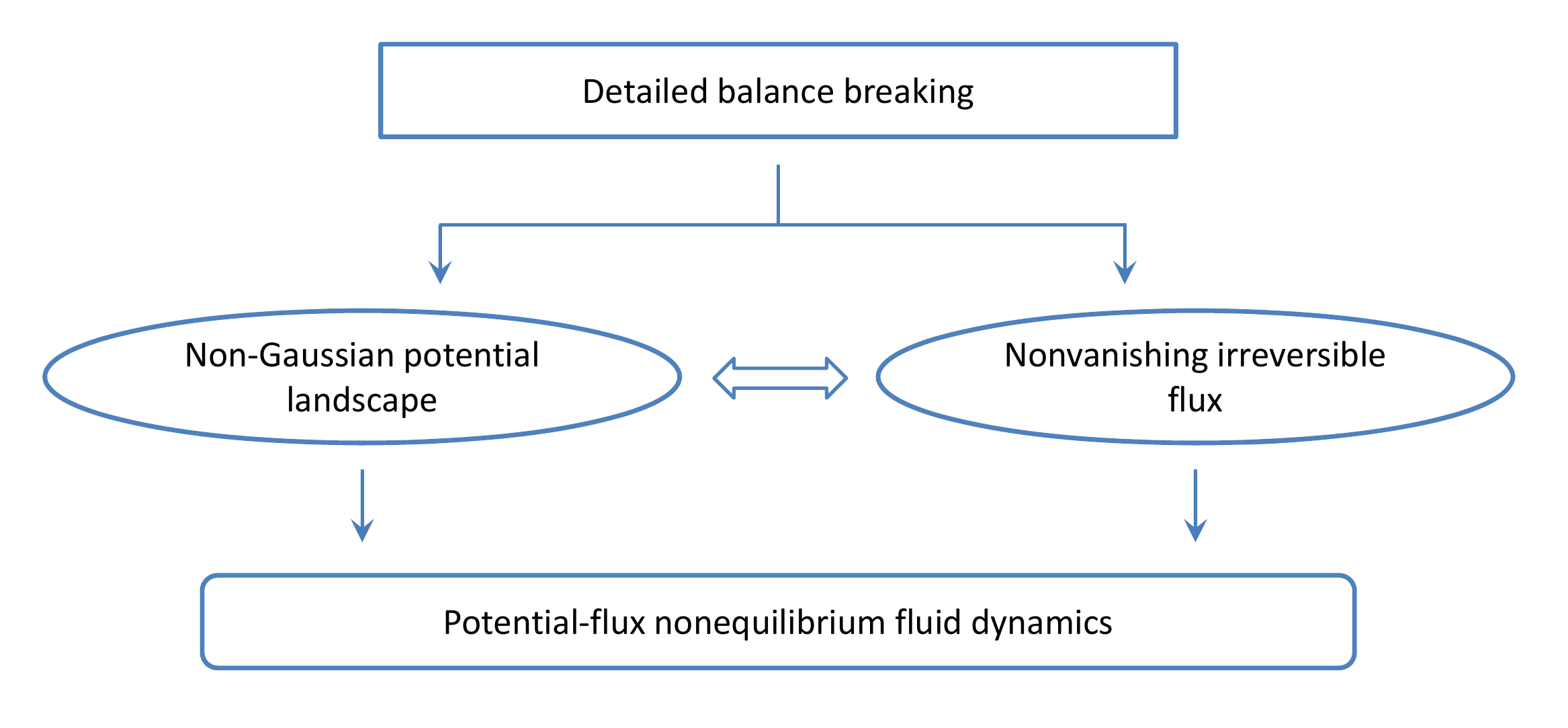}
\caption{The manifestation of the nonequilibrium trinity in nonequilibrium fluid dynamics.}\label{figure2}
\end{figure}

An illustration of the manifestation of the nonequilibrium trinity in the nonequilibrium fluid dynamics is shown in Fig. \ref{figure2}. The most important results in Section \ref{PFFLNSE} summarized in terms of equations  (in the logical sequence)  are the detailed balance constraint in Eq. (\ref{DBCENSE}), the nonequilibrium source equation in Eq. (\ref{DENSE}), the flux deviation relation in Eq. (\ref{FDRNSE}), the force decomposition equation in Eq. (\ref{FDENEQNSE}), and the potential-flux form of the stochastic Navier-Stokes equation in Eq. (\ref{PFFSNSE}).

\section{Energy balance, energy cascade and turbulence in the context of the potential landscape and flux field theory}\label{Energy}

Energy balance and energy cascade are most conveniently studied in the wavevector space. We first introduce the wavevector space representation and then discuss subjects related to energy balance, energy cascade and fully developed turbulence in the context of the potential landscape and flux field theory.

\subsection{The wavevector space representation}\label{wavenumberREP}

With the help of Fourier analysis, what has been formulated in the physical space can be translated into equivalent forms in the wavevector space and vice versa. For completeness we give the general rules of translation and some major results in the wavevector representation in preparation for later discussions.

\subsubsection{Dictionary for the translation}

The velocity field $\mathbf{u}(\mathbf{x})$ satisfying periodic boundary conditions can be expanded into a Fourier series $\mathbf{u}(\mathbf{x})=\sum_{\mathbf{k}}\mathbf{u}(\mathbf{k})e^{i\mathbf{k}\cdot\mathbf{x}}$, where the wavevector $\mathbf{k}=2\pi\mathbf{n}/L$ for $\mathbf{n}$ with integer components. The Fourier coefficient $\mathbf{u}(\mathbf{k})$, a vector-valued complex function of the wavevector $\mathbf{k}$, is given by the inverse relation $\mathbf{u}(\mathbf{k})=\int d\mathbf{x}\, \mathbf{u}(\mathbf{x})e^{-i\mathbf{k}\cdot\mathbf{x}} /L^3$, where the integral is over $\mathbf{T}^3$. The complex conjugate of $\mathbf{u}(\mathbf{k})$ is not independent since $\mathbf{u}^*(\mathbf{k})=\mathbf{u}(\mathbf{-k})$ due to the reality of $\mathbf{u}(\mathbf{x})$. The wavevector function $\mathbf{u}(\mathbf{k})$ is the wavevector representation of the velocity field $\mathbf{u}(\mathbf{x})$ in the physical space.

The state of the fluid system is described by solenoidal velocity fields with zero total momentum. Solenoidal fields satisfying  $\nabla\cdot\mathbf{u}(\mathbf{x})=0$  in the physical space  are characterized in the wavevector space by the condition $\mathbf{k}\cdot\mathbf{u}(\mathbf{k})=0$. The zero total momentum condition $\int \mathbf{u}(\mathbf{x})d\mathbf{x}=0$ is represented by $\mathbf{u}(\mathbf{k})|_{\mathbf{k}=\mathbf{0}}=0$. Hence, the state of the fluid system is represented in the wavevector space by $\mathbf{u}(\mathbf{k})$ satisfying $\mathbf{k}\cdot\mathbf{u}(\mathbf{k})=0$ and $\mathbf{u}(\mathbf{k})|_{\mathbf{k}=\mathbf{0}}=0$, in addition to $\mathbf{u}^*(\mathbf{k})=\mathbf{u}(\mathbf{-k})$. As a result of the condition $\mathbf{u}(\mathbf{k})|_{\mathbf{k}=\mathbf{0}}=0$, the $\mathbf{k}=\mathbf{0}$ term in the Fourier series no longer plays a role and the associated technical issues (e.g., the inverse of the Laplacian $\Delta$) can thus be avoided.

The gradient projection operator $\mathbf{\Pi}^{g}(\nabla)$ is represented in the wavevector space by the projection matrix $\mathbf{\Pi}^g(\mathbf{k})=\mathbf{k}\mathbf{k}/k^2$ along the $\mathbf{k}$ direction for $\mathbf{k}\neq \mathbf{0}$, so that $\mathbf{w}(\mathbf{x})=\mathbf{\Pi}^{g}(\nabla)\cdot\mathbf{v}(\mathbf{x})$ is represented by $\mathbf{w}(\mathbf{k})=\mathbf{\Pi}^{g}(\mathbf{k})\cdot\mathbf{v}(\mathbf{k})$. Similarly, the solenoidal projection operator $\mathbf{\Pi}^{s}(\nabla)$ is represented by the projection matrix $\mathbf{\Pi}^s(\mathbf{k})=\mathbf{I}-\mathbf{k}\mathbf{k}/k^2$ perpendicular to the $\mathbf{k}$ direction for $\mathbf{k}\neq \mathbf{0}$, so that $\mathbf{w}(\mathbf{x})=\mathbf{\Pi}^{s}(\nabla)\cdot\mathbf{v}(\mathbf{x})$ is represented by $\mathbf{w}(\mathbf{k})=\mathbf{\Pi}^{s}(\mathbf{k})\cdot\mathbf{v}(\mathbf{k})$.

The diffusion matrix $\mathbf{D}^s(\mathbf{x}-\mathbf{x}')$, which depends on $\mathbf{x}-\mathbf{x}'$, is represented by the matrix-valued wavevector function $\mathbf{D}^s(\mathbf{k})=\int d\mathbf{x}\, \mathbf{D}^s(\mathbf{x})e^{-i\mathbf{k}\cdot\mathbf{x}} /L^3$, with the inverse relation $\mathbf{D}^s(\mathbf{x})=\sum_{\mathbf{k}}\mathbf{D}^s(\mathbf{k})e^{i\mathbf{k}\cdot\mathbf{x}}$. It is easy to verify from the properties of $\mathbf{D}^s(\mathbf{x}-\mathbf{x}')$ that $\mathbf{k}\cdot\mathbf{D}^s(\mathbf{k})=\mathbf{0}$, $D^s_{ij}(\mathbf{k})=(D^{s}_{ji})^*(\mathbf{k})=D^{s}_{ji}(-\mathbf{k})$, and $\sum_{\mathbf{k}}\mathbf{v}^*(\mathbf{k})\cdot\mathbf{D}^s(\mathbf{k})\cdot\mathbf{v}(\mathbf{k})\geq 0$ for $\mathbf{v}(\mathbf{k})$ in the state space. Hence, $\mathbf{D}^s(\mathbf{k})$ is nonnegative-definite and Hermitian for $\mathbf{k}\neq\mathbf{0}$. Assuming that $\mathbf{D}^s(\mathbf{x}-\mathbf{x}')$ is invertible in the state space, $\mathbf{D}^s(\mathbf{k})$ further becomes positive-definite.

Functionals of $\mathbf{u}(\mathbf{x})$ in the physical space that do not depend on $\mathbf{x}$ explicitly, such as $P[\mathbf{u}]$ and $\Phi[\mathbf{u}]$, will be represented by the same notation in the wavevector space, with the functional dependence $[\mathbf{u}]$ reinterpreted as $[\mathbf{u}(\mathbf{k})]$. Functionals of $\mathbf{u}(\mathbf{x})$ that depend on $\mathbf{x}$ explicitly, such as $\mathbf{F}(\mathbf{x})[\mathbf{u}]$ and $\mathbf{J}(\mathbf{x})[\mathbf{u}]$, are represented by their Fourier coefficients in the wavevector space, as in $\mathbf{J}(\mathbf{k})[\mathbf{u}]=\int d\mathbf{x}\, \mathbf{J}(\mathbf{x})[\mathbf{u}]e^{-i\mathbf{k}\cdot\mathbf{x}}/L^3$, with the inverse relation $\mathbf{J}(\mathbf{x})[\mathbf{u}]=\sum_{\mathbf{k}}\mathbf{J}(\mathbf{k})[\mathbf{u}]e^{i\mathbf{k}\cdot\mathbf{x}}$.

The functional derivative in the physical space and the derivative in the wavevector space are related to each other by $\delta_{\mathbf{u}(\mathbf{x})}=\sum_{\mathbf{k}} (e^{-i\mathbf{k}\cdot\mathbf{x}}/L^3) \nabla_{\mathbf{u}(\mathbf{k})}$ and $\nabla_{\mathbf{u}(\mathbf{k})}=\int d\mathbf{x} \,e^{i\mathbf{k}\cdot\mathbf{x}}\delta_{\mathbf{u}(\mathbf{x})}$, where $\nabla_{\mathbf{u}(\mathbf{k})}$ is the vector-valued partial derivative $\partial/\partial \mathbf{u}(\mathbf{k})$ in the wavevector space. Note that because the components of $\mathbf{u}(\mathbf{k})$ are not independent due to the constraint $\mathbf{k}\cdot\mathbf{u}(\mathbf{k})=0$, the basic rule of differentiation in the wavevector space in terms of $\mathbf{u}(\mathbf{k})$ has the form $\partial u_i(\mathbf{k})/\partial u_j(\mathbf{k}')=\Pi^s_{ij}(\mathbf{k})\delta_{\mathbf{k}\mathbf{k}'}$, where $\Pi^s_{ij}(\mathbf{k})=\delta_{ij}-k_ik_j/k^2$, which plays the same role as the identity matrix as long as operations are restricted to the wavevector state space with the constraint $\mathbf{k}\cdot\mathbf{u}(\mathbf{k})=0$.

\subsubsection{Reformulation in the wavevector representation}

With the above dictionary in hand, we can translate the major results formulated in the physical space into the wavevector space. The FFPE in the physical space in Eq. (\ref{FFPENSE}) for the Navier-Stokes system, when transformed into the wavevector representation, has the form
\begin{equation}\label{FFPEBEGWNNSE}
\begin{split}
 \partial_t P_t[\mathbf{u}]=&-\sum_{\mathbf{k}}\nabla_{\mathbf{u}(\mathbf{k})}\cdot\left\{\left[\mathbf{\Pi}^s(\mathbf{k})\cdot\left(-i\mathbf{k}\cdot\sum_{\mathbf{k}'}\mathbf{u}(\mathbf{k}-\mathbf{k}')\mathbf{u}(\mathbf{k}')\right)\right]P_t[\mathbf{u}]\right\}\\
&-\sum_{\mathbf{k}}\nabla_{\mathbf{u}(\mathbf{k})}\cdot\left(-\nu k^2\mathbf{u}(\mathbf{k})P_t[\mathbf{u}]\right)+\sum_{\mathbf{k}}\nabla_{\mathbf{u}(\mathbf{k})}\cdot\left(\mathbf{D}^s(\mathbf{k})\cdot\nabla_{\mathbf{u}(-\mathbf{k})}P_t[\mathbf{u}]\right).
\end{split}
\end{equation}
This equation reduces to the Edwards-Fokker-Planck equation \cite{Edwards,McComb} when the diffusion matrix $\mathbf{D}^s(\mathbf{k})$ is specialized to $\mathbf{D}^s(\mathbf{k})=\mathbf{\Pi}^s(\mathbf{k})W(k)/2$ and the vector-matrix notations preferred in this article are spelled out with explicit index notations. In Eq. (\ref{FFPEBEGWNNSE}) we identify $\mathbf{F}^{rev}(\mathbf{k})[\mathbf{u}]=\mathbf{\Pi}^s(\mathbf{k})\cdot\left(-i\mathbf{k}\cdot\sum_{\mathbf{k}'}\mathbf{u}(\mathbf{k}-\mathbf{k}')\mathbf{u}(\mathbf{k}')\right)$ as the wavevector representation of the reversible solenoidal convective force $\mathbf{F}^{rev}(\mathbf{x})[\mathbf{u}]=\mathbf{\Pi}^s(\nabla)\cdot(-\mathbf{u}\cdot\nabla\mathbf{u})$ and $\mathbf{F}^{irr}(\mathbf{k})[\mathbf{u}]=-\nu k^2\mathbf{u}(\mathbf{k})$ as that of the irreversible viscous force $\mathbf{F}^{irr}(\mathbf{x})[\mathbf{u}]=\nu\Delta \mathbf{u}(\mathbf{x})$.

The FFPE in the wavevector representation in Eq. (\ref{FFPEBEGWNNSE}) has the form of the continuity equation in the wavevector state space: $\partial_t P_t[\mathbf{u}]=-\sum_{\mathbf{k}} \nabla_{\mathbf{u}(\mathbf{k})}\cdot\mathbf{J}_t(\mathbf{k})[\mathbf{u}]$. The steady-state probability flux in the wavevector representation, satisfying the functional-divergence-free condition $\sum_{\mathbf{k}} \nabla_{\mathbf{u}(\mathbf{k})}\cdot\mathbf{J}_s(\mathbf{k})[\mathbf{u}]=0$, reads
\begin{equation}\label{SSPFKNSE}
\begin{split}
\mathbf{J}_s(\mathbf{k})[\mathbf{u}]=\left[\mathbf{\Pi}^s(\mathbf{k})\cdot\left(-i\mathbf{k}\cdot\sum_{\mathbf{k}'}\mathbf{u}(\mathbf{k}-\mathbf{k}')\mathbf{u}(\mathbf{k}')\right)\right]P_s[\mathbf{u}]\\
-\nu k^2\mathbf{u}(\mathbf{k})P_s[\mathbf{u}]-\mathbf{D}^s(\mathbf{k})\cdot\nabla_{\mathbf{u}(-\mathbf{k})}P_s[\mathbf{u}].
\end{split}
\end{equation}
It can be decomposed, according to different time reversal behaviors, into the steady-state reversible probability flux
\begin{equation}\label{RSSPFKNSE}
\mathbf{J}_s^{rev}(\mathbf{k})[\mathbf{u}]=\left[\mathbf{\Pi}^s(\mathbf{k})\cdot\left(-i\mathbf{k}\cdot\sum_{\mathbf{k}'}\mathbf{u}(\mathbf{k}-\mathbf{k}')\mathbf{u}(\mathbf{k}')\right)\right]P_s[\mathbf{u}]
\end{equation}
and the steady-state irreversible probability flux
\begin{equation}\label{ISSPFKNSE}
\mathbf{J}_s^{irr}(\mathbf{k})[\mathbf{u}]=-\nu k^2\mathbf{u}(\mathbf{k})P_s[\mathbf{u}]-\mathbf{D}^s(\mathbf{k})\cdot\nabla_{\mathbf{u}(-\mathbf{k})}P_s[\mathbf{u}].
\end{equation}
The latter consists of the steady-state viscous probability flux $\mathbf{J}_s^{vis}(\mathbf{k})[\mathbf{u}]=[-\nu k^2\mathbf{u}(\mathbf{k})]P_s[\mathbf{u}]$ and stochastic probability flux $\mathbf{J}_s^{sto}(\mathbf{k})[\mathbf{u}]=-\mathbf{D}^s(\mathbf{k})\cdot\nabla_{\mathbf{u}(-\mathbf{k})}P_s[\mathbf{u}]$. The expressions of the transient probability fluxes, $\mathbf{J}_t(\mathbf{k})[\mathbf{u}]$, $\mathbf{J}_t^{rev}(\mathbf{k})[\mathbf{u}]$, $\mathbf{J}_t^{irr}(\mathbf{k})[\mathbf{u}]$, $\mathbf{J}_t^{vis}(\mathbf{k})[\mathbf{u}]$, and $\mathbf{J}_t^{sto}(\mathbf{k})[\mathbf{u}]$, can be obtained by replacing $P_s[\mathbf{u}]$ with $P_t[\mathbf{u}]$.

The fluid system has an equilibrium steady state with time reversal symmetry when the system obeys detailed balance. Reformulated in the wavevector representation, the \emph{detailed balance constraint} in Eq. (\ref{DBCENSE}) characterizing the detailed balance condition for fluid systems reads
\begin{equation}\label{DBCWNNSE}
\sum_{\mathbf{k}} \left[\mathbf{\Pi}^s(\mathbf{k})\cdot \left(-i\mathbf{k}\cdot\sum_{\mathbf{k}'}\mathbf{u}(\mathbf{k}-\mathbf{k}')\mathbf{u}(\mathbf{k}')\right)\right]\cdot\mathbf{D}_s^{-1}(\mathbf{k})\cdot\left[-\nu k^2\mathbf{u}(-\mathbf{k})\right]=0,
\end{equation}
where $\mathbf{D}_s^{-1}(\mathbf{k})=L^3\int \mathbf{D}_s^{-1}(\mathbf{x})e^{-i\mathbf{k}\cdot\mathbf{x}}d\mathbf{x}$ and it satisfies $\mathbf{D}^s(\mathbf{k})\cdot\mathbf{D}_s^{-1}(\mathbf{k})=\mathbf{\Pi}^s(\mathbf{k})$. As a consequence of detailed balance (see the discussions in Section \ref{ESS}), the equilibrium potential landscape has the Gaussian quadratic form $\Phi_{eq}[\mathbf{u}]=(\nu/2)\sum_{\mathbf{k}}\mathbf{u}(-\mathbf{k})\cdot[k^2\mathbf{D}_s^{-1}(\mathbf{k})]\cdot\mathbf{u}(\mathbf{k})$; the corresponding equilibrium probability distribution is the Gaussian distribution $P_{eq}[\mathbf{u}]=\mathcal{N}\exp\{-(\nu/2)\sum_{\mathbf{k}}\mathbf{u}(-\mathbf{k})\cdot[k^2\mathbf{D}_s^{-1}(\mathbf{k})]\cdot\mathbf{u}(\mathbf{k})\}$ . Accordingly, the steady-state irreversible probability flux velocity $\mathbf{V}_s^{irr}(\mathbf{k})[\mathbf{u}]$ and irreversible probability flux $\mathbf{J}_s^{irr}(\mathbf{k})[\mathbf{u}]$ vanish completely for all $\mathbf{k}$ and $\mathbf{u}(\mathbf{k})$. As a consequence, the irreversible viscous force has the potential gradient form  at equilibrium, $-\nu k^2\mathbf{u}(\mathbf{k})=-\mathbf{D}^s(\mathbf{k})\cdot\nabla_{\mathbf{u}(-\mathbf{k})}\Phi_{eq}[\mathbf{u}]$, which signifies the reversible character of the equilibrium stochastic dynamics. The particular example we studied in Section \ref{Example} corresponds to the special form of the diffusion matrix $\mathbf{D}^s(\mathbf{k})=(\nu k_B T k^2/L^3)\mathbf{\Pi}^s(\mathbf{k})$, with the equilibrium potential landscape $\Phi_{eq}[\mathbf{u}]=(L^3/2k_BT)\sum_{\mathbf{k}}\mathbf{u}(-\mathbf{k})\cdot\mathbf{u}(\mathbf{k})$ and the equilibrium probability distribution $P_{eq}[\mathbf{u}]=\mathcal{N}\exp\{-(L^3/2k_BT)\sum_{\mathbf{k}}\mathbf{u}(-\mathbf{k})\cdot\mathbf{u}(\mathbf{k})\}$.

When detailed balance is broken, the fluid system has a nonequilibrium steady state with intrinsic time irreversibility, characterized by the nonequilibrium trinity. According to the nonequilibrium source equation in Eq. (\ref{DENSE}), whose wavevector representation is complicated and will not be spelled out here, detailed balance breaking (i.e., violation of the detailed balance constraint) is the source of the deviation of the nonequilibrium potential landscape $\Phi[\mathbf{u}]$ from the Gaussian quadratic form $\Phi_{0}[\mathbf{u}]=(\nu/2)\sum_{\mathbf{k}}\mathbf{u}(-\mathbf{k})\cdot[k^2\mathbf{D}_s^{-1}(\mathbf{k})]\cdot\mathbf{u}(\mathbf{k})$, or equivalently, the deviation  of the nonequilibrium steady-state probability distribution $P_s[\mathbf{u}]$ from the Gaussian distribution $P_0[\mathbf{u}]=\mathcal{N}\exp\{-(\nu/2)\sum_{\mathbf{k}}\mathbf{u}(-\mathbf{k})\cdot[k^2\mathbf{D}_s^{-1}(\mathbf{k})]\cdot\mathbf{u}(\mathbf{k})\}$. The other consequence of detailed balance breaking is the nonvanishing steady-state irreversible probability flux velocity $\mathbf{V}_s^{irr}(\mathbf{k})[\mathbf{u}]$, or equivalently, the nonvanishing steady-state irreversible probability flux $\mathbf{J}_s^{irr}(\mathbf{k})[\mathbf{u}]$. These two consequential aspects of detailed balance breaking are connected to each other by the \emph{flux deviation relation} in the wavevector representation
\begin{equation}\label{FDRWVNSE}
\mathbf{V}_s^{irr}(\mathbf{k})[\mathbf{u}]=-\mathbf{D}^s(\mathbf{k})\cdot\nabla_{\mathbf{u}(-\mathbf{k})}\Lambda[\mathbf{u}],
\end{equation}
which relates the deviated potential landscape $\Lambda[\mathbf{u}]$ to the irreversible probability flux velocity $\mathbf{V}_s^{irr}(\mathbf{k})[\mathbf{u}]$. As a manifestation of the nonequilibrium trinity in the structure of the driving force, we have the \emph{force decomposition equation} in the wavevector representation
\begin{equation}
-\nu k^2\mathbf{u}(\mathbf{k})=-\mathbf{D}^s(\mathbf{k})\cdot\nabla_{\mathbf{u}(-\mathbf{k})}\Phi[\mathbf{u}]+\mathbf{V}_s^{irr}(\mathbf{k})[\mathbf{u}],
\end{equation}
which decomposes the irreversible viscous force into the potential-flux form. As a consequence, the nonequilibrium stochastic fluid dynamics has the following potential-flux form in the wavevector space:
\begin{equation}\label{PFFNSFDWS}
\partial_t \mathbf{u}(\mathbf{k},t)=\mathbf{F}^{rev}(\mathbf{k})[\mathbf{u}]-\mathbf{D}^s(\mathbf{k})\cdot\nabla_{\mathbf{u}(-\mathbf{k})}\Phi[\mathbf{u}]+\mathbf{V}_s^{irr}(\mathbf{k})[\mathbf{u}]+\bm{\xi}^s(\mathbf{k},t),
\end{equation}
where the irreversible probability flux velocity $\mathbf{V}_s^{irr}(\mathbf{k})[\mathbf{u}]$, related to the non-Gaussian potential landscape through Eq. (\ref{FDRWVNSE}), is the driving force that originates from detailed balance breaking and signifies the irreversible nature of the nonequilibrium stochastic fluid dynamics.

\subsection{Energy balance in relation to probability fluxes}

We now investigate energy balance in the fluid system, and, in particular, its relation to probability fluxes in the context of the potential landscape and flux field theory. The total kinetic energy of the fluid in the velocity field $\mathbf{u}(\mathbf{x})$ has the expression $E_{tot}=(1/2)\int |\mathbf{u}(\mathbf{x})|^2d\mathbf{x}=(L^3/2)\sum_{\mathbf{k}} \mathbf{u}(-\mathbf{k})\cdot\mathbf{u}(\mathbf{k})$. Hence, $\mathbf{u}(-\mathbf{k})\cdot\mathbf{u}(\mathbf{k})/2$ can be interpreted as the kinetic energy per unit mass at mode $\mathbf{k}$ in the velocity field $\mathbf{u}(\mathbf{x})$. The ensemble-averaged kinetic energy (per unit mass) at mode $\mathbf{k}$ is defined as $\mathcal{E}(\mathbf{k},t)=\langle \mathbf{u}(-\mathbf{k})\cdot\mathbf{u}(\mathbf{k})/2 \rangle_{t}=\int [\mathbf{u}(-\mathbf{k})\cdot\mathbf{u}(\mathbf{k})/2] P_t[\mathbf{u}] \delta \mathbf{u}$, where the ensemble probability distribution functional $P_t[\mathbf{u}]$ is governed by the FFPE in Eq. (\ref{FFPEBEGWNNSE}) and $\int \delta \mathbf{u}$ represents integration over all independent velocity field configurations in the wavevector state space.

\subsubsection{Energy balance in the transient state}
The rate of change of $\mathcal{E}(\mathbf{k},t)$ can be deduced from that of $P_t[\mathbf{u}]$ as follows:
\begin{eqnarray}\label{TEBPreNSE}
\partial_t \mathcal{E}(\mathbf{k},t)&=&\int \left(\DF{\mathbf{u}(-\mathbf{k})\cdot\mathbf{u}(\mathbf{k})}{2}\right)\left(-\sum_{\mathbf{k}'}\nabla_{\mathbf{u}(\mathbf{k}')}\cdot\mathbf{J}_t(\mathbf{k}')[\mathbf{u}]\right)\delta \mathbf{u}\nonumber\\
&=&\int \sum_{\mathbf{k}'}\left[\nabla_{\mathbf{u}(\mathbf{k}')}\left(\DF{\mathbf{u}(-\mathbf{k})\cdot\mathbf{u}(\mathbf{k})}{2}\right)\right]\cdot\mathbf{J}_t(\mathbf{k}')[\mathbf{u}]\delta \mathbf{u}\nonumber\\
&=&\int\left\{\DF{1}{2}\mathbf{u}(-\mathbf{k})\cdot\mathbf{J}_t(\mathbf{k})[\mathbf{u}]+\DF{1}{2}\mathbf{u}(\mathbf{k})\cdot\mathbf{J}_t(-\mathbf{k})[\mathbf{u}]\right\}\delta \mathbf{u},
\end{eqnarray}
where we have used Eq. (\ref{FFPEBEGWNNSE}) in the form of the continuity equation $\partial_tP_t[\mathbf{u}]=-\sum_{\mathbf{k}'}\nabla_{\mathbf{u}(\mathbf{k}')}\cdot\mathbf{J}_t(\mathbf{k}')[\mathbf{u}]$ and integration by parts in the wavevector state space. Eq. (\ref{TEBPreNSE}) can also be written as $\partial_t \mathcal{E}(\mathbf{k},t)=\mathcal{R}\left\{\int \mathbf{u}^*(\mathbf{k})\cdot\mathbf{J}_t(\mathbf{k})[\mathbf{u}]\delta \mathbf{u}\right\}$, where $\mathcal{R}\{\cdots\}$ denotes taking the real part.

The probability flux $\mathbf{J}_t(\mathbf{k})[\mathbf{u}]$ consists of a reversible part and an irreversible part, where the reversible flux comes from the solenoidal convective force and the irreversible flux consists of contributions from the viscous force and the stochastic force, respectively. Mathematically, we have $\mathbf{J}_t(\mathbf{k})[\mathbf{u}]=\mathbf{J}_t^{rev}(\mathbf{k})[\mathbf{u}]+\mathbf{J}_t^{irr}(\mathbf{k})[\mathbf{u}]$ and $\mathbf{J}_t^{irr}(\mathbf{k})[\mathbf{u}]=\mathbf{J}_t^{vis}(\mathbf{k})[\mathbf{u}]+\mathbf{J}_t^{sto}(\mathbf{k})[\mathbf{u}]$, with their expressions given by (see Eqs. (\ref{SSPFKNSE})-(\ref{ISSPFKNSE})): $\mathbf{J}_t^{rev}(\mathbf{k})[\mathbf{u}]=[\mathbf{\Pi}^s(\mathbf{k})\cdot(-i\mathbf{k}\cdot\sum_{\mathbf{k}'}\mathbf{u}(\mathbf{k}-\mathbf{k}')\mathbf{u}(\mathbf{k}'))]P_t[\mathbf{u}]$
, $\mathbf{J}_t^{vis}(\mathbf{k})[\mathbf{u}]=[-\nu k^2\mathbf{u}(\mathbf{k})]P_t[\mathbf{u}]$ and $\mathbf{J}_t^{sto}(\mathbf{k})[\mathbf{u}]=-\mathbf{D}^s(\mathbf{k})\cdot\nabla_{\mathbf{u}(-\mathbf{k})}P_t[\mathbf{u}]$.

Plugging the decomposition of the probability flux into Eq. (\ref{TEBPreNSE}), we obtain the energy balance equation
\begin{equation}\label{TEBENSE}
\partial_t \mathcal{E}(\mathbf{k},t)=\mathcal{T}(\mathbf{k},t)-\mathcal{D}(\mathbf{k},t)+\mathcal{I}(\mathbf{k},t),
\end{equation}
together with its relation to probability fluxes. Here $\mathcal{T}(\mathbf{k},t)$ is the energy transfer rate at mode $\mathbf{k}$, due to the nonlinear solenoidal convective force, associated with the reversible probability flux, with the expression
\begin{equation}\label{ETSNSE}
\begin{split}
\mathcal{T}(\mathbf{k},t)&=\mathcal{R}\left\{\int \mathbf{u}^*(\mathbf{k})\cdot\mathbf{J}_t^{rev}(\mathbf{k})[\mathbf{u}]\delta \mathbf{u}\right\}\\
&=\mathcal{R}\left\{-i\mathbf{k}\cdot\sum_{\mathbf{k}'}\left\langle\mathbf{u}(\mathbf{k}-\mathbf{k}')\mathbf{u}(\mathbf{k}')\cdot\mathbf{u}^*(\mathbf{k})\right\rangle_t\right\}.
\end{split}
\end{equation}
The solenoidal projection matrix $\mathbf{\Pi}^s(\mathbf{k})$ does not appear in the final expression of $\mathcal{T}(\mathbf{k},t)$, as it has been absorbed by the property $\mathbf{u}^*(\mathbf{k})\cdot \mathbf{\Pi}^s(\mathbf{k})=\mathbf{u}^*(\mathbf{k})$ due to the velocity field being solenoidal. The energy transfer rate satisfies $\sum_{\mathbf{k}} \mathcal{T}(\mathbf{k},t)=0$, which means the reversible convective force only redistributes the kinetic energy of the fluid among different modes (including modes at different scales) without changing its total amount.

$\mathcal{D}(\mathbf{k},t)$ in Eq. (\ref{TEBENSE}) is the energy dissipation rate at mode $\mathbf{k}$, due to the dissipative viscous force, associated with the irreversible viscous probability flux, with the expression
\begin{equation}\label{EDSNSE}
\mathcal{D}(\mathbf{k},t)=-\mathcal{R}\left\{\int \mathbf{u}^*(\mathbf{k})\cdot\mathbf{J}_t^{vis}(\mathbf{k})[\mathbf{u}]\delta \mathbf{u}\right\}=2\nu k^2\mathcal{E}(\mathbf{k},t).
\end{equation}
It is nonnegative and thus, according to Eq. (\ref{TEBENSE}), with a minus sign in front, dissipates (does not increase) the kinetic energy at mode $\mathbf{k}$.

$\mathcal{I}(\mathbf{k},t)$ in Eq. (\ref{TEBENSE}) is the energy injection rate at mode $\mathbf{k}$, due to the stochastic force, associated with the irreversible stochastic probability flux, with the expression
\begin{equation}\label{EISNSE}
\mathcal{I}(\mathbf{k},t)=\mathcal{R}\left\{\int \mathbf{u}^*(\mathbf{k})\cdot\mathbf{J}_t^{sto}(\mathbf{k})[\mathbf{u}]\delta \mathbf{u}\right\}=W(\mathbf{k}),
\end{equation}
where $W(\mathbf{k})=\text{Tr}[\mathbf{D}^s(\mathbf{k})]$ and $\text{Tr}[\cdots]$ means the trace of the matrix. In deriving the latter expression of $\mathcal{I}(\mathbf{k},t)$, we have used integration by parts in the state space and the normalization of $P_t[\mathbf{u}]$. $W(\mathbf{k})$ can also be determined directly from the statistical property of the stochastic force in the wavevector space: $\langle \bm{\xi}^s(\mathbf{k},t)\cdot\bm{\xi}^s(-\mathbf{k},t')\rangle=2W(\mathbf{k})\delta(t-t')$. Since $\mathbf{D}^s(\mathbf{k})$ is a nonnegative-definite Hermitian matrix, $W(\mathbf{k})$ as its trace is nonnegative and thus, according to Eq. (\ref{TEBENSE}), with a positive sign in front of $\mathcal{I}(\mathbf{k},t)$, increases (does not decrease) the kinetic energy at mode $\mathbf{k}$. Eq. (\ref{EISNSE}) shows that the energy injection rate $\mathcal{I}(\mathbf{k},t)$ does not depend on time or any specific information of the ensemble probability distribution; it is completely determined by the statistical property  $W(\mathbf{k})$ of the stochastic force. This also clarifies the physical meaning of $W(\mathbf{k})$ as the energy injection rate at wavevector $\mathbf{k}$.

\subsubsection{Energy balance in the steady state}

We are particularly interested in the steady-state regime, within which the probability distribution $P_s[\mathbf{u}]$ does  not change with time. As a result, $\mathcal{E}(\mathbf{k})$, $\mathcal{T}(\mathbf{k})$, $\mathcal{D}(\mathbf{k})$, and $\mathcal{I}(\mathbf{k})$ are all time-independent in the steady state. Since the energy at mode $\mathbf{k}$, $\mathcal{E}(\mathbf{k})=\langle \mathbf{u}(-\mathbf{k})\cdot\mathbf{u}(\mathbf{k})/2 \rangle_{s}$, now satisfies $\partial_t\mathcal{E}(\mathbf{k})=0$, the energy balance equation in the steady state reduces to
\begin{equation}\label{SSEBENSE}
\mathcal{T}(\mathbf{k})=\mathcal{D}(\mathbf{k})-\mathcal{I}(\mathbf{k}),
\end{equation}
where the energy transfer rate at mode $\mathbf{k}$ in the steady state reads
\begin{equation}\label{SSETSNSE}
\begin{split}
\mathcal{T}(\mathbf{k})&=\mathcal{R}\left\{\int \mathbf{u}^*(\mathbf{k})\cdot\mathbf{J}_s^{rev}(\mathbf{k})[\mathbf{u}]\delta \mathbf{u}\right\}\\
&=\mathcal{R}\left\{-i\mathbf{k}\cdot\sum_{\mathbf{k}'}\left\langle\mathbf{u}(\mathbf{k}-\mathbf{k}')\mathbf{u}(\mathbf{k}')\cdot\mathbf{u}^*(\mathbf{k})\right\rangle_s\right\};
\end{split}
\end{equation}
the energy dissipation rate at mode $\mathbf{k}$ in the steady state becomes
\begin{equation}\label{SSEDSNSE}
\mathcal{D}(\mathbf{k})=-\mathcal{R}\left\{\int \mathbf{u}^*(\mathbf{k})\cdot\mathbf{J}_s^{vis}(\mathbf{k})[\mathbf{u}]\delta \mathbf{u}\right\}=2\nu k^2\mathcal{E}(\mathbf{k});
\end{equation}
and the energy injection rate at mode $\mathbf{k}$ in the steady state is given by
\begin{equation}\label{SSEISNSE}
\mathcal{I}(\mathbf{k})=\mathcal{R}\left\{\int \mathbf{u}^*(\mathbf{k})\cdot\mathbf{J}_s^{sto}(\mathbf{k})[\mathbf{u}]\delta \mathbf{u}\right\}=W(\mathbf{k}).
\end{equation}

Summing over $\mathbf{k}$ in Eq. (\ref{SSEBENSE}) and noticing that $\sum_{\mathbf{k}}\mathcal{T}(\mathbf{k})=0$, we see that in the steady state the total energy dissipation rate (per unit mass) balances out the total energy injection rate:
\begin{equation}\label{OAEBNSE}
\varepsilon=\sum_{\mathbf{k}} 2\nu k^2\mathcal{E}(\mathbf{k})=\sum_{\mathbf{k}}W(\mathbf{k}).
\end{equation}
If the stochastic force that injects energy has an external origin, then $W(\mathbf{k})$ determined by the statistical property of the external stochastic force does not depend on properties inherent to the fluid, in particular, the viscosity $\nu$. This in turn implies that in the steady-state ensemble (in the long time limit $t\rightarrow \infty$), the total energy dissipation rate $\varepsilon$, which equals the total energy injection rate, does not depend on the viscosity $\nu$. But it also implies that $\mathcal{E}(\mathbf{k})$ in the steady state must be a function of $\nu$ for $\sum_{\mathbf{k}} 2\nu k^2\mathcal{E}(\mathbf{k})$ to be independent of $\nu$. These implications are relevant for our discussions of energy cascade in the inertial subrange later.

\subsection{Energy flux in relation to the nonequilibrium trinity}

The energy transfer rate $\mathcal{T}(\mathbf{k})$, which may be considered as an energy flux in the wavevector space, plays an important role in the energy cascade picture of turbulence. We investigate the relation between this energy flux and the nonequilibrium trinity as well as the implications thereof.

\subsubsection{Connection of the energy flux to the nonequilibrium trinity}
Combining Eqs. (\ref{SSEBENSE}), (\ref{SSEDSNSE}) and (\ref{SSEISNSE}) of energy balance in the steady state, with $\mathbf{J}_s^{irr}(\mathbf{k})[\mathbf{u}]=\mathbf{J}_s^{vis}(\mathbf{k})[\mathbf{u}]+\mathbf{J}_s^{sto}(\mathbf{k})[\mathbf{u}]$, we arrive at the following important \emph{energy-flux-irreversible-flux relation}:
\begin{equation}\label{EFPFRNSE}
\mathcal{T}(\mathbf{k})=-\mathcal{R}\left\{\int \mathbf{u}^*(\mathbf{k})\cdot\mathbf{J}_s^{irr}(\mathbf{k})[\mathbf{u}]\delta \mathbf{u}\right\}=-\mathcal{R}\left\{\langle \mathbf{u}^*(\mathbf{k})\cdot\mathbf{V}_s^{irr}(\mathbf{k})[\mathbf{u}]\rangle_s\right\},
\end{equation}
where the last step comes from $\mathbf{J}_s^{irr}(\mathbf{k})[\mathbf{u}]=\mathbf{V}_s^{irr}(\mathbf{k})[\mathbf{u}]P_s[\mathbf{u}]$ and $\langle\cdots\rangle_s$ denotes the ensemble average over the steady-state probability distribution functional $P_s[\mathbf{u}]$. The significance of Eq. (\ref{EFPFRNSE}) is that it has established a quantitative connection between the energy flux $\mathcal{T}(\mathbf{k})$, which is an essential quantity in the study of energy cascade in turbulence, and the irreversible probability flux $\mathbf{J}_s^{irr}(\mathbf{k})[\mathbf{u}]$ (or $\mathbf{V}_s^{irr}(\mathbf{k})[\mathbf{u}]$), which is an indispensable component in the nonequilibrium trinity of the potential landscape and flux field theory for fluid systems. This relation shows that the energy flux essential for energy cascade in turbulence is deeply connected to the irreversible probability flux that arises from detailed balance breaking. It also exemplifies the general perspective mentioned in the Introduction that the steady-state irreversible probability flux is a reflection, on the dynamical level, of the steady-state flux of matter, energy or information.

Moreover, since the irreversible probability flux and the non-Gaussian potential landscape are tightly coupled to each other through the flux deviation relation, the energy-flux-irreversible-flux relation also implies a close connection between the energy flux and the non-Gaussian potential landscape. In fact, combining the energy-flux-irreversible-flux relation in Eq. (\ref{EFPFRNSE}) with the flux deviation relation in Eq. (\ref{FDRWVNSE}), we obtain the following \emph{energy-flux-deviated-potential relation}:
\begin{equation}\label{PLEFRNSE}
\mathcal{T}(\mathbf{k})=\mathcal{R}\left\{\langle \mathbf{u}^*(\mathbf{k})\cdot\mathbf{D}^s(\mathbf{k})\cdot\nabla_{\mathbf{u}(-\mathbf{k})}\Lambda[\mathbf{u}]\rangle_s\right\}.
\end{equation}
This equation connects the energy flux $\mathcal{T}(\mathbf{k})$ to the deviation of the nonequilibrium potential landscape (steady-state probability distribution) from the Gaussian potential landscape $\Phi_0[\mathbf{u}]$ (the Gaussian distribution $P_0[\mathbf{u}]$).  Actually, the expression of  $\mathcal{T}(\mathbf{k})$ in Eq. (\ref{SSETSNSE}) in the form of third order moments testifies such a connection from another perspective, since the third order moments of Gaussian distributions (with zero mean) vanish. Therefore, the energy flux  essential to energy cascade in turbulence is also intimately related to the non-Gaussian characteristic of the potential landscape (probability distribution), which is the other consequence of detailed balance breaking in the nonequilibrium trinity.

Eqs. (\ref{EFPFRNSE}) and (\ref{PLEFRNSE}) have revealed an underlying connection of the energy flux in turbulence energy cascade to the nonequilibrium trinity that characterizes the nonequilibrium irreversible nature of the fluid system, through the non-Gaussian potential landscape and the irreversible probability flux, both of which are a direct consequence of detailed balance breaking that represents a form of mechanical imbalance in the driving forces of the fluid system. See Fig. \ref{figure3} (the upper part) for an illustration of the connection between the energy flux and the nonequilibrium trinity.

\subsubsection{Implications of the nonequilibrium trinity on energy flux and energy cascade}

When the detailed balance constraint in Eq. (\ref{DBCWNNSE}) characterizing equilibrium fluid systems with time reversal symmetry is satisfied, the irreversible probability flux velocity $\mathbf{V}_s^{irr}(\mathbf{k})[\mathbf{u}]$ (the irreversible probability flux $\mathbf{J}_s^{irr}(\mathbf{k})[\mathbf{u}]$) vanishes, and the equilibrium potential landscape $\Phi_{eq}[\mathbf{u}]$ (the equilibrium probability distribution $P_{eq}[\mathbf{u}]$) coincides with the Gaussian potential landscape $\Phi_0[\mathbf{u}]$ (the Gaussian probability distribution $P_0[\mathbf{u}]$). As a consequence, according to the energy-flux-irreversible-flux relation in Eq. (\ref{EFPFRNSE}) as well as the energy-flux-deviated-potential relation in Eq. (\ref{PLEFRNSE}), the energy flux $\mathcal{T}(\mathbf{k})$ vanishes completely for all the wavevectors. Therefore, there is no energy flux in the wavevector space associated with the nonlinear convective force that redistributes energy among different modes and couples them together. In other words, different Fourier modes become decoupled as a consequence of detailed balance.

This decoupling of modes is also reflected in the steady-state energy balance equation, which now simply reads $\mathcal{D}(\mathbf{k})=\mathcal{I}(\mathbf{k})$ since $\mathcal{T}(\mathbf{k})=0$. That is, the energy injection rate at wavevector $\mathbf{k}$ is exactly balanced by the energy dissipation rate at the \emph{same} wavevector. The balance between energy injection and energy dissipation is no longer merely on the overall level of all the wavevectors as in Eq. (\ref{OAEBNSE}), but on the detailed level of each wavevector as a manifestation of `detailed balance' in energy balance. This detailed energy balance is a consequence of and thus a necessary condition for detailed balance characterized by the detailed balance constraint that represents a specific form of mechanical balance of the driving forces in the fluid system. As a highly nonequilibrium phenomenon, turbulence of course cannot exist in such an equilibrium scenario with detailed balance and detailed energy balance, as is evident from the complete vanishing of the energy flux and thus lacking of energy cascade.

For energy cascade in turbulence to take place, detailed balance characterizing time reversal symmetry at equilibrium must be broken. That means the detailed balance constraint in Eq. (\ref{DBCWNNSE}) representing the mechanical balance of the driving forces in the fluid system must be violated. As a result, the steady state of the system is out of equilibrium without time reversal symmetry, which is indicated by the irreversible probability flux velocity $\mathbf{V}_s^{irr}(\mathbf{k})[\mathbf{u}]$ (irreversible flux $\mathbf{J}_s^{irr}(\mathbf{k})[\mathbf{u}]$) and the deviation of the potential landscape $\Phi[\mathbf{u}]$ (probability distribution $P_s[\mathbf{u}]$) from being Gaussian. Consequently, the energy flux $\mathcal{T}(\mathbf{k})$, in general (although not necessarily always), does not vanish (at least for some wavevectors), given the relation between the energy flux and the nonequilibrium trinity in Eqs. (\ref{EFPFRNSE}) and (\ref{PLEFRNSE}).

\begin{figure}[!ht]
\centering
\includegraphics[width=4in]{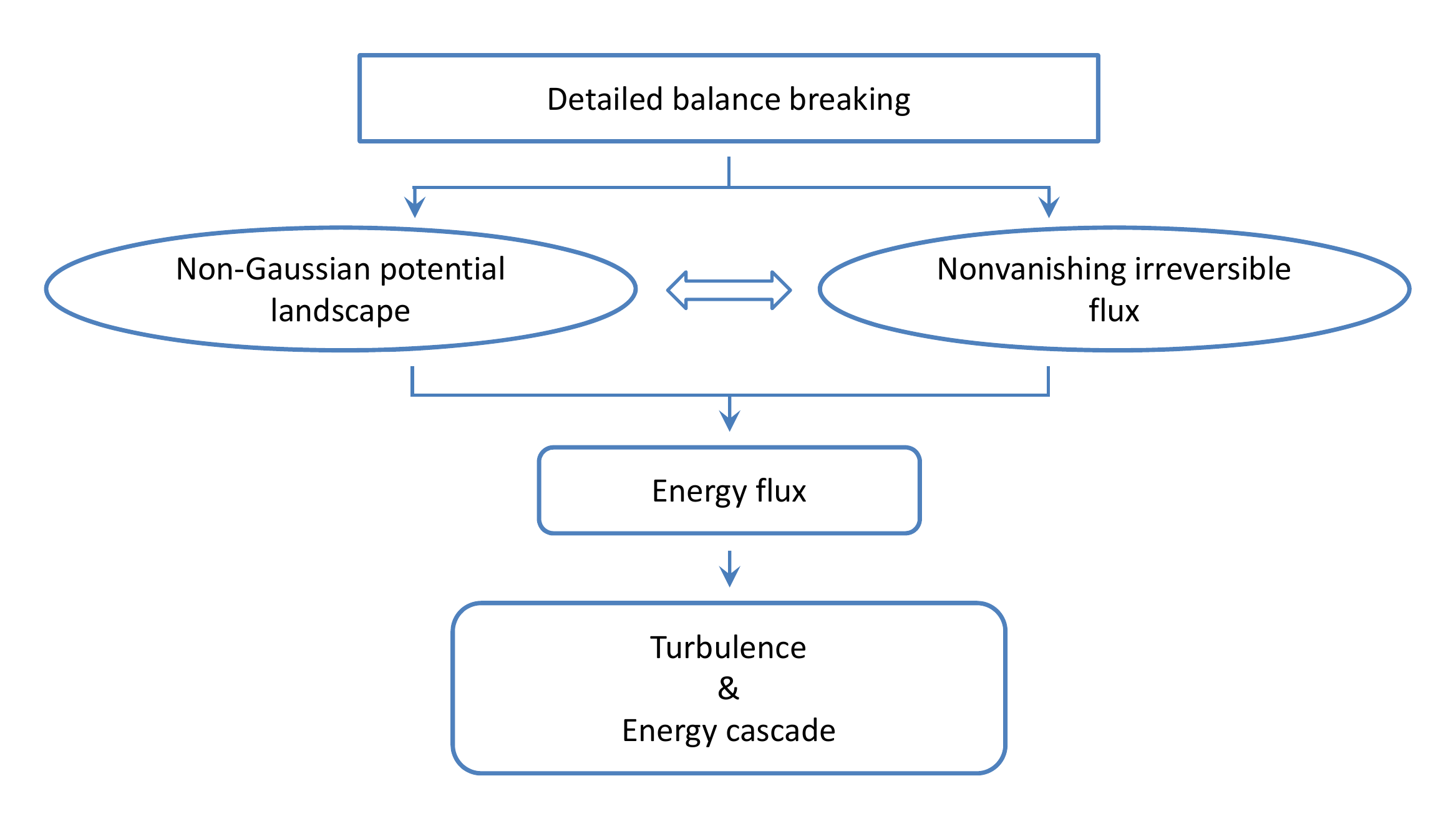}
\caption{The nonequilibrium trinity accommodating and generating the energy flux and the turbulence energy cascade.}\label{figure3}
\end{figure}

Nonvanishing energy flux $\mathcal{T}(\mathbf{k})$ couples different modes together and breaks the detailed energy balance between energy injection and energy dissipation. Now energy injection and energy dissipation are in general only balanced on the overall level of all the modes together rather than on the detailed level of each individual mode. The imbalance between energy injection and energy dissipation at a mode $\mathbf{k}$ is then transferred out and redistributed by the energy flux $\mathcal{T}(\mathbf{k})$ to other modes, allowing for the energy cascade in turbulence to occur. Without detailed balance breaking, there is no irreversible probability flux and no deviation from the Gaussian distribution, and therefore no energy flux or energy cascade, which then means no turbulence. Therefore, detailed balance breaking is at the heart of the very existence of turbulence, imbedded into the very nature of turbulence as a nonequilibrium irreversible phenomenon. It is the power source that drives the potential landscape (probability distribution) to deviate from Gaussian characteristics and generates the irreversible probability flux velocity (probability flux), which accommodates the energy flux and energy cascade in turbulence. The connection of the nonequilibrium trinity to energy flux and turbulence energy cascade is illustrated in Fig. \ref{figure3}, which will be further supported below.

\subsection{Turbulence energy cascade in the inertial subrange generated by detailed balance breaking}\label{inertialrangeNSE}

In the above we have shown that detailed balance breaking is \emph{necessary} for energy cascade in turbulence. We further demonstrate that a specific mechanism of detailed balance breaking is also \emph{sufficient} to generate energy cascade in fully developed turbulence. We first consider the cumulative energy balance equation and the assumptions related to the inertial subrange, in preparation for the discussion of the causal relationship between detailed balance breaking and turbulence energy cascade.

\subsubsection{Cumulative energy balance}

We first introduce the large system size limit and the continuous spectrum. In the large system size limit $L\rightarrow \infty$, the discrete wavevector $\mathbf{k}=2\pi \mathbf{n}/L$ becomes continuous. Accordingly, the discrete sum $\sum_{\mathbf{k}}$ is replaced by the continuous integral $(L/2\pi)^3\int d\mathbf{k}$. We use $\kappa$ to denote the norm of the continuous wavevector, i.e., $\kappa=|\mathbf{k}|$, with the limit $L\rightarrow \infty$ implied. We also remove the directional information of the wavevector by integration over a sphere in the wavevector space. With these considerations the (continuous) energy spectrum is defined as $\mathcal{E}(\kappa)=(L/2\pi)^3\oint_{|\mathbf{k}|=\kappa}\mathcal{E}(\mathbf{k})dS $, where $L\rightarrow\infty$ is implied and the integration is over the sphere $|\mathbf{k}|=\kappa$. Similarly, we have the energy transfer spectrum $\mathcal{T}(\kappa)=(L/2\pi)^3\oint_{|\mathbf{k}|=\kappa}\mathcal{T}(\mathbf{k})dS $, the energy injection spectrum $\mathcal{I}(\kappa)=(L/2\pi)^3\oint_{|\mathbf{k}|=\kappa}W(\mathbf{k})dS $, and the energy dissipation spectrum in relation to the energy spectrum $\mathcal{D}(\kappa)=2\nu\kappa^2\mathcal{E}(\kappa)$. The steady-state energy balance equation in the large system size limit for the continuous spectrum reads
\begin{equation}\label{CSSEBE}
\mathcal{T}(\kappa)=\mathcal{D}(\kappa)-\mathcal{I}(\kappa).
\end{equation}

The energy balance equation can also be expressed in terms of the cumulative quantities convenient for the discussions of the inertial subrange. The cumulative energy balance equation in the steady state has the following form (notice the sign difference from Eq. (\ref{CSSEBE})) \cite{Legacy}:
\begin{equation}\label{CCSSEBE}
\mathcal{T}^{c}(\kappa)=\mathcal{I}^{c}(\kappa)-\mathcal{D}^{c}(\kappa).
\end{equation}
Here $\mathcal{T}^{c}(\kappa)=\int_{\kappa}^{\infty}\mathcal{T}(\kappa')d\kappa'=-\int_{0}^{\kappa}\mathcal{T}(\kappa')d\kappa'=-(L/2\pi)^3\int_{|\mathbf{k}|\leq \kappa}\mathcal{T}(\mathbf{k})d\mathbf{k}$ characterizes the rate at which energy is transferred out of modes $|\mathbf{k}|<\kappa$ (above a length scale) through modes on the sphere $|\mathbf{k}|=\kappa$ (at that length scale) into modes $|\mathbf{k}|>\kappa$ (below that length scale) by the nonlinear inertial convective force. $\mathcal{T}^{c}(\kappa)$ is called the transport power in Ref. \cite{FDT2} and referred to as the energy flux in Ref. \cite{Legacy}. Given that we have termed $\mathcal{T}(\mathbf{k})$ the energy flux in the wavevector space, we shall call $\mathcal{T}^{c}(\kappa)$ the cumulative energy flux (or cumulative energy transfer spectrum), in order to avoid confusion. $\mathcal{I}^{c}(\kappa)$  is the cumulative energy injection spectrum defined by $\mathcal{I}^{c}(\kappa)=\int_{0}^{\kappa}\mathcal{I}(\kappa')d\kappa'=(L/2\pi)^3\int_{|\mathbf{k}|\leq\kappa}W(\mathbf{k})d\mathbf{k}$, which represents the rate at which energy is injected into modes $|\mathbf{k}|\leq \kappa$ by the stochastic force. $\mathcal{D}^{c}(\kappa)$ is the cumulative energy dissipation spectrum defined as $\mathcal{D}^{c}(\kappa)=\int_{0}^{\kappa}2\nu\kappa'^2\mathcal{E}(\kappa')d\kappa'=(L/2\pi)^3\int_{|\mathbf{k}|\leq\kappa}2\nu k^2\mathcal{E}(\mathbf{k})d\mathbf{k}$, which represents the rate at which energy in modes $|\mathbf{k}|\leq\kappa$ is dissipated by the irreversible viscous force. The overall energy balance now takes the form $\mathcal{I}^{c}(\infty)=\mathcal{D}^{c}(\infty)=\varepsilon$, with $\mathcal{T}^{c}(\infty)=0$.

\subsubsection{Assumptions for the inertial subrange}

Under certain assumptions there exists an inertial subrange of energy cascade in fully developed turbulence with high Reynolds number, throughout which the cumulative energy flux $\mathcal{T}^{c}(\kappa)$ is constant and equals the total dissipation rate per unit mass $\varepsilon$ \cite{Legacy,McComb}. The three assumptions we propose below are adapted to the setting of stochastically forced steady-state turbulence. They are inspired by those in Ref. \cite{Legacy}, but have differences (two subtle ones and a prominent one) which we will comment on. These assumptions are as follows: (1) The stochastic force acts \emph{mainly} at the large scale comparable to the integral length scale $L_0$. (2) The statistically steady state of the system reached in the long time limit $t\rightarrow \infty$ has a finite mean energy per unit mass \emph{in the limit $\nu\rightarrow 0$}. (3) \emph{The stochastic force has an external origin.}

The first assumption, which can be termed large-scale stochastic forcing, has no essential difference from that in Ref. \cite{Legacy}, but contains a subtle difference reflected in the word `mainly'. Basically, it requires that $\bm{\xi}^s(\mathbf{k},t)$ is negligibly small for $|\mathbf{k}|\gg \kappa_0$ $(\sim1/L_0)$, which also applies to $\mathbf{D}^s(\mathbf{k})$ and $W(\mathbf{k})$ characterizing the statistical properties of $\bm{\xi}^s(\mathbf{k},t)$. But for technical reasons of ensuring the invertibility of the diffusion matrix $\mathbf{D}^s(\mathbf{k})$, we require a smooth rather than a sharp cutoff at the scale $\kappa_0$ (this is what the word `mainly' refers to). That means $\mathbf{D}^s(\mathbf{k})$ and $W(\mathbf{k})$ decrease sufficiently rapidly with $\kappa=|\mathbf{k}|$ so that they are negligibly small yet remains nonzero for $\kappa \gg \kappa_0$. The consequence of this assumption is that $\mathcal{I}^{c}(\kappa)=(L/2\pi)^3\int_{|\mathbf{k}|\leq\kappa}W(\mathbf{k})d\mathbf{k}\simeq \mathcal{I}^{c}(\infty)=\varepsilon$. That is, for sufficiently large $\kappa$ the cumulative energy injection rate up to the wavenumber $\kappa$ is almost the total energy injection rate (which, in the steady state, also equals the total energy dissipation rate $\varepsilon$). In other words, energy injection by the stochastic force is concentrated at the large spatial scale (small wavenumber).

The second assumption, finite mean energy, is an amendment to that in Ref. \cite{Legacy} by adding the qualifier `in the limit $\nu\rightarrow 0$', expressed mathematically as $\lim\limits_{\nu\rightarrow 0}\int_{0}^{\infty}\mathcal{E}(\kappa,\nu)d\kappa<\infty$. (The limit $\nu\rightarrow 0$ is equivalent to the infinite Reynolds number limit.) The reason for this amendment can be seen from the intended consequence of this assumption. That is, the cumulative dissipation rate up to any finite wavenumber vanishes in the limit $\nu\rightarrow 0$. Mathematically, for any fixed finite $\kappa$, $\mathcal{D}^{c}(\kappa)=\int_{0}^{\kappa}2\nu\kappa'^2\mathcal{E}(\kappa',\nu)d\kappa'\le 2\nu\kappa^2\int_{0}^{\kappa}\mathcal{E}(\kappa',\nu)d\kappa'\le 2\nu\kappa^2\int_{0}^{\infty}\mathcal{E}(\kappa',\nu)d\kappa'\rightarrow 0$ as $\nu\rightarrow 0$. However, the last step, $2\nu\kappa^2\int_{0}^{\infty}\mathcal{E}(\kappa',\nu)d\kappa'\rightarrow 0$ as $\nu\rightarrow 0$, cannot be guaranteed if one only assumes $\int_{0}^{\infty}\mathcal{E}(\kappa',\nu)d\kappa'<\infty$ for any fixed $\nu$. That is the reason why we assume $\lim\limits_{\nu\rightarrow 0}\int_{0}^{\infty}\mathcal{E}(\kappa,\nu)d\kappa<\infty$ instead, so that $\mathcal{D}^{c}(\kappa)\rightarrow 0$ as $\nu\rightarrow 0$ for any finite $\kappa$. This assumption implies that, in the infinite Reynolds number limit, energy dissipation by the viscous force, if any, can only exist at the infinitesimal spatial scale (infinitely large wavenumber), since it vanishes for all finite wavenumbers.

The third assumption we propose, external stochastic forcing, appears quite different from the finite dissipation rate assumption usually made (as in Ref. \cite{Legacy}) which assumes that the total energy dissipation rate per unit mass $\varepsilon$ remains finite (larger than zero) in the limit $\nu\rightarrow 0$. In the setup of stochastically forced steady-state turbulence, the assumption of external stochastic forcing, which seems less nontrivial, reduces the finite dissipation rate assumption to a consequence. More specifically, external stochastic forcing implies that the total energy injection rate, determined solely by the stochastic force, is independent of the viscosity $\nu$ of the fluid. Consequently, at the steady state (with the long time limit $t\rightarrow \infty$ taken), the total energy dissipation rate $\varepsilon$, which must balance out the total energy injection rate, is also independent of the viscosity $\nu$. Therefore, our assumption implies that, in the infinite Reynolds number limit $\nu\rightarrow 0$, the energy dissipation rate $\varepsilon$ at the steady state remains finite, because it simply does not depend on $\nu$.

However, it should be noted that in this setting the long time limit $t\rightarrow \infty$ for the system (ensemble, more precisely) to reach the steady state (solution to the steady-state FFPE) should be taken first, before the infinite Reynolds number limit $\nu\rightarrow 0$ is taken. Effectively, this means the limit $\nu\rightarrow 0$ is taken on a family of steady states parameterized by $\nu$, with a constant energy dissipation rate $\varepsilon$ determined by the external stochastic force that is independent of $\nu$. In this setting, no matter how small the viscosity $\nu$ is, it cannot be completely removed (set to zero), because the very existence of a steady state presupposes a nonzero $\nu$ for the viscous force to dissipate energy in order to achieve steady-state energy balance.

We also remark that this third assumption can be relaxed. If the stochastic force is not entirely external, the total energy injection rate may depend on the viscosity $\nu$. But as long as it remains finite in the limit $\nu\rightarrow 0$, so is the total dissipation rate $\varepsilon$, guaranteed by the steady-state energy balance.

The second and third assumptions combined together imply that, in the infinite Reynolds number limit, energy dissipation by the viscous force still exists and is concentrated at the infinitesimal spatial scale (infinitely large wavenumber). Mathematically, $\mathcal{D}^c(\infty)=\varepsilon >0$ and $\mathcal{D}^{c}(\kappa)\rightarrow 0$ for any $\kappa<\infty$ as $\nu\rightarrow 0$. This agrees with the fact that the Kolmogorov length scale $\eta=(\nu^3/\varepsilon)^{1/4}$, characteristic of the dissipation range, is squeezed to an arbitrarily small value in the limit $\nu\rightarrow 0$ for finite $\varepsilon$.

Under the above three assumptions, the steady-state cumulative energy balance equation in Eq. (\ref{CCSSEBE}) shows that $\mathcal{T}^{c}(\kappa)\simeq\varepsilon$ for any fixed finite $\kappa\gg \kappa_0$ in the limit $\nu\rightarrow 0$. That is, the cumulative energy flux is almost constant and equal to the energy dissipation rate throughout the inertial subrange $\kappa_0 \ll\kappa<\infty$ in the infinite Reynolds number limit. In these three assumptions, the first and the last associated with the stochastic force can be controlled and implemented to some extent, but the second one is generally not. If the second assumption is violated, the fluid system does not necessarily have an inertial subrange with constant cumulative energy flux.

\subsubsection{Turbulence energy cascade as a consequence of detailed balance breaking}

We now discuss the relation of the constant cumulative energy flux in the inertial subrange characterizing turbulence energy cascade to detailed balance breaking. The fact that the cumulative energy flux $\mathcal{T}^c(\kappa)$ is (almost) constant throughout the inertial subrange implies, seemingly peculiar, that the energy flux $\mathcal{T}(\kappa)$ or $\mathcal{T}(\mathbf{k})$ (almost) vanishes throughout the inertial subrange. This is a direct result of the steady-state energy balance equation and the fact that the energy injection rate and the energy dissipation rate both (almost) vanish in the inertial subrange. One may wonder whether this means the inertial subrange obeys detailed balance. We would like to stress, however, that the concept of detailed balance, at least in the context of what we have established, is a global one, which does not apply to merely a segment of the spectrum such as the inertial subrange, but has to be applied to the entire spectrum, all the degrees of freedom, and over the whole state space. More specifically, a fluid system with detailed balance is indicated by vanishing steady-state irreversible probability flux field functional $\mathbf{J}_s(\mathbf{k})[\mathbf{u}]$, which should be understood as $\mathbf{J}_s(\mathbf{k})[\mathbf{u}]$ vanishing at all the wavevectors $\mathbf{k}$ and for all the velocity field configurations $\mathbf{u}(\mathbf{k})$ in the state space. According to the energy-flux-irreversible-flux relation in Eq. (\ref{EFPFRNSE}), this means the energy flux $\mathcal{T}(\mathbf{k})$ vanishes at all the wavevectors when detailed balance holds. (An alternative argument can be made to reach the same conclusion from the energy-flux-deviated-potential relation in Eq. (\ref{PLEFRNSE}), given that detailed balance in the fluid system also leads to the Gaussian form of the potential landscape and probability distribution.) As a result, the cumulative energy flux $\mathcal{T}^c(\kappa)$ vanishes completely over the entire range of the spectrum $\kappa\in[0,+\infty)$ if the system obeys detailed balance. This is obviously different from the case in the inertial subrange of fully developed turbulence where the cumulative energy flux $\mathcal{T}^c(\kappa)$ is a nonzero constant $\varepsilon$.

The fact that in the inertial subrange $\mathcal{T}^c(\kappa)$ does not vanish even though $\mathcal{T}(\mathbf{k})$ almost vanishes is related to the fact that $\mathcal{T}(\mathbf{k})$ cannot and does not vanish at the lower and upper ends of the spectrum, dictated by the steady-state energy balance equation. In the lower end of the spectrum close to the energy injection scale $\kappa\sim \kappa_0$ (large spatial scale), the energy injection rate does not vanish, while the energy dissipation rate vanishes in the limit $\nu\rightarrow 0$, which implies that the energy transfer rate $\mathcal{T}(\mathbf{k})$ does not vanish in this range of the spectrum according to the energy balance equation. In the upper end of the spectrum (small spatial scale) close to the inverse Kolmogorov scale $\kappa\sim 1/\eta$ (which becomes infinity in the limit $\nu\rightarrow 0$), the energy injection rate vanishes, while the energy dissipation rate does not (in the limit $\nu\rightarrow 0$ the dissipation is concentrated at $\kappa=\infty$), which implies that the energy transfer rate $\mathcal{T}(\mathbf{k})$ does not vanish in this range of the spectrum either according to the energy balance equation. The nonvanishing $\mathcal{T}(\mathbf{k})$ in the lower and upper ends of the spectrum, together with its almost vanishing value throughout the inertial subrange, results in the constant nonzero cumulative energy flux $\mathcal{T}^c(\kappa)$ throughout the inertial subrange. Without the nonvanishing behaviors of $\mathcal{T}(\mathbf{k})$ in the lower and upper ends of the spectrum, a constant nonzero $\mathcal{T}^c(\kappa)$ throughout the inertial subrange is impossible.

The nonvanishing $\mathcal{T}(\mathbf{k})$ at the two ends of the spectrum and the nonvanishing constant $\mathcal{T}^c(\kappa)$ through the inertial subrange indicate a constant flow of energy from the large spatial scale (small wavenumber) to the small spatial scale (large wavenumber) as a manifestation of energy cascade in turbulence. In contrast, there is no such cascade of energy flowing through different scales in the equilibrium fluid system with detailed balance, as is evident from the vanishing of $\mathcal{T}(\mathbf{k})$ and $\mathcal{T}^c(\kappa)$ throughout the entire spectrum. This distinct difference between these two types of systems demonstrates that energy cascade of turbulence in the inertial subrange is inherently a nonequilibrium irreversible process with detailed balance breaking, which also implies non-Gaussian potential landscape and nonvanishing irreversible probability flux according to the connection of the energy flux to the nonequilibrium trinity.

The mechanism of detailed balance breaking in this fully developed turbulence scenario is the wide separation of the energy injection scale and the energy dissipation scale, created by the conditions of large-scale stochastic forcing and high Reynolds number (small viscosity). Large-scale stochastic forcing (the first assumption) concentrates energy injection at the large spatial scale, while high Reynolds number (under the second and third assumptions) concentrates energy dissipation at the small spatial scale, thus separating the energy injection and dissipation scales to the two opposite ends of the spectrum. This condition reflects a highly imbalanced relation between the stochastic force responsible for energy injection and the viscous force responsible for energy dissipation, which now depends on the nonlinear convective force responsible for energy transfer to mediate between these two widely separated scales, by going through a process of energy cascade with a constant cumulative energy flux through the intermediate scales, to manage the overall balance of energy injection and energy dissipation in the steady state. This demonstrates how this specific mechanism of detailed balance breaking, namely wide separation of the energy injection and energy dissipation scales by large-scale stochastic forcing and high Reynolds number, is the source of the constant cumulative energy flux in the inertial subrange and the power for the energy cascade in fully developed turbulence. See Fig. \ref{figure3} for an illustration of the consequential outflows of detailed balance breaking generating the energy flux and turbulence energy cascade.

An analogy can be made between this mechanical imbalance scenario for fully developed turbulence and a thermal imbalance scenario for heat conduction. Consider two heat baths, one with a high temperature and the other with a low temperature, connected by a heat-conducting rod. Due to the thermal imbalance indicated by the temperature difference in these two heat baths, there is a heat flux conducted by the rod from the high temperature bath  to the low temperature bath. The high temperature bath is the analog of the stochastic forcing at the large spatial scale, and the low temperature bath is that of the viscous dissipation at the small spatial scale. The temperature difference of the two heat baths is in analogy with the separation of the energy injection and energy dissipation scales. The heat flux conducted by the rod serves as an analog of the cumulative energy flux in turbulence energy cascade mediated by the nonlinear convective force.

\subsection{The nonequilibrium trinity in the four-fifths law of fully developed turbulence}\label{FourFifthsLaw}

The four-fifths law for homogeneous isotropic turbulence is one of the most significant results in fully developed turbulence, which relates the third order structure function with the mean energy dissipation rate in an exact form \cite{Kolmogorov1941a}. As is known this law can be derived from the K\'arm\'an-Howarth equation \cite{Kolmogorov1941a,McComb,Pope}. In order to reveal the underlying connection of the four-fifths law to the nonequilibrium trinity and gain a deeper understanding of the origin of this law, here we adopt a different route to derive this law.

\subsubsection{Outline of our approach}

\begin{figure}[!ht]
\centering
\includegraphics[width=4in]{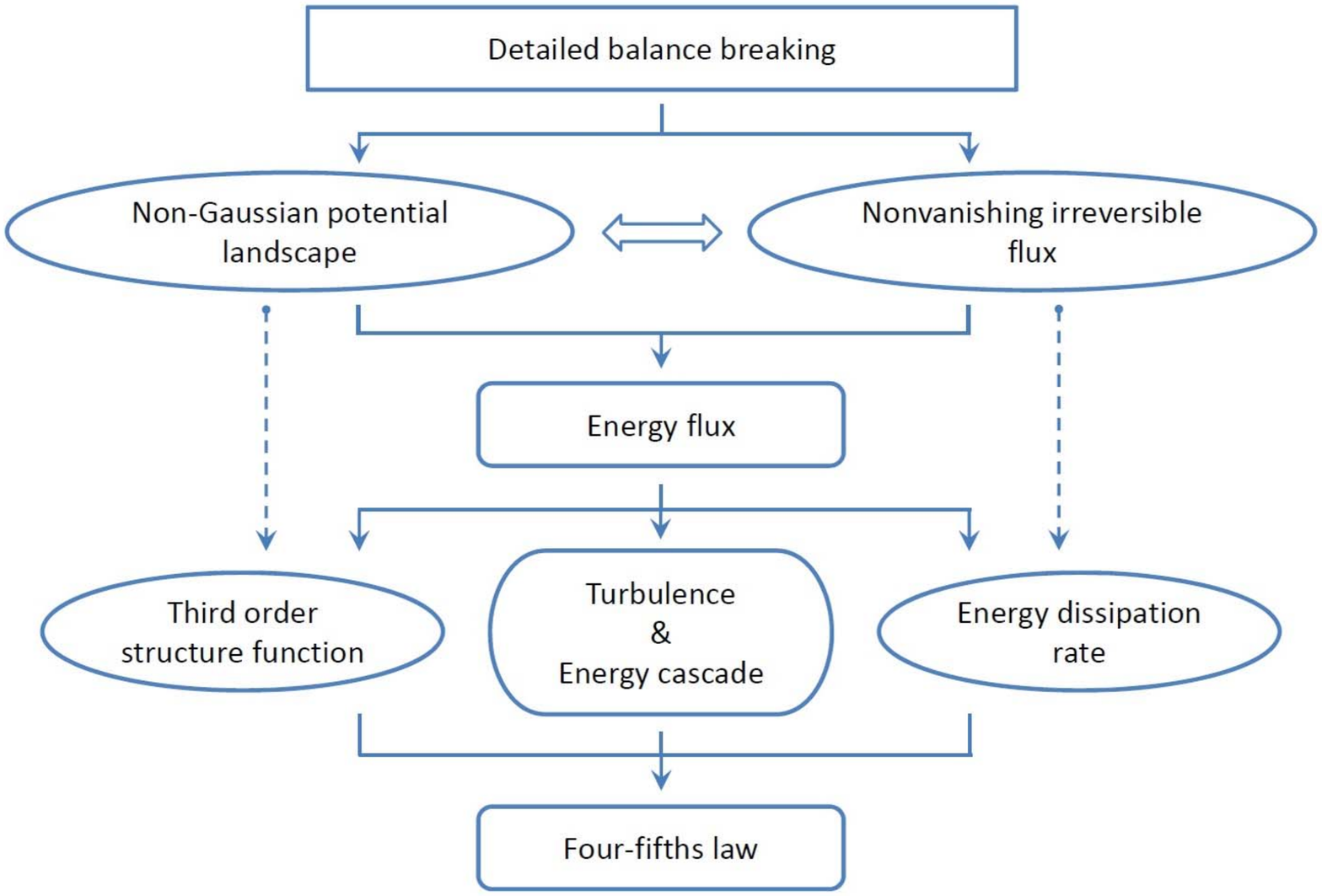}
\caption{The manifestation of the nonequilibrium trinity in the four-fifths law for fully developed turbulence.}\label{figure4}
\end{figure}

A schematic map of our approach can be found in Fig. \ref{figure4} and is outlined as follows. The central quantity in our approach is the energy flux $\mathcal{T}(\mathbf{k})$. On the one hand, we have shown that the energy flux $\mathcal{T}(\mathbf{k})$ is connected to the nonequilibrium trinity, through its connection to the non-Gaussian potential landscape in Eq. (\ref{PLEFRNSE}) and to the irreversible probability flux in Eq. (\ref{EFPFRNSE}). On the other hand, from the third-moment expression of $\mathcal{T}(\mathbf{k})$ in Eq. (\ref{SSETSNSE}), it can be expected that the energy flux is closely related to the third order structure function. Besides, it is also directly related to the cumulative energy flux which equals the dissipation rate $\varepsilon$ in the inertial subrange. Therefore, the energy flux plays a pivotal role in connecting the non-Gaussian potential landscape, the irreversible probability flux, the third order structure function, and the dissipation rate together (the four items in the circles in Fig. \ref{figure4} arranged at the four corners around the energy flux). With $\mathcal{T}(\mathbf{k})$ serving as a center of connection, the relation between the non-Gaussian potential landscape and the third order order structure function as well as the connection between the irreversible flux and the energy dissipation rate can be established (the two vertical dashed arrows in Fig. \ref{figure4}). With the flux deviation relation connecting the non-Gaussian potential landscape and the irreversible flux (the horizontal double-headed arrow in Fig. \ref{figure4}), we can then obtain the four-fifths law relating the third order structure function and the dissipation rate.

A few cautionary remarks are appropriate here. The way we derive the four-fifths law as outlined above may appear to be less than straightforward for readers familiar with other approaches to derive the four-fifths law. This is determined by our objective here, which is not to obtain the four-fifths law in the most efficient manner, but to uncover how the four-fifths law is deeply connected to the nonequilibrium trinity and trace back to the source and origin of the law. Also, part of the mathematical results in the following derivation can be found in Ref. \cite{Legacy}. However, our strategy of using $\mathcal{T}(\mathbf{k})$ as a center of connection as well as our objective just stated are fundamentally different from there. What we contribute here are the connections we reveal and the approach we use, not so much in the efficiency and mathematics to get the end result.

\subsubsection{Derivation of the four-fifths law}

In accord with our approach sketched above, we follow a three-step derivation of the four-fifths law.

\textbf{Step 1:} We relate the non-Gaussian potential landscape to the third order structure function through their connections to the energy flux $\mathcal{T}(\mathbf{k})$. (Note that the wavevector $\mathbf{k}$ is still discrete for now; we delay taking the large system size limit.) The connection of the deviated potential landscape to the energy flux has been given in Eq. (\ref{PLEFRNSE}), which we reproduce below for the reader's convenience
\begin{equation}\label{EFPLR}
\mathcal{T}(\mathbf{k})=\mathcal{R}\{\langle \mathbf{u}^*(\mathbf{k})\cdot\mathbf{D}^s(\mathbf{k})\cdot\nabla_{\mathbf{u}(-\mathbf{k})}\Lambda[\mathbf{u}]\rangle_s\},
\end{equation}
where $\Lambda[\mathbf{u}]=\Phi_0[\mathbf{u}]-\Phi[\mathbf{u}]$ characterizes the deviation of the nonequilibrium potential landscape from the Gaussian potential landscape.

The connection of the energy flux $\mathcal{T}(\mathbf{k})$ to the third order structure function is more involved. Since $\mathcal{T}(\mathbf{k})$ is in the wavevector space, while the structure function is in the physical space, we transform $\mathcal{T}(\mathbf{k})$ into the physical space. Define the physical-space energy flux $\mathcal{T}(\mathbf{r})=\sum_{\mathbf{k}}\mathcal{T}(\mathbf{k})e^{i\mathbf{k}\cdot\mathbf{r}}$, with the inverse relation $\mathcal{T}(\mathbf{k})=\int \mathcal{T}(\mathbf{r})e^{-i\mathbf{k}\cdot\mathbf{r}}d\mathbf{r}/L^3$. Notice that $\mathcal{T}(\mathbf{k})$ in Eq. (\ref{SSETSNSE}) can also be written as $\mathcal{T}(\mathbf{k})=\mathcal{R}\{\langle \mathbf{u}^*(\mathbf{k})\cdot\mathbf{F}^{con}(\mathbf{k})[\mathbf{u}]\rangle_{s}\}$, where $\mathbf{F}^{con}(\mathbf{k})[\mathbf{u}]$ is the Fourier amplitude of the convective force $\mathbf{F}^{con}(\mathbf{x})[\mathbf{u}]=-\mathbf{u}\cdot\nabla\mathbf{u}$. We derive the expression of the physical-space energy flux as follows:
\begin{eqnarray}\label{EFSF1NSE}
\mathcal{T}(\mathbf{r})&=&\DF{1}{2L^3}\int d\mathbf{x}\left\langle \mathbf{u}(\mathbf{x})\cdot\mathbf{F}^{con}(\mathbf{x}+\mathbf{r})[\mathbf{u}]+\mathbf{u}(\mathbf{x})\cdot\mathbf{F}^{con}(\mathbf{x}-\mathbf{r})[\mathbf{u}]\right\rangle_{s}\nonumber\\
&=&-\DF{1}{2L^3}\nabla_{\mathbf{r}}\cdot\int d\mathbf{x} \left\langle \mathbf{u}(\mathbf{x}+\mathbf{r})\mathbf{u}(\mathbf{x}+\mathbf{r})\cdot\mathbf{u}(\mathbf{x})-\mathbf{u}(\mathbf{x})\mathbf{u}(\mathbf{x})\cdot\mathbf{u}(\mathbf{x}+\mathbf{r})\right\rangle_{s}\nonumber\\
&=&\DF{1}{4}\nabla_{\mathbf{r}}\cdot \DF{1}{L^3}\int d\mathbf{x}\left\langle\left|\mathbf{u}(\mathbf{x}+\mathbf{r})-\mathbf{u}(\mathbf{x})\right|^2\left[\mathbf{u}(\mathbf{x}+\mathbf{r})-\mathbf{u}(\mathbf{x})\right]\right\rangle_{s}\nonumber\\
&=&\DF{1}{4}\nabla_{\mathbf{r}}\cdot\overline{\langle|\delta\mathbf{u}(\mathbf{x},\mathbf{r})|^2\delta\mathbf{u}(\mathbf{x},\mathbf{r})\rangle}_{s},
\end{eqnarray}
where $\delta\mathbf{u}(\mathbf{x},\mathbf{r})\equiv \mathbf{u}(\mathbf{x}+\mathbf{r})-\mathbf{u}(\mathbf{x})$ is the velocity increment and  $\overline{\langle|\delta\mathbf{u}(\mathbf{x},\mathbf{r})|^2\delta\mathbf{u}(\mathbf{x},\mathbf{r})\rangle}_{s}$ is the spatial average of $\langle|\delta\mathbf{u}(\mathbf{x},\mathbf{r})|^2\delta\mathbf{u}(\mathbf{x},\mathbf{r})\rangle_s$ for possibly inhomogeneous velocity fields. We have not made any approximations or additional assumptions in obtaining Eq. (\ref{EFSF1NSE}); it is an exact result of the definitions of the quantities involved.

If we assume that the velocity field is statistically \emph{homogeneous} in space, then the spatial average and the dependence on $\mathbf{x}$ in Eq. (\ref{EFSF1NSE}) can be dropped, which gives
\begin{equation}\label{EFSF2NSE}
\mathcal{T}(\mathbf{r})=\DF{1}{4}\nabla_{\mathbf{r}}\cdot\langle|\delta\mathbf{u}(\mathbf{r})|^2\delta\mathbf{u}(\mathbf{r})\rangle_{s}
\end{equation}
for statistically homogeneous (not necessarily isotropic) velocity fields in space. The form $\langle|\delta\mathbf{u}(\mathbf{r})|^2\delta\mathbf{u}(\mathbf{r})\rangle_{s}$ suggests its close connection to the third order structure function. In fact, by further assuming statistical \emph{isotropy} of the velocity fields, $\langle|\delta\mathbf{u}(\mathbf{r})|^2\delta\mathbf{u}(\mathbf{r})\rangle_{s}$ can be related to the third order longitudinal structure function $S_3(r)=\left\langle[\delta\mathbf{u}(\mathbf{r})\cdot\hat{\mathbf{r}}]^3\right\rangle_s$ as follows \cite{Landau,Legacy}:
\begin{equation}\label{RTTOSFNSE}
\langle|\delta\mathbf{u}(\mathbf{r})|^2\delta\mathbf{u}(\mathbf{r})\rangle_{s}=\DF{1}{3}[(4+r\partial_r)S_3(r)]\hat{\mathbf{r}},
\end{equation}
where $\hat{\mathbf{r}}=\mathbf{r}/r$ is the unit vector in the direction of $\mathbf{r}$ and $\partial_r=\partial/\partial r$. Combining Eqs. (\ref{EFSF2NSE}) and (\ref{RTTOSFNSE}) leads to the following relation between the physical-space energy flux and the third order structure function \cite{Legacy}:
\begin{equation}\label{TRENSE}
\mathcal{T}(r)=\DF{1}{12}(3+r\partial_r)(5+r\partial_r)\DF{S_3(r)}{r},
\end{equation}
where $\mathcal{T}(r)$ is now a function of $r=|\mathbf{r}|$. Eq. (\ref{TRENSE}) is obtained under the assumptions of statistical homogeneity and isotropy of the velocity field.

With the energy flux $\mathcal{T}(\mathbf{k})$ as a common connection, we obtain from Eqs. (\ref{EFPLR}) and (\ref{TRENSE}) the following relation between the non-Gaussian potential landscape and the third order structure function
\begin{equation}\label{SFNGPL}
\sum_{\mathbf{k}}e^{i\mathbf{k}\cdot\mathbf{r}}\,\mathcal{R}\left\{\left\langle \mathbf{u}^*(\mathbf{k})\cdot\mathbf{D}^s(\mathbf{k})\cdot\nabla_{\mathbf{u}(-\mathbf{k})}\Lambda[\mathbf{u}]\right\rangle_s\right\}=\DF{1}{12}(3+r\partial_r)(5+r\partial_r)\DF{S_3(r)}{r},
\end{equation}
for statistically homogeneous and isotropic velocity fields. This equation shows explicitly how detailed balance breaking, through the non-Gaussian potential landscape (its direct consequence), is manifested in the third order structure function, under the assumptions of statistical homogeneity and isotropy of the velocity field.

\textbf{Step 2:} We relate the irreversible probability flux to the dissipation rate through their connections to the energy flux $\mathcal{T}(\mathbf{k})$. The connection of the irreversible probability flux to the energy flux has been given in Eq. (\ref{EFPFRNSE}), which is reproduced below for the reader's convenience
\begin{equation}\label{EFPFRNSE2}
\mathcal{T}(\mathbf{k})=-\mathcal{R}\left\{\langle \mathbf{u}^*(\mathbf{k})\cdot\mathbf{V}_s^{irr}(\mathbf{k})[\mathbf{u}]\rangle_s\right\}.
\end{equation}

The connection of the energy flux to the dissipation rate is derived as follows. We take the large system size limit and work with the continuous spectrum. We know that the energy flux is related to the cumulative energy flux by $\mathcal{T}^{c}(\kappa)=-\int_{0}^{\kappa}\mathcal{T}(\kappa')d\kappa'=-(L/2\pi)^3\int_{|\mathbf{k}|\leq \kappa}\mathcal{T}(\mathbf{k})d\mathbf{k}$. Under the same three assumptions that lead to the existence of an inertial subrange with constant cumulative energy flux discussed in Section \ref{inertialrangeNSE}, we obtain the following relation between the energy flux and the dissipation rate:
\begin{equation}\label{EFDR}
-\left(\DF{L}{2\pi}\right)^3\int_{|\mathbf{k}|\leq \kappa}\mathcal{T}(\mathbf{k})d\mathbf{k}=\varepsilon
\end{equation}
for $\kappa\gg \kappa_0$ in the limit $\nu\rightarrow 0$.

With the energy flux $\mathcal{T}(\mathbf{k})$ as a common connection, we obtain from Eqs. (\ref{EFPFRNSE2}) and (\ref{EFDR}) the relation between the irreversible probability flux and the dissipation rate:
\begin{equation}\label{IFDR}
\left(\DF{L}{2\pi}\right)^3\int_{|\mathbf{k}|\leq \kappa}d\mathbf{k}\,\mathcal{R}\left\{\left\langle \mathbf{u}^*(\mathbf{k})\cdot\mathbf{V}_s^{irr}(\mathbf{k})[\mathbf{u}]\right\rangle_s\right\}=\varepsilon.
\end{equation}
This equation shows explicitly how detailed balance breaking, through its connection to the irreversible probability flux (its direct consequence), is manifested in the dissipation rate, under the assumptions ensuring the existence of an inertial subrange.

\textbf{Step 3:} Now we invoke the flux deviation relation in Eq. (\ref{FDRWVNSE}), $\mathbf{V}_s^{irr}(\mathbf{k})[\mathbf{u}]=-\mathbf{D}^s(\mathbf{k})\cdot\nabla_{\mathbf{u}(-\mathbf{k})}\Lambda[\mathbf{u}]$, which relates the non-Gaussian potential landscape with the irreversible probability flux. It then follows that
\begin{equation}
\mathcal{R}\{\langle \mathbf{u}^*(\mathbf{k})\cdot\mathbf{D}^s(\mathbf{k})\cdot\nabla_{\mathbf{u}(-\mathbf{k})}\Lambda[\mathbf{u}]\rangle_s\}=-\mathcal{R}\{\langle \mathbf{u}^*(\mathbf{k})\cdot\mathbf{V}_s^{irr}(\mathbf{k})[\mathbf{u}]\rangle_s\},
\end{equation}
which relates the two major terms in Eqs. (\ref{SFNGPL}) and (\ref{IFDR}). Inverting the Fourier series in Eq. (\ref{SFNGPL}) and plugging it (with a negative sign) into Eq. (\ref{IFDR}), we obtain
\begin{equation}\label{PRETOSFDPR1}
-\DF{1}{(2\pi)^3}\int_{|\mathbf{k}|\leq \kappa}d\mathbf{k}\int d\mathbf{r}\,e^{-i\mathbf{k}\cdot\mathbf{r}}\left[\DF{1}{12}(3+r\partial_r)(5+r\partial_r)\DF{S_3(r)}{r}\right]=\varepsilon,
\end{equation}
where $\kappa\gg \kappa_0$ in the limit $\nu\rightarrow 0$. This gives a preliminary relation between the third order structure function and the dissipation rate. The double integral can be carried out, which gives
\begin{equation}\label{PRETOSFDPR2}
-\DF{2}{\pi}\int_{0}^{\infty}\DF{\sin(\kappa r)}{r}(1+r\partial_r)\left[\DF{1}{12}(3+r\partial_r)(5+r\partial_r)\DF{S_3(r)}{r}\right]dr =\varepsilon,
\end{equation}
where integration by parts has been used to produce the form $(1+r\partial_r)$ in Eq. (\ref{PRETOSFDPR2}) from Eq. (\ref{PRETOSFDPR1}). Rewrite Eq. (\ref{PRETOSFDPR2}) in the following form:
\begin{equation}\label{RNSE}
-\DF{2}{\pi}\int_{0}^{\infty}\DF{\sin \theta}{\theta}\mathcal{F}\left(\theta/\kappa\right)d\theta=\varepsilon,
\end{equation}
where $\theta/\kappa=r$ and
\begin{equation}
\mathcal{F}(r)=\DF{1}{12}(1+r\partial_r)(3+r\partial_r)(5+r\partial_r)\DF{S_3(r)}{r}.
\end{equation}
The large $\kappa$ range ($\kappa\gg \kappa_0$) corresponds to the small $r$ range since $\theta/\kappa=r$. Hence, the integration in Eq. (\ref{RNSE}) is dominated by the behavior of $\mathcal{F}(r)$ for small $r$. With the help of the result $\int_{0}^{\infty}\sin \theta /\theta d\theta=\pi/2$, the following equation for the third order structure function in relation to the dissipation rate can be obtained \cite{Legacy}:
\begin{equation}\label{EFDRRNSE}
\mathcal{F}(r)=\DF{1}{12}(1+r\partial_r)(3+r\partial_r)(5+r\partial_r)\DF{S_3(r)}{r}\simeq -\varepsilon
\end{equation}
for $r\ll 1/\kappa_0(\sim L_0)$ in the limit $\nu\rightarrow 0$. The only solution which vanishes at $r=0$ is given by:
\begin{equation}\label{TOSFEFNSE}
S_3(r)=-\DF{4}{5}\varepsilon r
\end{equation}
for $r\ll L_0$ in the limit $\nu\rightarrow 0$.

We have thus derived the four-fifths law in Eq. (\ref{TOSFEFNSE}) which relates the third order structure function with the dissipation rate, together with the relation between the non-Gaussian potential landscape and the third order structure function in Eq. (\ref{SFNGPL}), and the relation between the irreversible probability flux and the dissipation rate in Eq. (\ref{IFDR}), by using the energy flux $\mathcal{T}(\mathbf{k})$ as the center of connection. In obtaining these relations what have been assumed are the statistical homogeneity and isotropy of the velocity field and the same three assumptions for the existence of an inertial subrange discussed in Section \ref{inertialrangeNSE}. No hypothesis of self-similarity or universality was invoked.

\subsubsection{Implications}

It is clear from the derivation as well as the connections revealed  that, under the assumptions made, the four-fifths law for fully developed turbulence is a manifestation and reflection of the non-Gaussian potential landscape and the irreversible probability flux that arise from detailed balance breaking which characterizes the nonequilibrium nature and drives the irreversible dynamics of turbulence. More specifically, the third order structure function on the l.h.s. of the law reflects the nonequilibrium nature of turbulence from its connection to the non-Gaussian potential landscape (probability distribution) quantified by Eq. (\ref{SFNGPL}). If the potential landscape (probability distribution) were Gaussian, the third order structure function would have vanished. On the other hand, the energy dissipation rate on the r.h.s. of the law reflects the nonequilibrium nature of turbulence from its connection to the irreversible probability flux quantified by Eq. (\ref{IFDR}). If the irreversible probability flux vanished completely, there would be no nonvanishing constant cumulative energy flux equal to the energy dissipation rate $\varepsilon$ in the inertial subrange. Thus in the equilibrium case with detailed balance, where the steady-state probability distribution is Gaussian and the steady-state irreversible probability flux vanishes completely, the four-fifths law in Eq. (\ref{TOSFEFNSE}) for fully developed turbulence reduces to the utterly trivial identity $0=0$. Therefore, the very existence of the four-fifths law is founded upon the nonequilibrium irreversible nature of turbulence, characterized by detailed balance breaking that is created by large-scale forcing and high Reynolds number, manifested in the non-Gaussian potential landscape and the irreversible probability flux, leading to the directional energy flux, which on the one hand is connected to the third order structure function, and on the other hand connected to the dissipation rate, bringing together these two different aspects that reflect the common nonequilibrium irreversible nature of turbulence into one law, namely, the four-fifths law for fully developed turbulence. See Fig. \ref{figure4} for an illustration of the causal flow from detailed balance breaking to the four-fifths law.

\subsection{The connection of the nonequilibrium trinity to the scaling laws in turbulence}

We now discuss the possible connection of the nonequilibrium trinity to the scaling laws in fully developed turbulence. Some simple arguments could give a hint. If the self-similarity \emph{hypothesis} is invoked that $\delta \mathbf{u}(\mathbf{x},\lambda\mathbf{r})=\lambda^h\delta \mathbf{u}(\mathbf{x},\mathbf{r})$ for all $|\mathbf{r}|\ll L_0$ and $|\lambda \mathbf{r}|\ll L_0$, then the four-fifths law (the third order structure function scales linearly with $r$) indicates that the scaling exponent $h=1/3$. This further implies that the $n$-th order structure function $S_n(r)=\langle [\delta \mathbf{u}(\mathbf{r})\cdot \hat{\mathbf{r}}]^n\rangle$ scales with $r$ as $S_n(r)\propto r^{n/3}$. In particular, the second order structure function $S_2(r)\propto r^{2/3}$ gives the $2/3$ law for the average kinetic energy, which implies the $5/3$ law for the energy spectrum $\mathcal{E}(\kappa)\propto \kappa^{-5/3}$ \cite{Legacy,Pope}. Therefore, under the self-similarity hypothesis, these scaling laws can be `derived' from the four-fifths law that results from the nonequilibrium trinity characterizing the nonequilibrium irreversible essence of turbulence.

However, experimental evidence indicates that although the structure function follows a scaling law of the form $S_n(r)\propto r^{\zeta_n}$, the scaling exponent $\zeta_n$ deviates from $n/3$ for $n\neq 3$ (significantly so for $n>3$). This deviation is referred to as anomalous scaling, which implies the breaking of scale invariance at small scales of the turbulent flow and is a signature of intermittency in turbulence. To accommodate intermittency and anomalous scaling, it is desirable in the context of this work to investigate more fundamental ways of connection between the scaling laws and the nonequilibrium trinity, which goes beyond the arguments above based on the self-similarity hypothesis. The following is such an attempt.

The scaling law associated with the behavior of a statistical quantity is fundamentally connected to the particular forms and properties of the ensemble probability distribution functional $P_s[\mathbf{u}]$ associated with fully developed turbulence. For specificity, we consider the $n$-th order structure function in the steady state $S_n(r)=\langle [\delta \mathbf{u}(\mathbf{r})\cdot \hat{\mathbf{r}}]^n\rangle_s$, with its expression spelled out as follows:
\begin{equation}\label{SFN}
S_n(r)=\langle [\delta \mathbf{u}(\mathbf{r})\cdot \hat{\mathbf{r}}]^n\rangle_s=\int [\delta \mathbf{u}(\mathbf{r})\cdot \hat{\mathbf{r}}]^n P_s[\mathbf{u}]\delta\mathbf{u}.
\end{equation}
It is clear from this expression that the behavior of $S_n(r)$ is dependent on the properties of the ensemble probability distribution functional $P_s[\mathbf{u}]$. The scaling law that $S_n(r)$ obeys in the inertial subrange can be related to the specific forms and properties of $P_s[\mathbf{u}]$ for fully developed turbulence. For instance, the anomalous scaling behavior of $S_n(r)$ implies that $P_s[\mathbf{u}]\delta\mathbf{u}$ is not scale invariant.

Now we relate the probability distribution functional $P_s[\mathbf{u}]$ to the non-Gaussian potential landscape and the irreversible probability flux which directly stem from detailed balance breaking. The relation between $P_s[\mathbf{u}]$ and the deviated potential landscape $\Lambda[\mathbf{u}]$ is given by $P_s[\mathbf{u}]=P_0[\mathbf{u}]e^{\Lambda[\mathbf{u}]}$. Plugging it into Eq. (\ref{SFN}), we obtain
\begin{equation}\label{SFN2}
S_n(r)=\int [\delta \mathbf{u}(\mathbf{r})\cdot \hat{\mathbf{r}}]^n e^{\Lambda[\mathbf{u}]} P_0[\mathbf{u}]\delta\mathbf{u}\equiv\left\langle[\delta \mathbf{u}(\mathbf{r})\cdot \hat{\mathbf{r}}]^n e^{\Lambda[\mathbf{u}]} \right\rangle_0,
\end{equation}
where $\langle\cdots\rangle_0$ denotes the ensemble average over the Gaussian distribution functional $P_0[\mathbf{u}]$. The above equation relates the deviated potential landscape $\Lambda[\mathbf{u}]$ to the $n$-th order structure function $S_n(r)$ explicitly, which shows how detailed balance breaking affects the form of $S_n(r)$ through the deviated potential landscape $\Lambda[\mathbf{u}]$.

Furthermore, the deviated potential landscape that arises from detailed balance breaking is related to the irreversible probability flux through the flux deviation relation in Eq. (\ref{FDRNSE}): $\mathbf{V}_s^{irr}(\mathbf{x})[\mathbf{u}]=-\int d\mathbf{x}'\, \mathbf{D}^s(\mathbf{x}-\mathbf{x}')\cdot\delta_{\mathbf{u}(\mathbf{x}')}\Lambda[\mathbf{u}]$. Inverting this equation gives $\delta_{\mathbf{u}(\mathbf{x})}\Lambda[\mathbf{u}]=-\int d\mathbf{x}'\,\mathbf{D}_s^{-1}(\mathbf{x}-\mathbf{x}')\cdot\mathbf{V}_s^{irr}(\mathbf{x})[\mathbf{u}]$. Formally integrating this equation (a functional integral), we obtain
\begin{equation}\label{SFN3}
\Lambda[\mathbf{u}]=\int^{\mathbf{u}}\delta\mathbf{w}\bullet\left\{-\int d\mathbf{x}'\,\mathbf{D}_s^{-1}(\mathbf{x}-\mathbf{x}')\cdot\mathbf{V}_s^{irr}(\mathbf{x})[\mathbf{w}]\right\},
\end{equation}
where `$\bullet$' represents the inner product in the state space. This equation expresses the deviated potential landscape in terms of the irreversible probability flux explicitly. Plugging Eq. (\ref{SFN3}) into Eq. (\ref{SFN2}), we obtain
\begin{equation}\label{SFN4}
S_n(r)=\Bigg\langle[\delta \mathbf{u}(\mathbf{r})\cdot \hat{\mathbf{r}}]^n \exp\left\{\int^{\mathbf{u}}\delta\mathbf{w}\bullet\left(-\int d\mathbf{x}'\,\mathbf{D}_s^{-1}(\mathbf{x}-\mathbf{x}')\cdot\mathbf{V}_s^{irr}(\mathbf{x})[\mathbf{w}]\right)\right\} \Bigg\rangle_0,
\end{equation}
which relates the irreversible probability flux to the $n$-th order structure function explicitly. Eqs. (\ref{SFN2}) and (\ref{SFN4}) show how detailed balance breaking can affect the scaling behavior of the $n$-th order structure function through the non-Gaussian potential landscape and the irreversible probability flux.

Examining how the above relations have been obtained, one will find these results are not dependent on the form of the $n$-th order structure function; they can actually be applied to any statistical quantity. More specifically, consider an observable $O[\mathbf{u}]$ which is a functional of the velocity field (note that $O[\mathbf{u}]$ can be dependent on $\mathbf{x}$ explicitly; it can also be dependent on multiple points). Following the same procedures as above, we obtain the following relations
\begin{equation}\label{SFN5}
\begin{split}
\langle O[\mathbf{u}]\rangle_s&=\left\langle O[\mathbf{u}]e^{\Lambda[\mathbf{u}]} \right\rangle_0\\
&=\left\langle O[\mathbf{u}]\exp\left\{\int^{\mathbf{u}}\delta\mathbf{w}\bullet\left(-\int d\mathbf{x}'\,\mathbf{D}_s^{-1}(\mathbf{x}-\mathbf{x}')\cdot\mathbf{V}_s^{irr}(\mathbf{x})[\mathbf{w}]\right)\right\} \right\rangle_0,
\end{split}
\end{equation}
where $\langle\cdots\rangle_s$ and $\langle\cdots\rangle_0$ denote the ensemble average over $P_s[\mathbf{u}]$ and $P_0[\mathbf{u}]$, respectively. Eq. (\ref{SFN5}) is expressed in the physical space; its counterpart in the wavevector space reads as follows:
\begin{equation}\label{SFN6}
\begin{split}
\langle O[\mathbf{u}]\rangle_s&=\left\langle O[\mathbf{u}]e^{\Lambda[\mathbf{u}]} \right\rangle_0\\
&=\left\langle O[\mathbf{u}]\exp\left\{-\int^{\mathbf{u}}\sum_{\mathbf{k}}d\mathbf{w}(\mathbf{-k})\cdot\mathbf{D}_s^{-1}(\mathbf{k})\cdot\mathbf{V}_s^{irr}(\mathbf{k})[\mathbf{w}]\right\} \right\rangle_0.
\end{split}
\end{equation}
Eqs. (\ref{SFN5}) and (\ref{SFN6}) relate explicitly the non-Gaussian potential landscape and the irreversible probability flux that arise from detailed balance breaking to the statistical average of the observable.  It shows how detailed balance breaking can affect the scaling behaviors of the averaged observable $\mathcal{O}[\mathbf{u}]\equiv \langle O[\mathbf{u}]\rangle_s$.

A more specific quantity of particular interest is the energy spectrum $\mathcal{E}(\kappa)=(L/2\pi)^3\oint_{|\mathbf{k}|=\kappa}\mathcal{E}(\mathbf{k})dS=(L/2\pi)^3\oint_{|\mathbf{k}|=\kappa}\langle\mathbf{u}(-\mathbf{k})\cdot\mathbf{u}(\mathbf{k})/2\rangle dS$. With the help of Eq. (\ref{SFN6}), we obtain
\begin{equation}\label{SFN7}
\begin{split}
\mathcal{E}(\kappa)=&\DF{1}{2}\left(\DF{L}{2\pi}\right)^3\oint_{|\mathbf{k}|=\kappa}\left\langle \mathbf{u}(-\mathbf{k})\cdot\mathbf{u}(\mathbf{k}) e^{\Lambda[\mathbf{u}]} \right\rangle_0\\
=&\DF{1}{2}\left(\DF{L}{2\pi}\right)^3\oint_{|\mathbf{k}|=\kappa}\Bigg\langle \mathbf{u}(-\mathbf{k})\cdot\mathbf{u}(\mathbf{k})\\
&\hspace{35pt} \times\exp\left\{-\int^{\mathbf{u}}\sum_{\mathbf{k}'}d\mathbf{w}(-\mathbf{k}')\cdot\mathbf{D}_s^{-1}(\mathbf{k}')\cdot\mathbf{V}_s^{irr}(\mathbf{k}')[\mathbf{w}]\right\} \Bigg\rangle_0.
\end{split}
\end{equation}
This equation relates the energy spectrum $\mathcal{E}(\kappa)$ explicitly to the deviated potential landscape $\Lambda[\mathbf{u}]$ and the irreversible probability flux velocity $\mathbf{V}_s^{irr}(\mathbf{k})[\mathbf{w}]$ generated by detailed balance breaking. It shows how detailed balance breaking can affect the scaling behavior of the energy spectrum $\mathcal{E}(\kappa)$.

In the above, we have related the structure functions, the energy spectrum, and more general ensemble-averaged observables to the nonequilibrium trinity. These relations show how the properties and behaviors of these statistical observables can be affected by detailed balance breaking through the non-Gaussian potential landscape and the irreversible probability flux. In particular, the scaling laws
of these quantities in the inertial range of fully developed turbulence can be expected to be related to the
specific forms and properties of the non-Gaussian potential landscape and the irreversible probability flux
that are generated by the specific detailed balance breaking mechanism for fully developed turbulence. It may be worth noting that turbulent systems are nonequilibrium fluid  systems with detailed balance breaking; yet not all fluid  systems with detailed balance breaking are turbulent systems. The energy cascade process in the inertial range of fully developed turbulence is generated by a specific detailed balance breaking mechanism, that is, the separation of the energy injection and energy dissipation scales through large-scale forcing and high Reynolds number, which represents
a highly imbalanced relation between the stochastic force and the viscous force. We speculate that the particular character of this detailed balance breaking mechanism for fully developed turbulence, in contrast with
general nonequilibrium fluid  systems with other forms of detailed balance breaking, will generate specific forms of non-Gaussian potential landscape and irreversible probability flux, which determine the scaling behaviors of the various statistical observables in fully developed turbulence.

We remark that previous studies on the statistical properties of the turbulence have focused extensively on two-point and three-point correlation functions. In contrast, our approach to turbulent fluids and general nonequilibrium fluids has a distinctive global characteristic that goes beyond that. The non-Gaussian potential landscape and the irreversible probability flux that arise from detailed balance breaking are both global characterizations of the nonequilibrium aspects of the statistical properties of turbulence and nonequilibrium fluids. They contain the nonequilibrium statistical information of all orders of correlation functions considering their relations to the steady-state probability functional defined on the entire field configuration space. However, this also makes it a very challenging problem to solve. Although we have not yet obtained quantitative mathematical results on the scaling laws along this line of thought, it seems worthy to investigate further in this direction. We hope our potential landscape and flux field theory presented here, when combined with other field theoretical approaches, could offer quantitative new insights into the phenomenon of anomalous scaling and intermittency in turbulence in the future.

\section{Conclusion}

In this article we have established the potential landscape and flux field theory for incompressible fluids governed by stochastic Navier-Stokes equations. We have found that detailed balance, which characterizes the equilibrium nature of the fluid system, is quantified by the detailed balance constraint that relates the solenoidal convective force, the viscous force, and the stochastic force together in a very specific manner, representing mechanical balance in the driving forces of the fluid system. As a result of detailed balance, the equilibrium potential landscape is Gaussian and the irreversible probability flux vanishes completely. Consequently, the stochastic Navier-Stokes equation has a potential form in the state space, reflecting the reversible nature of the equilibrium stochastic fluid dynamics with detailed balance. As a manifestation of detailed balance on the level of energy balance, the energy injection rate at a Fourier mode is exactly balanced by the energy dissipation rate at the same mode, indicating the absence of energy transfer between different modes that couples them together. As a result, the energy flux in the wavevector space vanishes entirely. No turbulence energy cascade through different scales can take place in such equilibrium fluid systems with detailed balance.

We have shown that for turbulence to exist, detailed balance characterized by the detailed balance constraint must be violated, which represents a form of mechanical imbalance in the driving forces of the fluid system. Detailed balance breaking, according to the nonequilibrium source equation, leads directly to the non-Gaussian potential landscape and the irreversible probability flux, with the latter two aspects related to each other by the flux deviation relation. Detailed balance breaking, non-Gaussian potential landscape, and irreversible probability flux form a nonequilibrium trinity that captures the essential characteristics common in nonequilibrium fluid systems governed by stochastic Navier-Stokes equations, including turbulent fluids. Therefore, this nonequilibrium trinity is manifested in various facets of nonequilibrium fluid systems. The nonequilibrium stochastic fluid dynamics is driven by both the  gradient of the potential landscape and the irreversible probability flux velocity, together with the solenoidal convective force and stochastic force, as a result of the potential-flux form of the irreversible driving force that arises from detailed balance breaking. We have established a quantitative connection of the energy flux essential to energy cascade in turbulence to the irreversible probability flux and the non-Gaussian potential landscape, both arising directly from detailed balance breaking. We have shown that detailed balance breaking is able to accommodate turbulence energy cascade, through breaking the detailed energy balance between energy injection and energy dissipation at each mode, so that energy can be transferred among different modes to allow the energy cascade in turbulence to take place. We have further demonstrated that a specific mechanism of detailed balance breaking, which is created by the conditions of large-scale forcing and high Reynolds number that separate the energy injection and energy dissipation scales far away from each other, acts as a power source that drives the energy cascade process through the intermediate scales, manifesting  a constant cumulative energy flux throughout the inertial subrange in fully developed turbulence. Furthermore, with the energy flux as a bridge of connection, the four-fifths law in fully developed turbulence has been derived, which we have shown to be a mirror of the two consequential aspects of detailed balance breaking, where the third order structure function reflects the non-Gaussian potential landscape and the energy dissipation rate mirrors the irreversible probability flux. We have also discussed the scaling laws in fully developed turbulence in connection to the nonequilibrium trinity, where we have given explicit mathematical forms relating the non-Gaussian potential landscape and the irreversible probability flux to the statistical observables of the fluid system.

In summary, our potential landscape and flux field theory for fluid systems has revealed that the nonequilibrium trinity (detailed balance breaking, non-Gaussian potential landscape, and irreversible probability flux) captures the nonequilibrium irreversible nature of turbulent fluids and general fluid systems with nonequilibrium steady states, which is manifested in various facets of these systems, including the nonequilibrium stochastic fluid dynamics, the energy flux associated with turbulence energy cascade, the four-fifths law in fully developed turbulence, and possibly the scaling laws.

This work lays the foundation for the further development of a more complete conceptual and mathematical theory for turbulence and nonequilibrium fluid systems in general. In future work we intend to investigate more specific connections of our theory to the scaling laws and intermittency in turbulence, as well as the implications of our framework on other important topics in fluid dynamics, including the transition from laminar to turbulent flow and the nonequilibrium thermodynamics for turbulence with the quantification of detailed balance breaking. Moreover, we speculate that a generalized form of the concept of nonequilibrium trinity born in this work has significant ramifications beyond fluid dynamics and can be applied to more general nonequilibrium stochastic dynamical systems. This line of work has already started and will be further explored in the future.

\section*{Acknowledgements}
WW, FZ and JW were supported by the National Natural Science Foundation of China (91430217). JW was supported by the National Science Foundation (NSF-PHYS-76066).

\section*{Appendix. Functional divergence of the solenoidal convective force}\label{FunctionalDivergence}

With the aid of the dictionary in Section \ref{wavenumberREP}, we calculate the functional divergence of the solenoidal convective force, $\int d\mathbf{x}\,\delta_{\mathbf{u}(\mathbf{x})}\cdot[\mathbf{\Pi}^{s}(\nabla)\cdot\left(-\mathbf{u}\cdot\nabla\mathbf{u}\right)]$, in the wavevector space:
\begin{equation}\label{FDCF}
\begin{split}
&\int d\mathbf{x}\,\delta_{\mathbf{u}(\mathbf{x})}\cdot[\mathbf{\Pi}^{s}(\nabla)\cdot\left(-\mathbf{u}\cdot\nabla\mathbf{u}\right)]\\
=&\sum_{\mathbf{k}}\nabla_{\mathbf{u}(\mathbf{k})}\cdot\left[\mathbf{\Pi}^s(\mathbf{k})\cdot\left(-i\mathbf{k}\cdot\sum_{\mathbf{k}'}\mathbf{u}(\mathbf{k}-\mathbf{k}')\mathbf{u}(\mathbf{k}')\right)\right]\\
=&\sum_{\mathbf{k}}\sum_{ijl}-i k_l\Pi^s_{ij}(\mathbf{k})\DF{\partial}{\partial u_i(\mathbf{k})}\sum_{\mathbf{k}'}u_l(\mathbf{k}-\mathbf{k}')u_j(\mathbf{k}')\\
=&\sum_{\mathbf{k}}\sum_{ijl}-i k_l\Pi^s_{ij}(\mathbf{k})[\Pi^s_{il}(\mathbf{k})u_j(\mathbf{0})+\Pi^s_{ij}(\mathbf{k})u_l(\mathbf{0})]\\
=&\sum_{\mathbf{k}}-i\mathbf{k}\cdot\mathbf{\Pi}^s(\mathbf{k})\cdot\mathbf{\Pi}^s(\mathbf{k})\cdot\mathbf{u}(\mathbf{0})-i\mathbf{k}\cdot\mathbf{u}(\mathbf{0})\text{Tr}[\mathbf{\Pi}^s(\mathbf{k})\cdot\mathbf{\Pi}^s(\mathbf{k})]\\
=&\sum_{\mathbf{k}}-i\mathbf{k}\cdot\left(\mathbf{I}-\DF{\mathbf{k}\mathbf{k}}{k^2}\right)\cdot\mathbf{u}(\mathbf{0})-i\mathbf{k}\cdot\mathbf{u}(\mathbf{0})\text{Tr}\left[\mathbf{I}-\DF{\mathbf{k}\mathbf{k}}{k^2}\right]\\
=&-2i\sum_{\mathbf{k}}\mathbf{k}\cdot\mathbf{u}(\mathbf{0})=0,
\end{split}
\end{equation}
where we have used the rule of differentiation $\partial u_i(\mathbf{k})/\partial u_j(\mathbf{k'})=\Pi^s_{ij}(\mathbf{k})\delta_{\mathbf{k}\mathbf{k}'}$ for $\mathbf{u}(\mathbf{k})$ under the constraint $\mathbf{k}\cdot\mathbf{u}(\mathbf{k})=0$, as well as the idempotence property of the projection matrix, $\mathbf{\Pi}^s(\mathbf{k})\cdot\mathbf{\Pi}^s(\mathbf{k})=\mathbf{\Pi}^s(\mathbf{k})$. The last step in Eq. (\ref{FDCF}) holds as the velocity fields in the state space have zero total momentum, i.e., $\mathbf{u}(\mathbf{0})\equiv \mathbf{u}(\mathbf{k})|_{\mathbf{k}=\mathbf{0}}=\mathbf{0}$. Otherwise this result would be contingent upon the vanishing of the series $\sum_{\mathbf{k}}\mathbf{k}$, which is conditionally true if summed pair by pair, with each wavevector $\mathbf{k}$ canceled by its opposite.

\bibliographystyle{ieeetr}

\bibliography{Ref}

\end{document}